\documentclass[11pt,a4paper]{article}
\pdfoutput=1
\usepackage{jheppub}
\usepackage{rotating}
\usepackage{array}
\usepackage{amsmath}
\usepackage{slashed,hyperref}
\usepackage[normalem]{ulem}
\usepackage{cancel}

\newenvironment{emergency}[1]{%
  \par
  \setlength{\emergencystretch}{#1}%
}{%
  \par
}

\newcommand{\MA}{m_{A}} 

\usepackage{etoolbox}
\makeatletter
\patchcmd{\@sect}{#8}{\boldmath #8}{}{}
\let\ori@chapter\@chapter
\def\@chapter[#1]#2{\ori@chapter[\boldmath#1]{\boldmath#2}}
\makeatother

\preprint{DESY 14-238\\ \null \hfill SLAC-PUB-16179}

\title{A taste of dark matter: Flavour constraints on pseudoscalar mediators}

\author[a]{Matthew J.~Dolan,}
\author[b]{Felix Kahlhoefer,}
\author[c]{Christopher McCabe}
\author[b]{and \mbox{Kai Schmidt-Hoberg}}

\affiliation[a]{Theory Group, SLAC National Accelerator Laboratory, Menlo Park, California 94025, USA}
\affiliation[b]{DESY, Notkestra\ss e 85, D-22607 Hamburg, Germany}
\affiliation[c]{GRAPPA, University of Amsterdam, Science Park 904, 1098 XH Amsterdam, Netherlands}

\emailAdd{mdolan@slac.stanford.edu}
\emailAdd{felix.kahlhoefer@desy.de}
\emailAdd{c.mccabe@uva.nl}
\emailAdd{kai.schmidt.hoberg@desy.de}

\abstract{Dark matter interacting via the exchange of  a light pseudoscalar can induce observable signals in indirect detection experiments and experience large self-interactions while evading the strong bounds from direct dark matter searches. The pseudoscalar mediator will however induce flavour-changing interactions in the Standard Model, providing a promising alternative way to test these models. We investigate in detail the constraints arising from rare meson decays and fixed target experiments for different coupling structures between the pseudoscalar and Standard Model fermions. The resulting bounds are highly complementary to the information inferred from the dark matter relic density and the constraints from primordial nucleosynthesis. We discuss the implications of our findings for the dark matter self-interaction cross section and the prospects of probing dark matter coupled to a light pseudoscalar with direct or indirect detection experiments. In particular, we find that a pseudoscalar mediator can only explain the  Galactic Centre excess if its mass is above that of the $B$ mesons, and that it is impossible to obtain a sufficiently large direct detection cross section to account for the DAMA modulation.}


\begin{document}
\maketitle

\section{Introduction}

One of the major objectives of modern particle physics is to determine the properties of particle dark matter (DM). A critical part of this task is understanding how the DM particle interacts with Standard Model (SM) states and how these interactions lead to the observed DM relic abundance. Stringent bounds from direct detection experiments~\cite{Agnese:2013jaa,Akerib:2013tjd,Agnese:2014aze,Angloher:2014myn} and recent constraints on the invisible width of the SM Higgs~\cite{Aad:2014iia,Chatrchyan:2014tja} make it increasingly difficult to understand these interactions in terms of the known interactions of the SM. It is therefore a well-motivated possibility to assume that the interactions of DM are mediated by an additional new particle that couples weakly to the visible sector.

Out of the various possibilities for this new particle, pseudoscalar mediators are particularly interesting for several reasons. First of all, with the discovery of the Higgs boson, there is now convincing evidence that fundamental scalars exist in nature. Many extensions of the Higgs sector (such as Two-Higgs Doublet Models~\cite{Gunion:1989we} e.g.\ in the context of Supersymmetry~\cite{Martin:1997ns}) naturally include additional pseudoscalar states, making searches for such particles a well-motivated and timely task. Pseudoscalar mediators are at the same time attractive from a purely phenomenological point of view, because they predict a strong suppression of the event rate in direct detection experiments, thus avoiding some of the most severe constraints on the interactions of DM~\cite{Freytsis:2010ne,Dienes:2013xya}. This consideration provides one reason why DM models with a pseudoscalar mediator (sometimes referred to as `Coy DM'~\cite{Boehm:2014hva, Hektor:2014kga,Arina:2014yna}) have received much attention in the context of explaining the diffuse GeV-energy 
excess of gamma-ray emission from the Galactic Centre observed with the Fermi-LAT instrument~\cite{Goodenough:2009gk,Hooper:2010mq,Hooper:2011ti,Abazajian:2012pn,Gordon:2013vta,Macias:2013vya,Daylan:2014rsa,Zhou:2014lva,Calore:2014xka,Agrawal:2014oha,Calore:2014nla} (see also~\cite{Alves:2014yha,Berlin:2014tja,Izaguirre:2014vva,Cerdeno:2014cda,Ipek:2014gua,Abdullah:2014lla,Martin:2014sxa,Berlin:2014pya,Han:2014nba,Cheung:2014lqa,Huang:2014cla,Ghorbani:2014qpa,Cahill-Rowley:2014ora,Guo:2014gra,Cao:2014efa,Freytsis:2014sua,Buckley:2014fba} for further model-building involving pseudoscalars).

An interesting possibility is that the pseudoscalar mediator mass is sub-GeV and is therefore light compared to Large Hadron Collider (LHC) energies. Such a set-up was for example advocated in the context of asymmetric DM~\cite{MarchRussell:2012hi} in order to avoid constraints from LHC monojet searches. At the same time, for light mediators the suppression of DM scattering in direct detection experiments is significantly reduced, so that it might be possible to obtain observable signals in present or future experiments~\cite{Boehm:2014hva} and possibly even explain the DAMA modulation signal~\cite{Arina:2014yna}. Moreover, if the pseudoscalar has a mass smaller than the DM mass, DM can annihilate into pairs of pseudoscalars, which subsequently decay into SM particles. This way, DM can obtain the required relic density even if the interactions between the pseudoscalar and SM particles are constrained to be rather weak.

Furthermore, the presence of a light mediator offers the interesting possibility to obtain large self-interactions in the dark sector. Such self-interactions have received much interest in the context of explaining the discrepancies between $N$-body simulations of collisionless cold DM and the observations of small-scale structures~\cite{Spergel:1999mh}. The central idea is that DM scattering can reduce both the central densities of DM halos and the size and number of Milky Way satellites. For velocity-independent self-interactions, however, there are important constraints from colliding galaxy clusters~\cite{Markevitch:2003at, Kahlhoefer:2013dca}, sub-halo evaporation~\cite{Gnedin:2000ea, Rocha:2012jg} and elliptical galaxies~\cite{Feng:2009mn,Buckley:2009in,Peter:2012jh}. These constraints can be evaded if self-interactions are suppressed for large relative velocities~\cite{Vogelsberger:2012ku,Zavala:2012us}. Such a velocity dependence can for example arise from a light mediator inducing a Yukawa potential~\cite{Feng:2009mn,Buckley:2009in,Loeb:2010gj,Aarssen:2012fx,Tulin:2013teo}. In addition, for a light mediator DM self-interactions may be additionally enhanced at low velocities by non-perturbative effects corresponding to the (temporary) formation of DM bound states~\cite{Buckley:2009in,Tulin:2013teo}.

However, there are stringent constraints on new light states coupling to SM particles. Of particular interest in this context are experimental searches for rare meson decays, because the presence of a new light pseudoscalar mediator $A$ will in general lead to a large enhancement in the rates of  flavour-changing processes such as $K \rightarrow \pi \, A$ or \mbox{$B \rightarrow K \, A$}~\cite{Hiller:2004ii, Andreas:2010ms}. Flavour observables therefore provide a unique opportunity to constrain the interactions of the dark sector with SM particles via a light mediator. While similar constraints have been studied for light vector mediators~\cite{Fayet:2006sp,Fayet:2007ua}, many other cases remain relatively unexplored, although there has been some interest in flavoured DM~\cite{Batell:2011tc,Agrawal:2014una}.

The topic of this paper is to explore in detail various constraints on the SM couplings of a new light pseudoscalar particle and infer the implications for the interactions between DM and SM particles. We will show that flavour constraints completely rule out the possibility to obtain an observable DM signal in direct detection experiments from scattering via the exchange of light pseudoscalars. Similarly, indirect detection signals can only be sizeable if the mediator mass is so large that it cannot be produced on-shell in the decay of $B$ mesons. In particular, it appears impossible to obtain both indirect detection signals and large DM self-interactions from the same pseudoscalar mediator.

Our paper is structured as follows. Sec.~\ref{sec:notation} contains the general set-up for our study and discusses the various ways in which flavour-changing processes can induce rare meson decays. In Sec.~\ref{sec:constraints} we use various experimental results to constrain a light pseudoscalar coupling to the SM. The resulting bounds are presented in Sec.~\ref{sec:results}. The focus of Sec.~\ref{sec:dark} is the connection to the dark sector and the resulting cosmology. Sec.~\ref{sec:detection} considers implications for possible DM signals, in particular concerning the interpretation of the DAMA annual modulation and the Galactic Centre excess. Various details of our calculations are provided in Appendices~\ref{sec:mesondecays}--\ref{sec:direct}.

\section{General set-up and conventions}
\label{sec:notation}

We are interested in the interactions of a light real pseudoscalar~$A$ with the DM particle~$\chi$, which we take to be a Dirac fermion, and with SM fermions. Neglecting $CP$-violating couplings, we write the DM-pseudoscalar coupling as
\begin{equation}
\mathcal{L}_\text{DM} = i \, g_\chi \, A \, \bar{\chi} \gamma^5 \chi \; ,
\label{eq:ADM}
\end{equation}
where we introduce a factor of $i$ so that the coupling $g_{\chi}$ is real.
For the interactions between $A$ and SM particles we write in general
\begin{equation}
\mathcal{L}_\text{SM} = \sum_{f = q,\ell,\nu} i \, g_f \, A \, \bar{f} \gamma^5 f \;,
\label{eq:AgSM}
\end{equation}
where $g_f$ is the effective coupling and $f$ refers to all SM quarks $q = \{u, d, s, c, b, t\}$, all charged SM leptons $\ell = \{e, \mu, \tau\}$ and all SM neutrinos~$\nu$.

In the following we will consider different cases for the coupling structure with the charged SM fermions; unless explicitly stated otherwise, we assume that $g_{\nu} \simeq 0$. 
\begin{itemize}

\item \emph{Yukawa-like} couplings: Arguably the most natural case is the one where the couplings to all charged SM fermions are proportional to the SM Yukawa couplings:
\begin{equation}
\mathcal{L}^{(Y)}_\text{SM} = i  \, g_Y \sum_{f = q,\ell} \frac{\sqrt{2} \, m_f}{v} A \, \bar{f} \gamma^5 f \; ,
\label{eq:LYukawa}
\end{equation}
where $m_f$ is the fermion mass and $v\simeq246\:\text{GeV}$ is the vacuum expectation value (vev) of the SM Higgs field. In this case $g_f = \sqrt{2} \, g_Y \, m_f / v$. This coupling structure is expected for pseudoscalars arising from an extended Higgs sector, because the couplings of the pseudoscalar to SM fermions arise from mixing with the SM Higgs boson and are therefore automatically proportional to the SM Yukawa couplings. Such extended Higgs sectors often contain additional $CP$-even and charged Higgs particles as well. Our results should apply in such theories as long as the effects of these particles decouple.

\item \emph{Quark Yukawa-like} couplings: As we shall see many experimental constraints assume that the pseudoscalar can decay into charged leptons. These constraints can be significantly relaxed~-- or even removed altogether~-- if the pseudoscalar is assumed to couple only to quarks i.e.\ $g_f = \sqrt{2} \, g_{Yq} \, m_f / v$ for $f=q$ and $g_f=0$ otherwise. Such a coupling structure can be expected for axion-like particles with a shift symmetry, which would have a coupling proportional to $e'_f \partial^\mu a \, \bar{f} \gamma_\mu \gamma^5 f$, where $e'_f$ is the charge of the fermion under the new global $U(1)$ symmetry. This coupling structure leads to Yukawa-like couplings after integrating by parts and using the equations of motion. If $e'_f = 0$ for leptons, such a particle would couple only to quarks (like the QCD-axion).

\item \emph{Quark universal} couplings: The assumption of Yukawa-like couplings for the pseudoscalar is consistent with the hypothesis of minimal flavour violation (MFV)~\cite{D'Ambrosio:2002ex}. Consequently, one would expect other (non-MFV) coupling structures to lead to significantly stronger experimental bounds. Nevertheless, it is interesting from the phenomenological point of view to consider the case that the pseudoscalar has universal couplings to all quarks and no couplings to leptons:
\begin{equation}
\mathcal{L}^{(q)}_\text{SM} = i \, g_q \sum_{q} A \, \bar{q} \gamma^5 q \; .
\label{eq:ASM}
\end{equation}
Interactions of this type have been proposed as an explanation for both the Fermi Galactic Centre excess and the DAMA signal simultaneously~\cite{Arina:2014yna}. 

\item \emph{Quark third generation} couplings: Finally, we will also comment on the case where the pseudoscalar couples only to the third family of quarks, assuming equal couplings ($g_{Q}$) to $b$ and $t$.

\end{itemize}

Experimental searches (to be discussed in the following section) typically look for rare decays of the form $K \rightarrow \pi + X$ or $B \rightarrow K + X$ where $X$ is a set of (potentially invisible) SM particles. In the context of our model, these processes can be decomposed into the production of a pseudoscalar in a flavour changing process, such as $K \rightarrow \pi \, A$ followed by the decay of $A$ into SM particles. We will therefore now discuss the theoretical predictions for both contributions.

\subsection{Effective flavour-changing interactions}
\label{sec:penguin}

\begin{figure}
\centering
\includegraphics[width=0.8\textwidth]{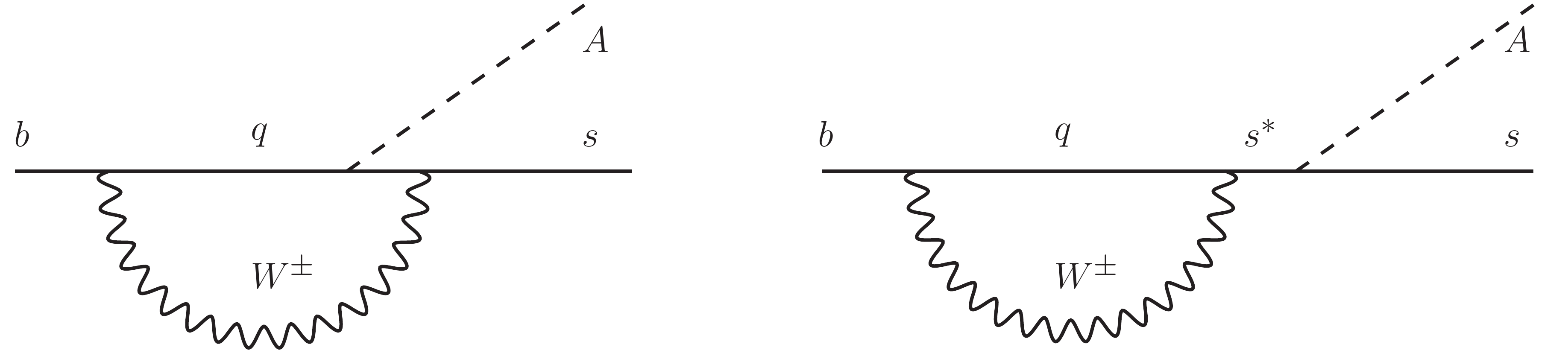}
\caption{Flavour-changing transitions such as $b \to s A$ (and also $s\to d A$ after relabelling the external lines) are generated by diagrams with heavy quarks and $W^{\pm}$-bosons.}
\label{fig:FlavLoop}
\end{figure}

Although the tree-level interactions of $A$ are assumed to be flavour-diagonal, flavour-changing neutral currents (FCNCs) arise at the one-loop level from diagrams with heavy quarks and $W$-bosons, such as those depicted in Fig.~\ref{fig:FlavLoop}.  We will be interested in the transitions $b \rightarrow s \, A$ and $s \rightarrow d \,  A$. The relevant flavour-changing terms are typically parameterised in the form~\cite{Deshpande:2005mb}:
\begin{equation}
\mathcal{L}_\text{FCNC} \supset A \, \bar{d} (h^S_{ds} + h^P_{ds} \gamma^5) s + A \, \bar{s} (h^S_{sb} + h^P_{sb} \gamma^5) b + \text{h.c.}
\end{equation}
where the coefficients $h^{S,P}_{qq'}$ are typically complex, so we do not include an extra factor~$i$ in front of the pseudoscalar coupling. To connect to various results in the literature, we note that this expression can also be written as
\begin{equation}
\mathcal{L}_\text{FCNC} \supset h^R_{ds} A \, \bar{d}_L s_R + h^L_{ds} A \, \bar{d}_R s_L  + 
h^R_{sb} A \, \bar{s}_L b_R + h^L_{sb} A \, \bar{s}_R b_L 
+ \text{h.c.}
\end{equation}
where $q_{R,L} = \frac{1}{2}(1\pm\gamma^5) q$ and the couplings are related by
\begin{equation}
h^S_{qq'} = (h^R_{qq'} + h^L_{qq'})/2 \qquad h^P_{qq'} = (h^R_{qq'} - h^L_{qq'})/2 \; .
\end{equation}

In order to calculate the loop-induced flavour-changing couplings, we first of all need to determine the quark field renormalisation constants. This can be done by calculating the loop-induced contribution to the quark two-point function and fixing the counterterms in such a way that all flavour changing transitions $q \rightarrow q'$ vanish for on-shell quarks~\cite{Hall:1981bc, Bobeth:2001sq}. 
Since we assume that the pseudoscalar has no flavour-changing interactions at tree-level, this requirement then fully determines the counterterm for the three-point vertex, which contributes to the processes $b \rightarrow s \, A$ and $s \rightarrow d \, A$.\footnote{We note that it is also possible to perform the same calculation by explicitly including self-energy diagrams for the external quark lines~\cite{Logan:2000iv}. We have checked that both approaches yield the same result.} 

Using FeynArts FormCalc and LoopTools~\cite{Hahn:1998yk,Hahn:2000kx}, we find that the the one-loop contribution to flavour-changing transitions is in general divergent. In dimensional regularisation we obtain a pole in $\epsilon = (4 - d)/2$ of the form
\begin{equation}
 h^R_{sb} = \sum_{q=u,c,t}
\frac{\alpha \, \left[3 \, m_s \, m_b \, m_q^2 \, g_s - (m_s^2 + 2 m_b^2) \, m_q^2 \, g_b + 2 \, m_b \, (m_s^2 - m_b^2) \, m_q \, g_q \right]}{16 \pi \, m_W^2 \, \sin(\theta_W)^2 \, (m_b^2 - m_s^2)} \, V_{qb} V_{qs}^\ast \times \frac{1}{\epsilon} \; ,
\label{eq:oneloop}
\end{equation}
where $m_W$ is the $W$-boson mass, $\alpha \equiv e^2 / (4 \pi)$, $\theta_W$ is the Weinberg angle and $V$ is the CKM matrix. The corresponding expression for $h^L_{sb}$ (up to an overall sign) is obtained by exchanging $s$ and $b$.

Since we have assumed flavour-diagonal couplings at tree-level, there is no freedom in the vertex counter-term to cancel this divergence. In other words, in order to render the theory renormalisable, one would need to include flavour-changing interactions already at tree-level, so that an additional counter-term can be introduced that cancels the divergence of the three-point functions. Clearly, such a model would be extremely tightly constrained by experiments and would therefore typically not be phenomenologically viable.\footnote{We thank Ulrich Haisch for raising this point.} The presence of such a divergence is surprising, given that the interaction in equation (\ref{eq:AgSM}) appears renormalisable. However, additional new states and interactions must be present in order to couple SM quarks to a pseudoscalar singlet in a gauge-invariant way, so that the couplings between the pseudoscalar and quarks actually arise from higher-dimensional operators. The divergence of the one-loop diagrams reflects the dependence of our results on the suppression scale $\Lambda$ of these operators, which corresponds to the scale where aditional new states appear.



We should therefore think of equation (\ref{eq:AgSM}) as an effective theory that arises in the low-energy limit of a more complete theory and assume that additional new physics at the scale $\Lambda$ cancels the divergences present in the effective theory. This new physics will in general induce additional higher-dimensional operators with flavour-changing interactions, but we assume that these effects are small compared to the effects that we consider here and that the coupling structure is not significantly changed by renormalisation group evolution over the energy range that we consider. Clearly, these assumptions are optimistic, but we will show that due to the loop-induced flavour-changing couplings models with general coupling structure are nevertheless tightly constrained. 
We note, however, that within a specific UV completion there can also been cancellations between different contributions, which can potentially weaken constraints considerably, as has been observed for example in two-Higgs-doublet models~\cite{Freytsis:2009ct}. 

Interpreting our model as an effective theory below some new physics scale $\Lambda$ corresponds to making the replacement $1/\epsilon + \log(\mu^2/m^2) \rightarrow \log(\Lambda^2 / m^2)$, where $m = m_t$ is the relevant mass scale for the process under consideration~\cite{Batell:2009jf}. We then obtain the following expressions in the limit $m_t \gg m_b \gg m_s$:\footnote{We note that our result for Yukawa-like couplings differs by a factor of 4 from the one obtained for a pseudoscalar with derivative coupling of the form $\partial^\mu a \, \bar{f} \gamma_\mu \gamma^5 f$~\cite{Batell:2009jf}. The reason is that these two interactions are not equivalent at the one-loop level unless one also includes additional dimension-5 operators involving a pseudoscalar and a Higgs boson. We have checked that, when including these additional interactions, we reproduce the results from~\cite{Batell:2009jf}.}

\begin{align}
 & \text{Yukawa-like couplings: } & & h^R_{sb} = - \frac{\alpha \, g_Y \, m_b \, m_t^2}{2 \sqrt{2} \pi \, m_W^2 \, \sin(\theta_W)^2 \, v} \, V_{tb} V_{ts}^\ast \, \log\left(\frac{\Lambda^2}{m_t^2}\right) \; , \\
 & \text{Universal couplings: } & & h^R_{sb} = - \frac{\alpha \, g_q \, m_t^2}{8 \pi \, m_W^2 \, \sin(\theta_W)^2} \, V_{tb} V_{ts}^\ast \, \log\left(\frac{\Lambda^2}{m_t^2}\right) \; . 
\end{align}
The corresponding expressions for $h^L_{sb}$ (up to an overall sign) are again obtained by replacing $m_b$ by $m_s$. The results for $h^R_{ds}$ and $h^L_{ds}$ are completely analogous, except that in this case a second term with $t$ replaced by $c$ gives a relevant contribution. Using the full expression, substituting the measured masses, couplings and mixing angles, and taking $\Lambda = 1\:\text{TeV}$, we find
\begin{equation}
\begin{aligned}
h^S_{ds} & = \left(3.5 \cdot 10^{-9} + 1.5 \cdot 10^{-9} i\right) g_Y \, , & h^P_{ds} & = \left(3.9 \cdot 10^{-9} + 1.7\cdot 10^{-9} i\right) g_Y\,, \\
h^S_{sb} & = \left(2.3 \cdot 10^{-5} - 4.2 \cdot 10^{-7} i \right) g_Y \, , & h^P_{sb} & = \left(2.3 \cdot 10^{-5} - 4.4 \cdot 10^{-7} i \right) g_Y \; .
\label{eq:hsyuk}
\end{aligned}
\end{equation}
for Yukawa-like couplings,
\begin{equation}
\begin{aligned}
h^S_{ds} & \approx \left(4.6 \cdot 10^{-6} + 2.0 \cdot 10^{-6} i \right) g_q \, , & h^P_{ds} & \approx \left(1.7 \cdot 10^{-6} + 7.3 \cdot 10^{-7} i\right) g_q\,, \\
h^S_{sb} & \approx \left(6.3 \cdot 10^{-4} - 1.2 \cdot 10^{-5} i \right) g_q \, , & h^P_{sb} & \approx \left(6.9 \cdot 10^{-4} - 1.3 \cdot 10^{-5} i \right) g_q\,,
\label{eq:hsuni}
\end{aligned}
\end{equation}
for quark-universal couplings, and 
\begin{equation}
\begin{aligned}
h^S_{ds} & \approx \left(1.7 \cdot 10^{-9} + 7.6 \cdot 10^{-10} i\right) g_Q \, , & h^P_{ds} & \approx \left(1.9 \cdot 10^{-9} + 8.4 \cdot 10^{-10} i\right) g_Q\,, \\
h^S_{sb} & \approx \left(4.3 \cdot 10^{-4} - 8.0 \cdot 10^{-6} i \right) g_Q \, , & h^P_{sb} & \approx \left(4.5 \cdot 10^{-4} - 8.5 \cdot 10^{-6} i \right) g_Q \,,
\label{eq:hsq3}
\end{aligned}
\end{equation}
for couplings only to the third generation.

Using these results, we can now calculate the partial decay widths for various flavour changing meson decays. The relevant formulae are provided in Appendix~\ref{sec:mesondecays}. To obtain the branching ratios (BR) into SM final states, we need to divide these partial widths by the total meson decay width and multiply with the branching ratio of the pseudoscalar into the appropriate final state. For example,
\begin{equation}
\begin{split}
 \mathrm{BR}(K^+ \rightarrow \pi^+ \gamma \gamma) &= \frac{\Gamma(K^+ \rightarrow \pi^+ \gamma \gamma)}{\Gamma_{K^+}} \\
& = \frac{\Gamma(K^+ \rightarrow \pi^+ A) \times \text{BR}(A \rightarrow \gamma \gamma) }{\Gamma_{K^+}} + \mathrm{BR}(K^+ \rightarrow \pi^+ \gamma \gamma)_\text{SM} \; ,
\end{split}
\end{equation}
where we have made use of the narrow width approximation in the second line. We will therefore now discuss the decays of the pseudoscalar mediator.

\subsection{Pseudoscalar decays}

\begin{figure}[!t]
\centering
\includegraphics[width=0.46\textwidth]{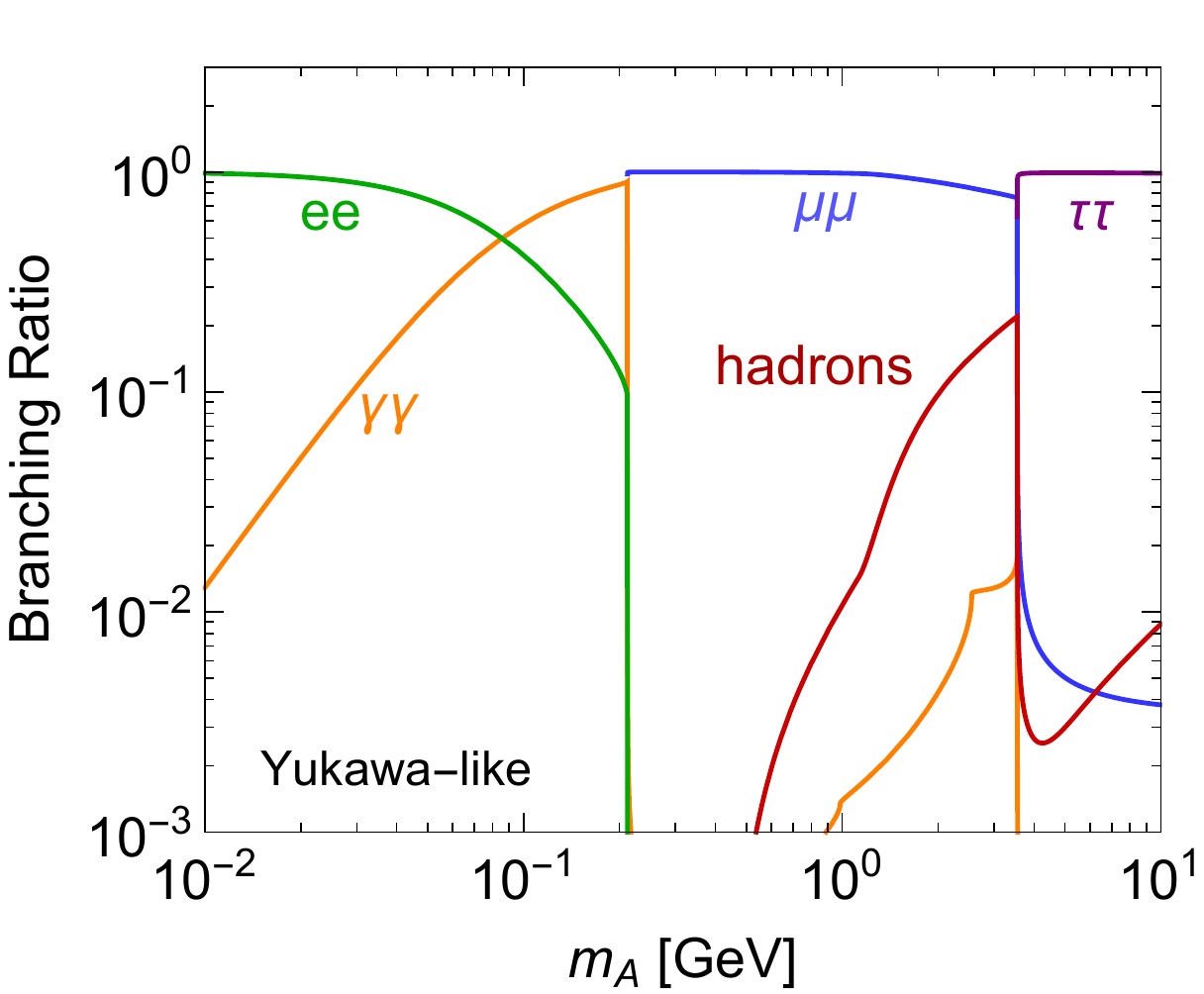}
\quad
\includegraphics[width=0.46\textwidth]{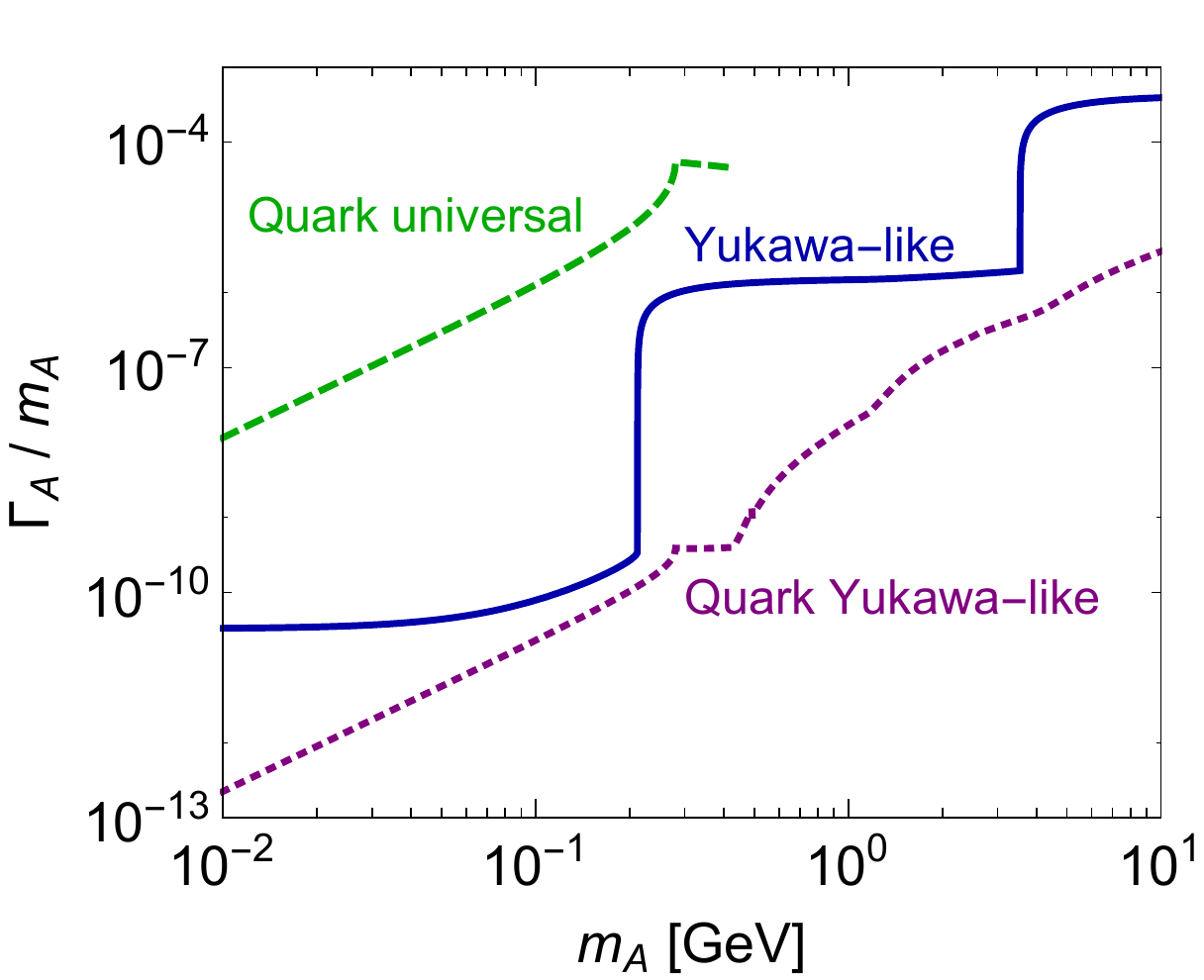}
\caption{Left: Branching ratios of the pseudoscalar for Yukawa-like couplings according to the perturbative spectator model (see Appendix~\ref{sec:decays} for further details). Right: Total width in units of the pseudoscalar mass for Yukawa-like couplings to all fermions (with $g_Y = 10$), Yukawa-like couplings only to quarks (with $g_{Yq} = 10$) and universal quark couplings (with $g_q = 0.1$). For the latter case, we do not attempt to estimate the total width once hadronic decay channels open up (when $m_A>3 m_{\pi}$). The narrow width approximation is valid for all parameter values that we consider. }
\label{fig:BRY}	
\end{figure}

In principle, the pseudoscalar $A$ can decay into SM states which are kinematically accessible, such as leptons, photons and hadrons. Our results for the various branching ratios in the case of Yukawa-like couplings~-- and the total width of the pseudoscalar~-- are shown in Fig.~\ref{fig:BRY}.\footnote{These results are in agreement with~\cite{Hiller:2004ii,McKeen:2008gd}, but disagree with the branching ratios presented in~\cite{Domingo:2008rr,Dermisek:2010mg}, where it has been neglected that the decay $A \to g g$ is kinematically forbidden for $m_A <2 \,m_\pi$ and that the decay of $A$ into two pseudoscalar mesons violates $CP$.} We provide details of our calculations of these quantities in Appendix~\ref{sec:decays}, being careful to provide a self-consistent estimation of the hadronic decay width. In particular, we point out that for all couplings and masses we consider, the pseudoscalar state is a narrow resonance (i.e.\ $\Gamma_A / m_A \ll 1$) so that the narrow width approximation introduced above is valid.

For \mbox{$m_A < 2 \, m_\pi$}, hadronic decays are kinematically forbidden. For $\MA > 2 \, m_\pi$, the decays $A \to \pi \pi$ and $A \to \pi \pi \gamma$ are kinematically allowed, but forbidden by $CP$. Consequently, sizeable hadronic decay channels only open up for $\MA > 3 \, m_\pi$ when $A \to \pi \pi \pi$ becomes possible. While the decay $A \to \pi \gamma \gamma$ is allowed for $m_A > m_\pi$, it is always suppressed compared to $A \to\gamma \gamma$ because of the smaller available phase-space. 

For $\MA \leq 3 \, m_{\pi}$ the pseudoscalar will therefore dominantly decay into pairs of leptons or photons. In the case of Yukawa-like couplings, decays into electrons dominate for pseudoscalar masses of about $(1\text{--}100)\:\text{MeV}$. As the pseudoscalar mass approaches the $\mu^+ \mu^-$ threshold, the branching ratio for $A \to \gamma \gamma$ becomes sizeable, while above the threshold the $\mu^+ \mu^-$ decay channel is the most important one. If couplings to leptons are absent, the only allowed two-body decay for $\MA < 3 \, m_\pi$ is $A \to \gamma \gamma$. Since loops with light quarks give an important contribution to this process, the total decay width depends on the precise matching of the light quark masses. We discuss this complication in more detail in Appendix~\ref{sec:decays}.

Predicting the hadronic decay width of $A$ for $\MA > 3 \, m_\pi$ is a notoriously difficult problem. Indeed, even for the SM-like Higgs with a mass $m_h \sim (0.5\text{--}1)\:\text{GeV}$ there is an unresolved debate about the ratio of $\Gamma(h \rightarrow \pi\pi)$ to $\Gamma(h \rightarrow \mu^+\mu^-)$~\cite{Donoghue:1990xh} (see also~\cite{Clarke:2013aya} for a recent review). To obtain an approximate expression for the case of Yukawa-like couplings to quarks, we employ the perturbative spectator model~\cite{Gunion:1989we}, even though this treatment is not expected to be very accurate for $\MA \lesssim 1 \: \text{GeV}$. Given that we find the partial decay width for hadronic final states to be significantly smaller than the corresponding width for leptons (due to the phase-space suppression for three-body final states), the total width and the leptonic branching fraction do not suffer significantly from these uncertainties and our results based on these quantities are robust. Still, it is important to bear in mind that in the presence of resonance effects, using the perturbative spectator model might significantly underestimate the branching fraction into hadrons, so that bounds based on leptonic decays of $A$ might be significantly suppressed for particular values of the pseudoscalar mass.

For the case of quark universal couplings we assume that the branching ratio to hadrons is 100\% for pseudoscalar masses above $3 \, m_{\pi}$. This is a reasonable approximation since the typical phase-space suppression for three-body decays is only $1/(32\pi^2)$, compared to a suppression factor $\alpha^2 / (16 \pi^2)$ for loop-induced decays into photons.
Nevertheless, since the tree-level couplings to light quarks give a significant contribution to hadronic decays, it is very difficult to reliably estimate the total decay width of the pseudoscalar. Fortunately, our results do not depend sensitively on this quantity since the total width is neither so small that one could obtain displaced vertices, nor so large that the narrow width approximation becomes invalid.

\section{Experimental constraints}
\label{sec:constraints}

To constrain the interactions of the pseudoscalar mediator with SM particles, we study a large set of flavour constraints and other experiments searching for rare processes. We provide a concise list of the searches we consider in Tab.~\ref{tab:PseudoSearches}. We will now discuss in more detail those measurements which are responsible for the best limits in pseudoscalar parameter space.

\begin{table}[t]
\setlength{\tabcolsep}{5pt}
\renewcommand{\arraystretch}{1.3}
\center
\begin{tabular}{ccccc} 
\hline \hline 
\textbf{Channel} & \textbf{Experiment} & \textbf{Mass range [MeV]} & \textbf{Ref.} & \textbf{Relevant for} \\[5pt]
\hline
$K^+ \rightarrow \pi^+ + \text{inv}$ & E949 & 0--110 & \cite{Anisimovsky:2004hr} & Long lifetime$^\ast$
\\
& & 150--260 & \cite{Artamonov:2009sz} & Long lifetime$^\ast$
\\
& E787 & 0--110 \& 150--260 & \cite{Adler:2004hp} & Long lifetime
\\
$K^+ \rightarrow \pi^+ \pi^0 \rightarrow \pi^+ \nu \bar{\nu}$ & E949 & 130--140 & \cite{Artamonov:2005cu} & Long lifetime$^\ast$
\\
\hline
$K^+ \rightarrow \pi^+ \, e^+ e^-$ & NA48/2 & 140--350 & \cite{Batley:2009aa} & Leptonic decays 
\\
$K_L \rightarrow \pi^0 \, e^+ e^-$ & KTeV/E799 & 140--350 & \cite{AlaviHarati:2003mr} & Leptonic decays$^\ast$
\\
$K^+ \rightarrow \pi^+ \, \mu^+ \mu^-$ & NA48/2 & 210--350 & \cite{Batley:2011zz} & Leptonic decays
\\
$K_L \rightarrow \pi^0 \, \mu^+ \mu^-$ & KTeV/E799 & 210--350 & \cite{AlaviHarati:2000hs} & Leptonic decays$^\ast$
\\
\hline
$K_L \rightarrow \pi^0 \, \gamma \gamma$ & KTeV & 40--100 \& 160--350 & \cite{Abouzaid:2008xm} & Photonic decays$^\ast$
\\
$K_L \rightarrow \pi^0 \pi^0 \rightarrow 4\gamma$ & KTeV & 130--140 & \cite{Alexopoulos:2004sx} & Photonic decays$^\ast$
\\
\hline
$K^+ \rightarrow \pi^+ \, A$ & $K_{\mu2}$ & 10--130 \& 140--300 & \cite{Yamazaki:1984vg} & All decay modes$^\ast$
\\
\hline
$B^0 \rightarrow K_S^0 + \text{inv}$ & CLEO & 0--1100 & \cite{Ammar:2001gi} & Long lifetime$^\ast$
\\
\hline
$B \rightarrow K \, \ell^+ \ell^-$ & BaBar & 30--3000 & \cite{Aubert:2003cm} & Leptonic decays
\\
& BELLE & 140--3000 & \cite{Wei:2009zv} & Leptonic decays
\\
& LHCb & 220--4690 & \cite{Aaij:2012vr} & Leptonic decays$^\ast$
\\
$B \rightarrow X_s \, \mu^+ \mu^-$ & BELLE & 210--3000 & \cite{Brodzicka:2012jm} & Leptonic decays
\\
\hline
$b \rightarrow s \, g$ & CLEO & $m_A < m_B - m_K$ & \cite{Coan:1997ye} & Hadronic decays$^\ast$
\\
\hline
$B_s \rightarrow \mu^+ \mu^- $ & LHCb/CMS & all masses & \cite{Aaij:2013aka,Chatrchyan:2013bka} & Lepton couplings$^\ast$
\\
\hline
$\Upsilon \rightarrow \gamma \, \tau^+ \tau^-$ & BaBar & 3500--9200 & \cite{Lees:2012te} & Leptonic decays$^\ast$
\\
$\Upsilon \rightarrow \gamma \, \mu^+ \mu^-$ & BaBar & 212--9200 & \cite{Lees:2012iw} & Leptonic decays$^\ast$
\\
$\Upsilon \rightarrow \gamma + \text{hadrons}$ & BaBar & 300--7000 & \cite{Lees:2011wb} & Hadronic decays$^\ast$
\\
\hline
$K,\,B \rightarrow A+X$ & CHARM & 0--4000 & \cite{Bergsma:1985qz} & Leptonic and
\\
& & & & photonic decays$^\ast$
\\
\hline \hline
\end{tabular}
\caption{Overview of experimental searches probing rare decays induced by the pseudoscalar $A$. The mass-range column indicate the range of $A$ masses constrained by the searches (rather than the cuts imposed by the experiments). The final column indicates which pseudoscalar decay channels these searches are relevant for. The superscript $^\ast$ indicates experimental results that we use for our final analysis.}
\label{tab:PseudoSearches}
\end{table}

\subsection{Flavour constraints}

The simplest constraints can be obtained by requiring that the partial widths $\Gamma(K \rightarrow \pi A)$ and $\Gamma(B \rightarrow X_s A)$ (where in the inclusive decay $X_s$ is any strange meson) do not exceed the experimentally measured total width of the $K$ and $B$ mesons, or in other words, by imposing $\Gamma(K \rightarrow \pi A)/\Gamma_K^{\rm{exp}}<1$ and $\Gamma(B \rightarrow X_s A)/\Gamma_{B_s}^{\rm{exp}}<1$. 
In principle, one could obtain stronger constraints by taking into account the partial widths of well-known (or well-measured) decay channels. For kaons, for example, the partial widths for leptonic decays can be accurately calculated and could therefore be subtracted when setting a bound on the pseudoscalar couplings. For $B^0$ mesons, on the other hand, the PDG quotes $\text{BR}(B^0 \rightarrow c \bar{c} + X) = 119 \pm 6 \%$~\cite{Agashe:2014kda}, which would imply that $\text{BR}(B \rightarrow X_s A)$ must be significantly smaller than unity. We do not attempt to include these contributions and therefore set a very conservative bound.

\subsection*{\boldmath{$K^+ \rightarrow \pi^+ + \text{inv}$}}

The pseudoscalar mediator introduced in Sec.~\ref{sec:notation} does not have any invisible decay modes (except for negligible loop-induced decays to neutrinos). However, as we discuss in more detail in Appendix~\ref{sec:lifetime}, its lifetime can be so long that it will leave the detector without decaying at all. If such a pseudoscalar is produced in a kaon decay, the resulting experimental signature (a single pion in association with missing energy) is identical to the one from the rare SM process $\text{BR}(K \rightarrow \pi^{+} \nu \bar{\nu})$, which has recently received much interest. Indeed, the E949 collaboration~\cite{Anisimovsky:2004hr, Artamonov:2009sz} has measured this branching ratio to be in good agreement with the theoretical SM prediction. 

To suppress backgrounds, however, these searches restrict the momentum range for the pion in the final state. Most searches have focussed on the range \mbox{$211\:\text{MeV} < p_\pi < 229\:\text{MeV}$}, corresponding to the maximum pion momentum allowed by kinematics for an off-shell mediator~\cite{Anisimovsky:2004hr}. For an on-shell mediator, the pion momentum will be reduced as $m_A$ increases. The search window above therefore ceases to be constraining for $m_A > 110\:\text{MeV}$. An additional search for low-momentum pions has been performed in Ref.~\cite{Artamonov:2009sz}, covering the momentum range $140\:\text{MeV} < p_\pi < 199\:\text{MeV}$, which corresponds to $150\:\text{MeV}
 < m_A < 260\:\text{MeV}$. In both search regions, the experiment places an upper bound of 
\begin{equation}
\text{BR}(K^+ \rightarrow \pi^+ + \text{inv}) \lesssim 4 \cdot 10^{-10} \; .
\label{eq:invwidth}
\end{equation}

The overwhelming SM background for $m_A \sim m_{\pi}$ makes it impossible to search in the intermediate region. Nevertheless, E949 has also measured the branching ratio for the process $K^+ \rightarrow \pi^+ \pi^0 \rightarrow \pi^+ \nu \bar{\nu}$, finding  
\begin{equation}
\text{BR}(K^+ \rightarrow \pi^+ \pi^0 \rightarrow \pi^+ \nu \bar{\nu}) \approx 6 \cdot 10^{-8} \; .
\end{equation}
For the mass range $110\:\text{MeV} < m_A < 150\:\text{MeV}$ we therefore require that the branching ratio for $K^+ \rightarrow \pi^+ A$ multiplied with the probability that $A$ escapes from the detector does not exceed this value. In practice, we require that the pseudoscalar travels at least 4 metres in the laboratory frame before decaying.

\subsection*{\boldmath{$K_L \rightarrow \pi^0 \ell^+ \ell^-$}}

For Yukawa-like couplings and $210\:\text{MeV} < m_A < 420\:\text{MeV}$, the pseudoscalar will dominantly decay into a pair of muons. The KTeV/E799 experiment constrains the decay of kaons into $\pi^0 \mu^+ \mu^-$~\cite{AlaviHarati:2000hs}:
\begin{align}
\text{BR}(K_L \rightarrow \pi^0 \mu^+ \mu^-) & \lesssim 4 \cdot 10^{-10} \; .
\end{align}
This search is inclusive in the sense that it does not depend on the invariant mass of the dilepton pair. 

Nevertheless, an important complication arises from the fact that experimental searches require the tracks from the three particles to originate from a common vertex. Consequently, if the lifetime of the pseudoscalar is too large,  its decays into muons will be vetoed because of the poor quality of the reconstructed vertex. We therefore have to multiply the theoretical prediction for the branching ratio with the probability that the pseudoscalar decays instantaneously (i.e.\ within the vertex resolution of the detector). We discuss the calculation of this probability, taking into account the boost factor of the pseudoscalar, in Appendix~\ref{sec:lifetime}. For our analysis we take the vertex resolution of KTeV/E799 to be $4\:\text{mm}$.

Similar searches exist for the decays of the pseudoscalar into a pair of electrons, which are relevant for $m_A < 210\:\text{MeV}$. These searches, however, typically require that the invariant mass of the electron pair satisfies $m_{ee} > 140\:\text{MeV}$. Consequently, these searches do not constrain pseudoscalars with very low mass, even though the branching ratio into electrons is largest at these masses. 

\subsection*{\boldmath{$K_L \rightarrow \pi^0 \gamma \gamma$}}

The most promising place to look for the process $K \rightarrow \pi A \rightarrow \pi \gamma \gamma$ is in decays of $K_L$, since the background from $K_L \rightarrow \pi^0 \pi^0$ violates $CP$ and is therefore suppressed. Nevertheless, this background still reduces the experimental sensitivity for $m_A \sim m_\pi$. For this reason, the KTeV experiment~\cite{Abouzaid:2008xm} excludes the mass range $100\:\text{MeV} < m_{\gamma \gamma} < 160\:\text{MeV}$. In the remaining mass range, the experiment finds 
\begin{equation}
 \text{BR}(K_L \rightarrow \pi^0 \gamma \gamma) = (1.29 \pm 0.03 \pm 0.05) \cdot 10^{-6} \; .
\end{equation}
This measurement is in agreement with the SM expectation~\cite{D'Ambrosio:1996sw}. The theoretical uncertainties for the SM prediction are of the order of $10^{-7}$, i.e.\ larger than experimental uncertainties. We will therefore take the experimental data to give a constraint on new physics contributions of the order of
\begin{equation}
\text{BR}(K_L \rightarrow \pi^0 \gamma \gamma) \lesssim 10^{-7} \; .
\end{equation}

As before, we cannot obtain such a strong constraint for $m_A \sim m_{\pi}$. Nevertheless, we can still obtain a relevant constraint by requiring that $\text{BR}(K_L \rightarrow \pi^0 A \rightarrow \pi^0 \gamma \gamma)$ does not exceed $\text{BR}(K_L \rightarrow \pi^0 \pi^0)$. This estimate yields~\cite{Alexopoulos:2004sx}
\begin{equation}
\text{BR}(K_L \rightarrow \pi^0 \gamma \gamma) < 9 \cdot 10^{-4} \; .
\end{equation}
As in the case of decays into charged leptons, we require that the pseudoscalar decays within the vertex resolution of the detector.

\subsection*{\boldmath{$K^+ \rightarrow \pi^+ + X$}}

The $K_{\mu2}$ experiment has studied the momentum distribution of charged pions produced in the decays of $K^+$~\cite{Yamazaki:1984vg}. In the presence of a new light pseudoscalar, the decay channel $K^+ \rightarrow \pi^+ \, A$ would lead to a bump in the momentum spectrum. The absence of such a bump allows $K_{\mu2}$ to constrain the branching ratio for $K^+ \rightarrow \pi^+ \, A$ independent of the consecutive decay channels of $A$. Consequently, this search is particularly interesting in the case where the decays of the pseudoscalar are such that they would be hard to see in most experiments (e.g.\ due to displaced vertices). To set a limit, we take the bound from Fig.~2 of~\cite{Yamazaki:1984vg}.

\subsection*{\boldmath{$B^0 \rightarrow K_S^0 + \text{inv}$}}

This search channel works in complete analogy to the case $K^+ \rightarrow \pi^+ + \text{inv}$. The strongest bound results from CLEO~\cite{Ammar:2001gi}. Again, we require that the pseudoscalar travels a distance of at least $4\:\text{m}$ before decaying.

\subsection*{\boldmath{$B \rightarrow K \ell^+ \ell^-$}}

This search channel proceeds in analogy to the case $K^+ \rightarrow \pi^+ \mu^+ \mu^-$, with the strongest constraint resulting from LHCb~\cite{Aaij:2013aka}. We use the Feldman-Cousins method~\cite{Feldman:1997qc} to obtain an upper bound on the new physics contribution in each of the bins considered by LHCb. We take the vertex resolution of LHCb to be $5\:\text{mm}$.\footnote{We have checked that this simplified treatment approximately reproduces the detector acceptance determined in~\cite{Schmidt-Hoberg:2013hba} using Monte Carlo simulations for the case of a scalar mediator.} 

As is usually done in the analysis of $B \rightarrow K \, \ell^+ \ell^-$, LHCb does not consider events where the invariant mass of the two muons is close to the charmonium resonances $J/\psi$ and $\psi(2S)$. To obtain approximate bounds in this regions, we require that the branching ratio for $B \rightarrow K \, A \rightarrow K \, \ell^+ \ell^-$ does not exceed the measured value for $B \rightarrow K \, J/\psi \rightarrow K \, \ell^+ \ell^-$ and $B \rightarrow K \, \psi(2S) \rightarrow K \, \ell^+ \ell^-$, respectively, giving~\cite{Agashe:2014kda}
\begin{align}
\text{BR}(B \rightarrow K \, \ell^+ \ell^-) & < 5 \cdot 10^{-5} \text{ for } 2.95\:\text{GeV} < m_A < 3.18\:\text{GeV} \\
\text{BR}(B \rightarrow K \, \ell^+ \ell^-) & < 5 \cdot 10^{-6} \text{ for } 3.59\:\text{GeV} < m_A < 3.77\:\text{GeV} \; .
\end{align}

\subsection*{\boldmath{$B \rightarrow K \gamma \gamma$}}

If the pseudoscalar couples only to quarks, the only allowed decay channel for $m_A < 3 \, m_\pi$ is $A \rightarrow \gamma \gamma$. For $m_A > m_K - m_\pi$, however, the pseudoscalar cannot be produced in kaon decays. An interesting way to constrain the mass range $350 \: \text{MeV} < m_A < 420 \: \text{MeV} $ (apart from the trivial constraint from the total $B$-width) would be to search for the process $B \rightarrow K \gamma \gamma$. Nevertheless, this decay mode is not listed in the PDG review. To indicate the potential impact of constraining this channel, we show the bound obtained from the assumption that the branching ratio is smaller than $10^{-2}$.

\subsection*{\boldmath{$b \rightarrow s \, g$}}

CLEO has performed a fully inclusive search for any $B$ decays involving an FCNC process, such as $b \rightarrow  s \, g$ or $b \rightarrow s q\bar{q}$, which are collectively denoted as $b \rightarrow s \, g$. The resulting bound $\text{BR}(b \rightarrow s \, g) < 6.8\%$ can be used to constrain the branching ratio for $B \rightarrow X_s \, A$ for the case that $A$ decays hadronically (see Appendix~\ref{sec:mesondecays}).

\subsection*{\boldmath{$B_s \rightarrow \mu^+ \mu^-$}}

LHCb~\cite{Aaij:2013aka} and CMS~\cite{Chatrchyan:2013bka, CMS:2014xfa} have determined the branching ratio for $B_s \rightarrow \mu^+ \mu^-$ to be in good agreement with the SM prediction. However, since the contribution from the pseudoscalar can in principle interfere with the SM processes, a simple subtraction of the SM prediction from the observed branching ratio is not possible~\cite{Buras:2013uqa}. We therefore take a more conservative approach and require that the contribution from the pseudoscalar should not exceed the SM prediction. It is clear that a very strong bound will be obtained for $m_A \approx m_{B_s}$, when the $B_s$ decay receives a resonant enhancement from the on-shell mediator. For other values of $m_A$ the pseudoscalar is forced to be off-shell, so that the branching ratio is proportional to $g_f^4$. Consequently, for very small couplings this search will not be competitive with decays that produce the pseudoscalar on-shell. On the other hand, the constraints do not vanish above the $B_s$ meson mass and are very relevant there.

\subsection{Fixed target experiments}

It has long been known that beam-dump experiments are sensitive to long-lived light new states. The majority of beam-dump experiments operate using electron beams which are insensitive to new light scalars or pseudoscalars with Yukawa-like couplings or couplings only to quarks. However, there are a number of proton beam-dump results in the literature including CHARM~\cite{Bergsma:1985qz}, NuCal~\cite{Blumlein:1990ay} and E613~\cite{Duffy:1988rw} as well as the recent $\rm{DAE\delta ALUS}$~\cite{Kahn:2014sra} proposal. We have calculated the constraints from CHARM, following~\cite{Bezrukov:2009yw,Clarke:2013aya} and estimated the implications of NuCal, following~\cite{Blumlein:2011mv,Blumlein:2013cua}. We find that  the CHARM reach is greater than NuCal, and so we present only results for CHARM here. Reproducing the calculations of~\cite{Soper:2014ska} in order to derive results from E613 is beyond the scope of this paper.

Constraints from the fixed target CHARM experiment~\cite{Bergsma:1985qz} are important when the pseudoscalar can be produced in $K_L$, $K^+$ and $B$ decays~\cite{Bezrukov:2009yw} or through mixing with $\pi^0$~\cite{Essig:2010gu} and subsequently decays into $e^+e^-$, $\mu^+\mu^-$ or $\gamma\gamma$.  Since CHARM observed 0~events, we set a bound at 90\% confidence level of $N_{\rm{det}}<2.3$~events~\cite{Clarke:2013aya}.
We assume that the $A$ production cross section is dominated by meson decays from either kaons or $B$ mesons (i.e.\ we neglect direct mediator production) such that the approximate pseudoscalar production cross section is
\begin{equation}
\sigma_A \sim \sigma_{pp} M_{pp} \left( \frac{1}{14}\text{BR}(K^+ \to \pi^+ A) +\frac{1}{28} \text{BR}(K_L \to \pi^0 A) + 3\cdot 10^{-8} \, \text{BR}(B \to A+X)    \right) \; ,
\label{eq:CHARM}
\end{equation}
where $\sigma_{pp}$ is the proton-proton cross section, $M_{pp}$ is the average hadron multiplicity and the numerical prefactors for the fraction of kaons and $B$ mesons are taken from~\cite{Bezrukov:2009yw}. Using the relation $\sigma_{\pi_0} \approx \sigma_{pp} \, M_{pp} / 3$, we can normalise this cross section to the neutral pion yield. Since this yield is known, we can then write the number of pseudoscalars produced in the solid angle of the detector as 
\begin{equation}
 N_A \approx 2.9 \cdot 10^{17} \frac{\sigma_A}{\sigma_{\pi_0}} \; .
\end{equation}
To obtain the number of decays in the detector region we need to multiply this number with the branching ratio of the pseudoscalar into electrons, muons and photons, and with the probability that the pseudoscalar decays inside the detector. The detector was placed 480 metres away from the beam-dump, and was 35 metres long. The number of events in the detector region is thus (see Appendix~\ref{sec:lifetime})
\begin{equation}
N_{\mathrm{det}} \sim N_A \left[ \exp\left( -\frac{480\:\text{m}}{\gamma \, \beta \, c \, \tau_A}\right)  - \exp\left( -\frac{(480+35)\:\text{m}}{\gamma \, \beta \, c \, \tau_A}\right) \right] \sum_{X=e,\mu,\gamma}\text{BR}(A\rightarrow XX) \; .
\end{equation} 
As long as the pseudoscalar can be produced in $K$ decays, we typically find $N_A \gg 1$ in the parameter region that we study. The expected number of events in CHARM therefore depends crucially on whether the lifetime of the pseudoscalar $\tau_A=\Gamma_A^{-1}$ is long enough for it to reach the detector.

\subsection{Radiative \texorpdfstring{$\Upsilon$}{Upsilon} decays}

Once $m_A > m_B - m_K$, constraints on the pseudoscalar couplings can only come from either processes where the pseudoscalar is off-shell (such as $B_s \rightarrow \mu^+ \mu^-$) or from $\Upsilon$ decays. In particular, there are strong constraints from BaBar on new states $A$ produced in the radiative decay $\Upsilon \rightarrow A \gamma$, which apply for a wide range of different final states. For Yukawa-like couplings the strongest bound comes from $A \rightarrow \mu^+ \mu^-$ for $m_A < 2 \, m_\tau$~\cite{Lees:2012te} and from $A \rightarrow \tau^+ \tau^-$ above the kinematic threshold~\cite{Lees:2012iw}. For universal quark couplings, strong bounds can still be obtained from hadronic decays of $A$ by searching for a bump in the momentum spectrum of the photon~\cite{Lees:2011wb}.

\section{Excluded parameter regions}
\label{sec:results}

\begin{figure}[!t]
\centering
\includegraphics[width=0.47\textwidth]{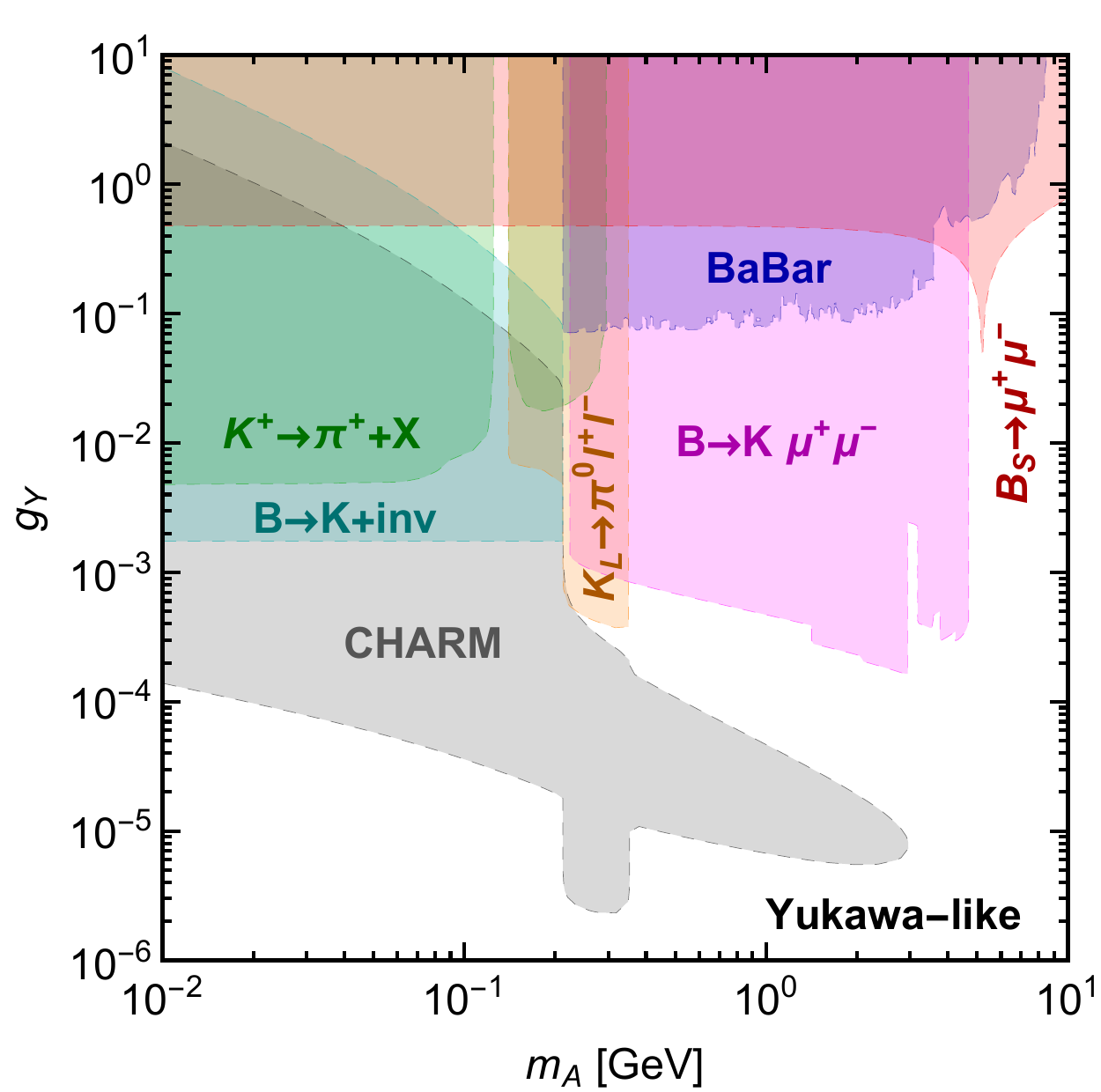}
\quad
\includegraphics[width=0.47\textwidth]{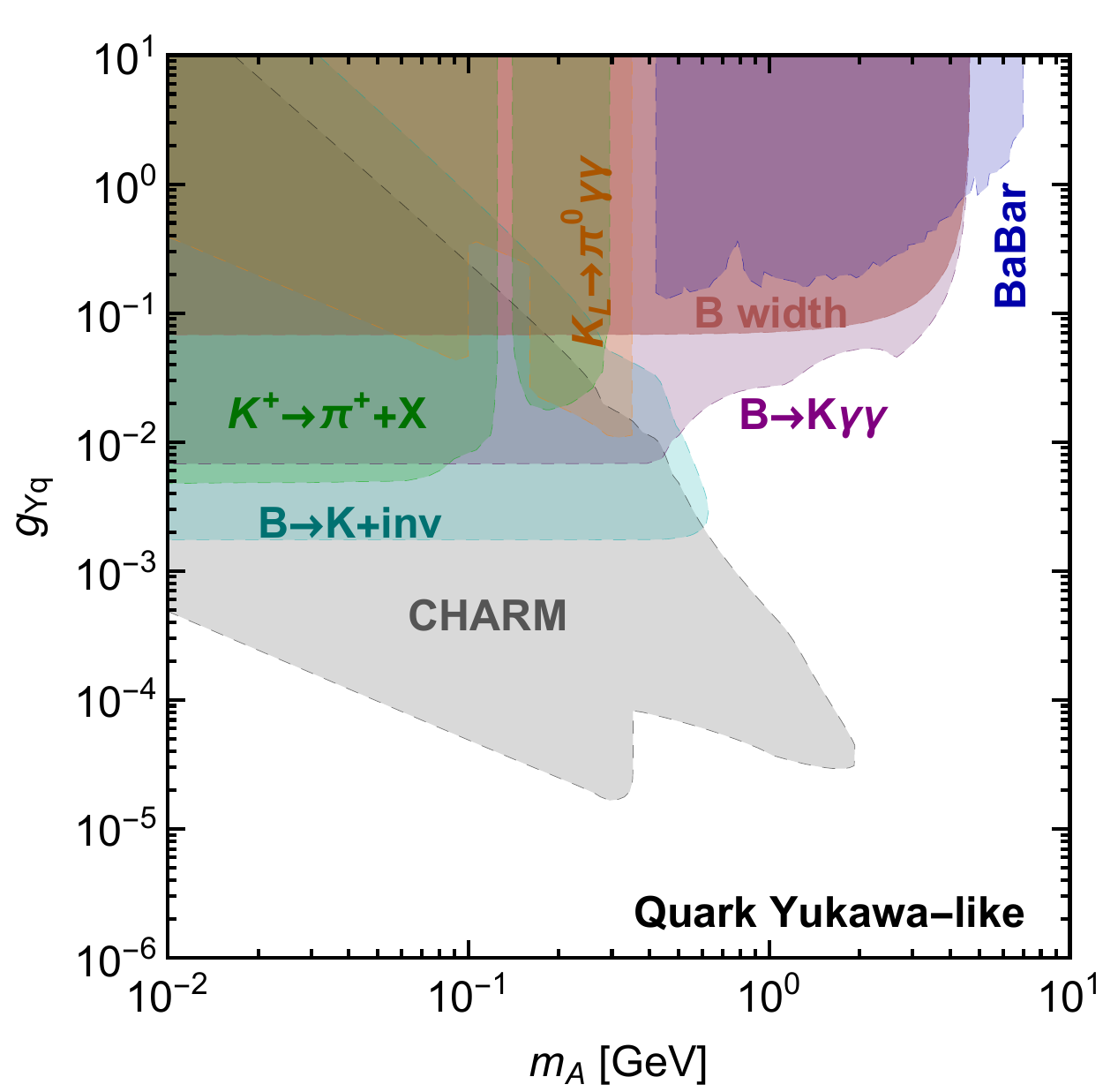}
\caption{Excluded parameter regions for a pseudoscalar $A$ with Yukawa-like couplings to all fermions (left) and Yukawa-like couplings only to quarks (right); the coupling $g_Y$ was defined in Eq.~\eqref{eq:LYukawa}.}
\label{fig:yuk}	
\end{figure}

The parameter regions excluded by the various experimental results discussed above are presented in Fig.~\ref{fig:yuk} for the case of Yukawa-like couplings and Yukawa-like quark couplings, and in Fig.~\ref{fig:univqu} for the case of universal quark couplings and third generation quark couplings. Let us briefly discuss the different cases in more detail.

\subsection{Yukawa-like couplings}

A straight-forward bound on $g_Y$ can be obtained from $K_{\mu2}$, which gives $\text{BR}(K^+ \rightarrow \pi^+ A) < 10^{-6}$ for $m_A \lesssim 100\:\text{MeV}$ independent of the decay modes of $A$. Substituting the value for $h^S_{ds}$ from Eq.~(\ref{eq:hsyuk}) into Eq.~(\ref{eq:KLwidth}), we obtain the prediction $\text{BR}(K_L \rightarrow \pi^0 A) \sim 0.06 \, g_Y^2$ in this mass region. Consequently, the bound from $K_{\mu2}$ implies $g_Y \lesssim 0.005$ for $m_A \sim 100\:\text{MeV}$. As many other searches, this bound is significantly weakened for $m_A \sim m_\pi$.\footnote{Indeed, there appears to be an allowed region for $m_A \approx m_\pi$ and $g_Y \sim 0.3$. However, for $m_A$ so close to the pion mass, the pseudoscalar mediator would significantly enhance the pion decay rate, disfavouring such a set-up.}

Most of the experimental constraints that we consider depend on the pseudoscalar branching ratios and its decay length. For example, the bound $\text{BR}(B \rightarrow K + \text{inv}) \lesssim 5 \cdot 10^{-5}$ together with the prediction $\text{BR}(B \rightarrow K + A) \sim 20 \, g_Y^2$ gives the bound $g_Y \lesssim 2 \cdot 10^{-3}$ provided the pseudoscalar escapes from the detector before decaying. This is indeed the case for $g_Y \sim 10^{-1}$ and  $m_A \lesssim 100 \: \text{MeV}$, but it is no longer true for larger couplings or larger mediator masses. In particular, searches for $B\rightarrow K + \text{inv}$ cannot exclude pseudoscalars with $m_A > 2 \, m_\mu$, which would typically decay within the detector. In a similar way the shape of the CHARM exclusion can be understood from the requirement that the pseudoscalar should neither decay too quickly nor too slowly in order to give an observable signal. The feature at $m_A \sim 210 \: \text{MeV}$ results from the rapidly decreasing lifetime of the pseudoscalar as decays into muons become allowed.

Searches for $K_L \rightarrow \pi^0 \ell^+ \ell^-$ or $B \rightarrow K \ell^+ \ell^-$, on the other hand, can only be applied to pseudoscalars that decay promptly within the detector. In other words, these searches lose sensitivity towards small couplings not because the number of pseudoscalars produced is too small, but because the pseudoscalar lifetime becomes so large that most decays occur from displaced vertices (see Appendix~\ref{sec:lifetime}). For example, the bound $\text{BR}(K_L \rightarrow \pi^0 \mu^+ \mu^-) \lesssim 10^{-10}$ cannot exclude couplings all the way down to $g_Y \sim 10^{-4}$, because the decay length of the pseudoscalar for such small couplings would be significantly larger than the vertex resolution of the detector. Consequently, there is a useful complementarity between searches for invisible decays and searches for leptonic decays. By implementing a search for displaced vertices, LHCb should be able to significantly improve its sensitivity and probe pseudoscalar couplings down to $g_Y \sim 10^{-4}$.

Finally, above the $B$ meson mass, the only sensitive searches are those for $\Upsilon$ decays at BaBar and for $B_s \rightarrow \mu^+ \mu^-$ at LHCb, both probing roughly $g_Y \sim 1$. Note that the process $B_s \rightarrow \mu^+ \mu^-$ is the only search channel that involves an off-shell pseudoscalar. Consequently, the resulting bound in principle extends up to arbitrarily large pseudoscalar mass. For $m_A \gg m_{B_s}$ it can be written as
\begin{equation}
g_Y \lesssim 0.9 \left(\frac{m_A}{10\:\text{GeV}}\right) \; .
\end{equation}

\subsection{Yukawa-like couplings to quarks}

For small pseudoscalar masses the case of Yukawa-like couplings only to quarks proceeds in complete analogy to the one including leptons, with the only differences being that the bound on $K_L \rightarrow \pi^0 \gamma \gamma$ replaces the search for $K_L \rightarrow \pi^0 \ell^+ \ell^-$ and that there is no special feature at $m_A = 2 \, m_\mu$. In the absence of the process $B \rightarrow K \mu^+ \mu^-$, however, it becomes very difficult to constrain the mass region $m_K - m_\pi < m_A < 3 \, m_\pi$, when the pseudoscalar can no longer be produced in kaon decays, but hadronic decays are still forbidden. A promising way to search for such pseudoscalars would be the decay $B \rightarrow K \gamma \gamma$, but no such search exists in the literature. To estimate experimental sensitivity we show the bound corresponding to $\text{BR}(B \rightarrow K \gamma \gamma) < 10^{-2}$. In the absence of such a search, we still obtain $g_{Yq} < 0.07$ from the total $B$ width.

For $m_A > 3 \, m_\pi$ hadronic decays become allowed and completely dominate over photonic decays. Consequently, the only relevant searches are $b \rightarrow s \, g$ and $\Upsilon \rightarrow \gamma + \text{hadrons}$, the latter one giving stronger constraints. Note that in the absence of couplings to muons, the lifetime of the pseudoscalar is significantly increased above $210 \: \text{MeV}$. Consequently, the CHARM bound extends up to larger pseudoscalar masses and includes a relevant contribution from $B$ meson decays.

\subsection{Universal couplings to quarks}

\begin{figure}[!t]
\centering
\includegraphics[width=0.47\textwidth]{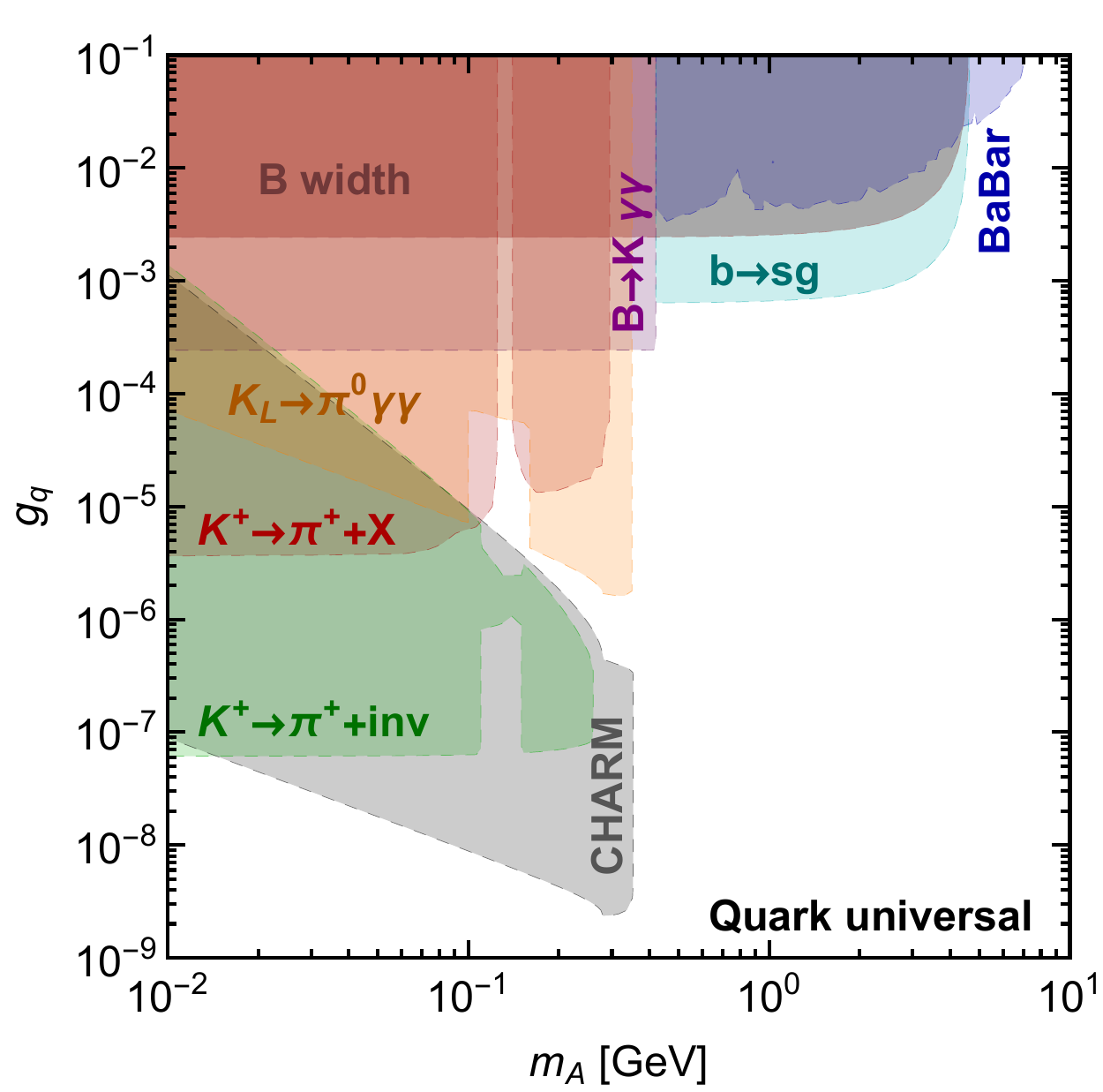}
\quad
\includegraphics[width=0.47\textwidth]{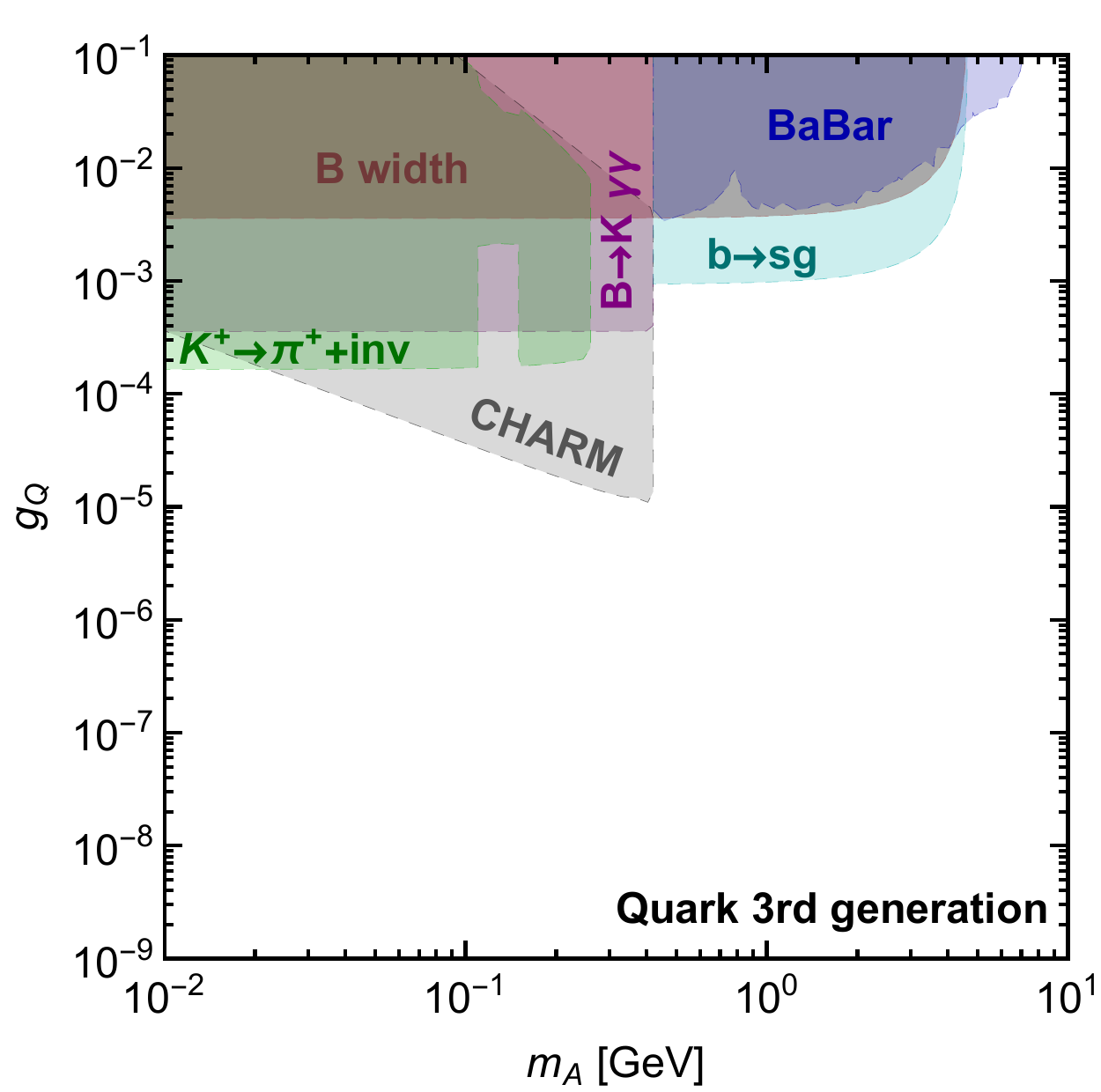}
\caption{Excluded parameter regions for a pseudoscalar $A$ with universal couplings to quarks; the coupling $g_q$ was defined in Eq.~\eqref{eq:ASM}. The left panel shows the constraints when $A$ couples to all quarks, while the right panel shows the constraints when $A$ couples only to third generation quarks~$Q=\{b,t\}$. }
\label{fig:univqu}
\end{figure}

For the case of universal quark couplings, the constraints on $g_q$ are considerably stronger than the corresponding ones for $g_Y$. This is partially due to the fact that there is no factor $\sqrt{2} \, m_f / v$ in the couplings, but more importantly due to larger flavour-changing effects resulting from the non-MFV coupling structure. The former reason also implies that experiments probing rare kaon decays become more important compared to experiments probing rare $B$ meson decays. The enhancement of flavour-changing effects, on the other hand, implies that bounds from processes like $b\rightarrow s g$ give stronger bounds than processes like $\Upsilon \rightarrow \gamma + \text{hadrons}$.

For small pseudoscalar masses, we again find a clear complementarity between searches for $K_L \rightarrow \pi^0 \gamma \gamma$ and searches for $K^+ \rightarrow \pi^+ + \text{inv}$  (see left panel of Fig.~\ref{fig:univqu}), ruling out the entire parameter region $m_A < m_K - m_\pi$ and $g_q \gtrsim 10^{-7}$. To close the gap between $m_A < m_K - m_\pi$ and $m_A > 3 \, m_\pi$, we again show the bound corresponding to $\text{BR}(B \rightarrow K \gamma \gamma) < 10^{-2}$. By assumption, we take photonic decays to be negligible above the hadronic decay threshold. Consequently, the dominant bound comes from $b \rightarrow s g$ for $m_A \lesssim m_B$, and from BaBar for $m_B \lesssim m_A \lesssim m_\Upsilon$.

\subsection{Universal couplings to \texorpdfstring{$b$}{b} and \texorpdfstring{$t$}{t} quarks only}

If we assume that the pseudoscalar couples only to bottom and top quarks with the same coupling strength $g_Q$, the effective flavour changing coupling $h^S_{ds}$ is reduced by more than three orders of magnitude, as can be seen by comparing Eq.~(\ref{eq:hsq3}) with Eq.~(\ref{eq:hsuni}). Consequently, any bounds on $g_Q$ from kaon decays are relaxed by more than three orders of magnitude (cf.~left and right panels of Fig.~\ref{fig:univqu}). At the same time, the lifetime of the pseudoscalar grows significantly, because for universal quark couplings, light quarks give the dominant contribution to the loop-induced decay of the pseudoscalar into photons. In this scenario, these light quark contributions are absent with the result that, even for $g_Q \sim 10^{-2}$, the pseudoscalar will almost always escape from the detector without decaying. Therefore searches for pions in association with missing momentum are constraining up to large couplings.

The effective coupling $h^S_{sb}$ responsible for flavour-changing $B$ meson decays receives its dominant contribution from the heavy-quark couplings of the pseudoscalar. Consequently, bounds from $B \rightarrow K \, A$ remain largely unaffected if we exclude couplings to light quarks. The same is true for radiative $\Upsilon$ decays, which only probe the bottom-quark coupling of~$A$.

\subsection{The case of an invisibly decaying pseudoscalar}

\begin{figure}[!t]
\centering
\includegraphics[width=0.47\textwidth]{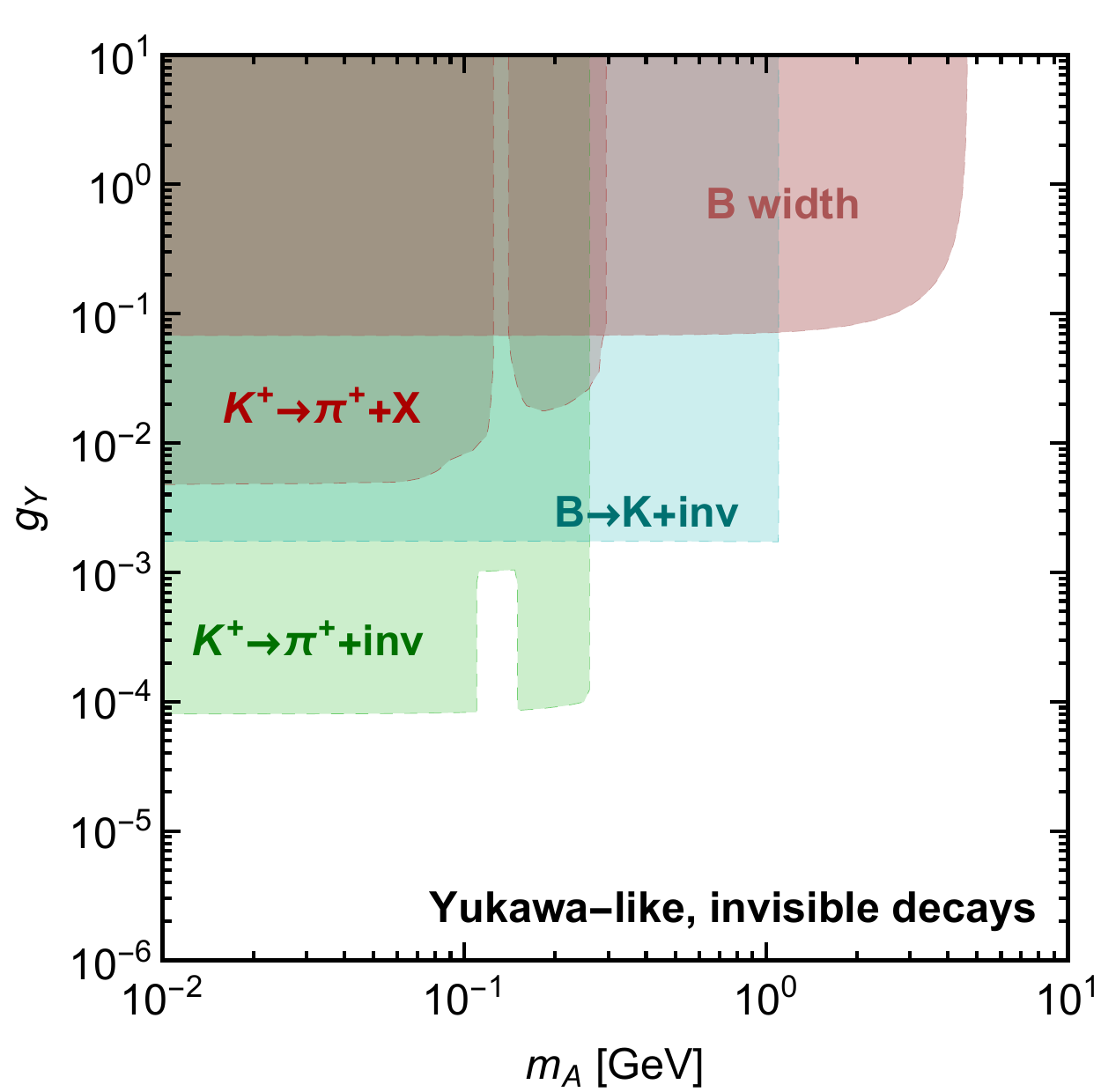}
\quad
\includegraphics[width=0.47\textwidth]{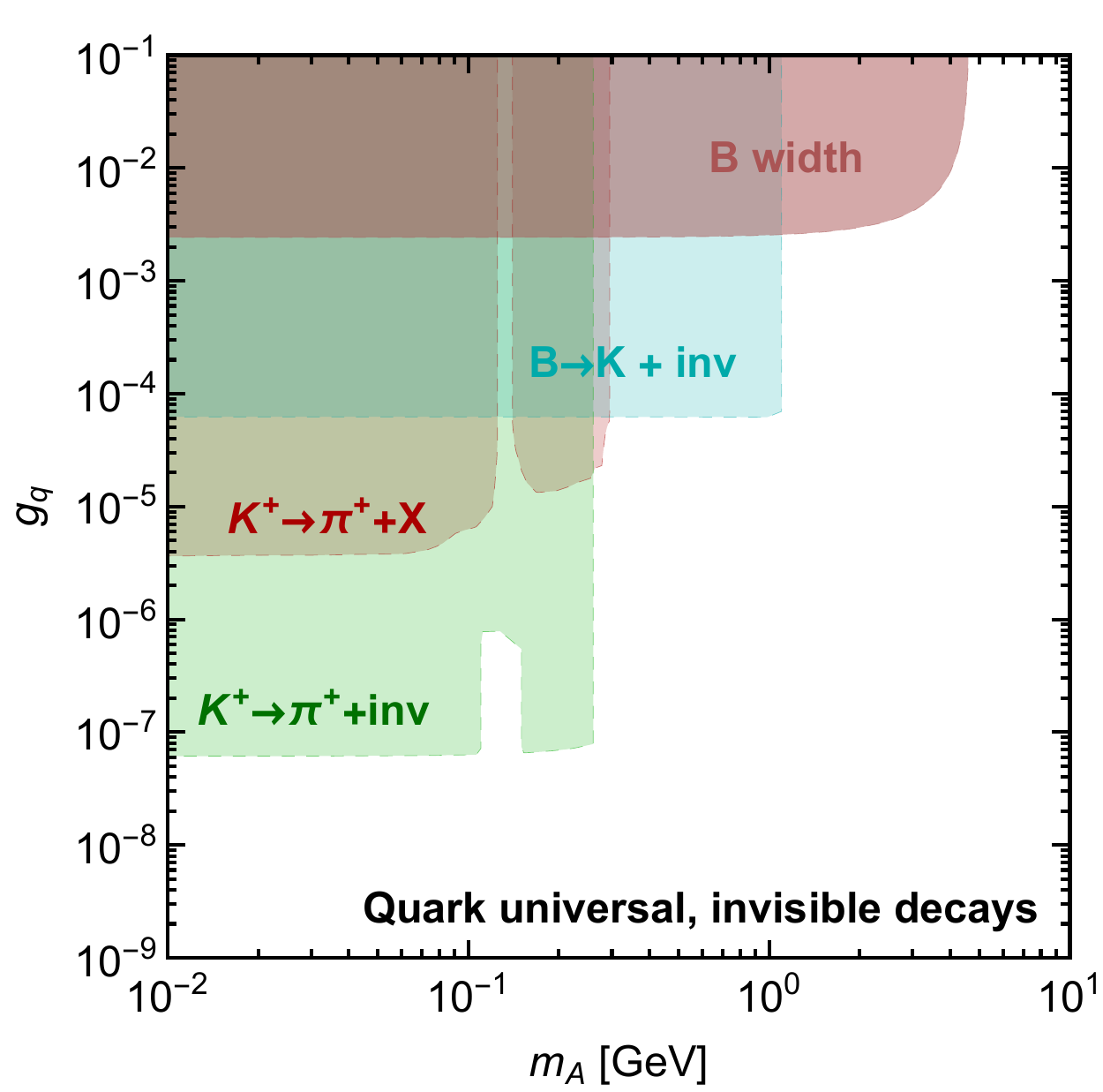}
\caption{Excluded parameter regions for an invisibly decaying pseudoscalar $A$. The left panel assumes Yukawa-like couplings to quarks, while the right panel considers universal quark couplings.}
\label{fig:invisible}
\end{figure}

One might be tempted to think that the constraints discussed above can be evaded if the pseudoscalar dominantly decays into neutrinos or other invisible states (such as a DM particle with $m_\chi < m_A / 2$). The experimental bounds on new contributions to $B \rightarrow K + \text{inv}$ and $K^+ \rightarrow \pi^+ + \text{inv}$, however, imply that these decay channels are in fact tightly constrained. We show these constraints in Fig.~\ref{fig:invisible}. For the case that the pseudoscalar is produced via Yukawa-like quark couplings, but decays dominantly into invisible final states, we find
\begin{equation}
g_Y \lesssim \left\{
      \begin{array}{ll}
        8 \cdot 10^{-5} & \text{ for } m_A \lesssim 100\:\text{MeV}\\
        2 \cdot 10^{-3} & \text{ for } 100\:\text{MeV} \lesssim m_A \lesssim 1\:\text{GeV} \, .\\
      \end{array}
    \right.
\end{equation}
If the pseudoscalar has universal couplings to quarks, the constraints are
\begin{equation}
g_q \lesssim \left\{
      \begin{array}{ll}
        6 \cdot 10^{-8} & \text{ for } m_A \lesssim 100\:\text{MeV}\\
        6 \cdot 10^{-5} & \text{ for } 100\:\text{MeV} \lesssim m_A \lesssim 1\:\text{GeV} \, . \\
      \end{array}
    \right.
\end{equation}

\section{The dark matter connection}
\label{sec:dark}

So far we have mainly discussed the couplings of the pseudoscalar to SM fields and found that they are strongly constrained by flavour observables. In the following we will concentrate on the pseudoscalar couplings to DM.  As we shall see, the presence of a new light particle that can communicate interactions between DM particles and the SM can  have important implications for both the generation of DM in the early Universe and the detection of DM in present and future experiments. In this section we discuss the general concepts, while more specific direct and indirect signals will be considered in the following section.

\subsection{Relic density constraints}

Two processes are relevant for the freeze-out of DM in the early Universe: Annihilation into pairs of pseudoscalars $\bar{\chi}\chi\to AA$ and annihilation into SM fermions $\bar{\chi}\chi\to \bar{f}f$. For a Dirac DM particle, the thermally-averaged annihilation cross section into SM fermions is given by
\begin{equation}
 \langle \sigma v \rangle_{\bar{\chi}\chi\to\bar{f}f} \simeq \sum_f \frac{N_c}{2 \pi} \frac{g_f^2 \, g_\chi^2 \, m_\chi^2}{(4 m_\chi^2 - m_A^2)^2} \sqrt{1-\frac{m_f^2}{m_\chi^2}} \; ,
\label{eq:annqq}
\end{equation}
while for the annihilation into pseudoscalars we find
\begin{equation}
\langle \sigma v \rangle_{\bar{\chi}\chi\to AA} \simeq \frac{g_\chi^4}{24 \pi} \frac{m_\chi (m_\chi^2 - m_A^2)^{5/2}}{(m_A^2 - 2 \, m_\chi^2)^4} \frac{6}{x} \; .
\label{eq:annAA}
\end{equation}
Here $x = m_\chi/T$, where $T$ is the temperature; we see that the annihilation into SM fermions is $s$-wave while the annihilation to pseudoscalars is $p$-wave.

\begin{figure}[tb]
\centering
\includegraphics[width=0.48\textwidth]{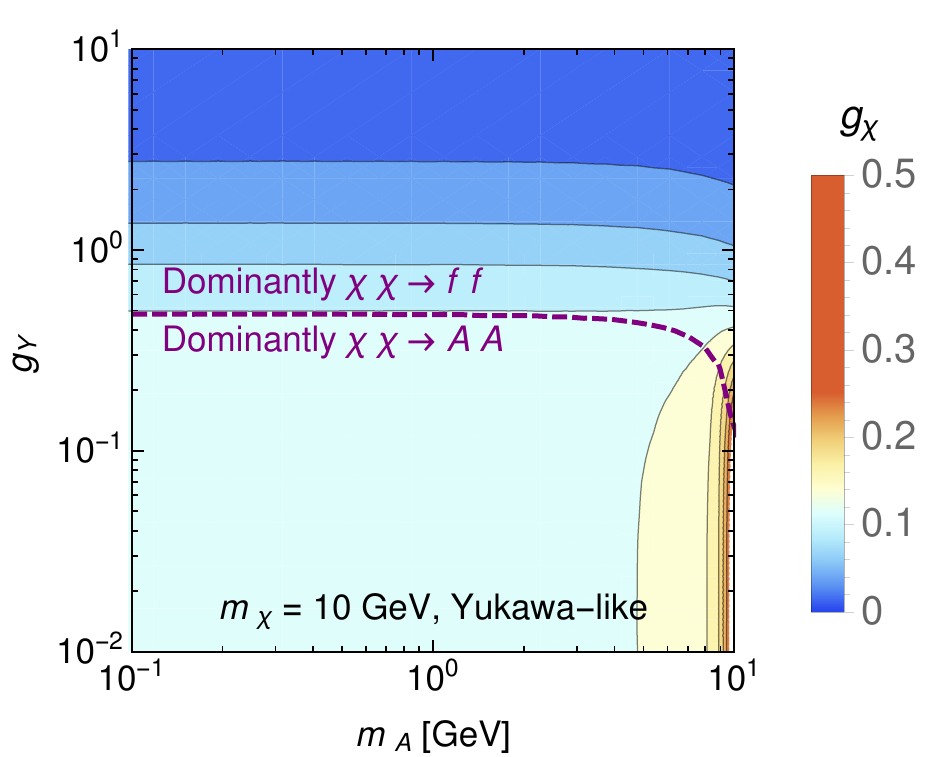}
\quad
\includegraphics[width=0.48\textwidth]{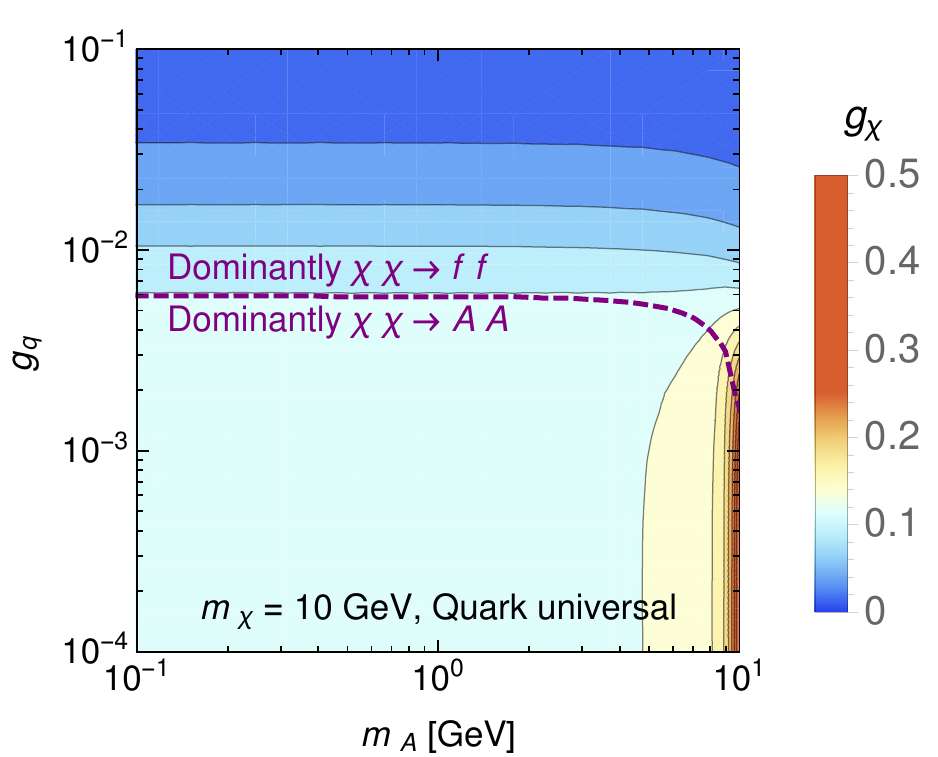}
\caption{Value of $g_\chi$ required to reproduce the observed DM density as a function of $m_A$ and $g_f$ for fixed DM mass $m_\chi = 10\:\text{GeV}$. The left (right) plot corresponds to the case of Yukawa-like couplings, $g_f=\sqrt{2}\,g_Y m_f/v$ (universal quark couplings, $g_f=g_q$). }
\label{fig:relic}
\end{figure}

Since both annihilation cross sections depend on $g_\chi$, we can always eliminate this parameter (for given $m_A$, $m_\chi$ and $g_f$) by imposing that our model reproduces the observed abundance of DM. In Fig.~\ref{fig:relic}, we show the dependence of $g_\chi$ on $m_A$ and $g_f$ for the fixed value $m_\chi=10$~GeV. While these results are obtained numerically from \mbox{micrOMEGAs}~\cite{Belanger:2013oya}, it is instructive to use an approximate freeze-out calculation to estimate the required value of $g_\chi$.

First, we consider the case $g_f \ll g_\chi$ so that $\bar{\chi}\chi\to AA$ dominates. When \mbox{$m_A \ll m_\chi$}, the relic density requirement reads
\begin{equation}
\frac{1}{2} \times\frac{1}{2} \times \frac{g_\chi^4}{64 \pi} \frac{1}{m_\chi^2} \frac{1}{x_\text{f}} \simeq 2.5 \cdot 10^{-26}\, \text{cm}^3/\text{s}
\end{equation}
where $x_\text{f}$ is the value of $x$ at freeze-out and there is one extra factor of 1/2 for Dirac DM and another since the annihilation is a $p$-wave process~\cite{Gondolo:1990dk}. Taking $x_\text{f} = 20$, we obtain
\begin{equation}
\frac{g_\chi^2}{ m_\chi} \approx 0.006 \: \text{GeV}^{-1} \; ,
\label{eq:estimate}
\end{equation}
which is in good agreement with the value in Fig.~\ref{fig:relic}. As $m_A \rightarrow m_\chi$, annihilation into pseudoscalars becomes suppressed due to the reduced phase space, so a larger value of $g_\chi$ is necessary to compensate.

For larger values of $g_f$ the annihilation to SM fermions becomes increasingly relevant. The purple lines in Fig.~\ref{fig:relic} indicate the transition from freeze-out dominantly into pseudoscalars to freeze-out dominantly into SM fermions. For $m_A \ll m_\chi$, we can estimate the corresponding value of $g_f$ by equating Eq.~(\ref{eq:annqq}) and Eq.~(\ref{eq:annAA}), again taking into account a factor of $1/2$ for the $p$-wave process. Substituting the estimate for $g_\chi$ from Eq.~(\ref{eq:estimate}), we find
\begin{equation}
\frac{\sum_f N_c \, g_f^2}{m_{\chi}}\approx 7 \cdot 10^{-5}  \: \text{GeV}^{-1} \; .
\end{equation}
Again, this estimate is in good agreement with the values shown in Fig.~\ref{fig:relic}. For larger values of $g_f$, the annihilation cross section into pseudoscalars (and hence the coupling $g_{\chi}$) must be reduced to avoid underproduction of DM.

As we have shown in Figs.~\ref{fig:yuk} and~\ref{fig:univqu}, the coupling $g_f$ is strongly constrained by precision measurements. Comparing these figures with Fig.~\ref{fig:relic}, we see that in most of the allowed parameter space the dominant contribution to DM annihilation will result from annihilation to pseudoscalars, which is independent of $g_f$. The fact that annihilation into pseudoscalars is $p$-wave only means that present-day indirect detection signals are unobservably small over much of the parameter space. We will, however, also consider parameter regions where the annihilation cross section into SM fermions is comparable to the thermal cross section during freeze-out and indirect detection signals may be observable.

\subsection{Thermal equilibrium}

As we have seen above, for very small values of $g_f$ the DM relic density becomes independent of the SM couplings of the pseudoscalar, since annihilation into pseudoscalars completely dominates for the freeze-out calculation. However, we cannot make the couplings $g_f$ arbitrarily small, because the calculation of the relic density relies crucially on the assumption that the dark sector and visible sector are at the same temperature. If the coupling between the two sectors becomes too weak, the energy transfer becomes so scarce that thermal equilibrium between the two sectors may not be achieved. Of course, the dark sector may still thermalise if the intra-sector coupling $g_\chi$ is large enough, but it will in general have a temperature different from the visible sector. DM can therefore still freeze out into pseudoscalars, but the resulting abundance becomes sensitive to the details of reheating~\cite{Chu:2011be}.\footnote{Note that even if $g_\chi$ is so weak that the dark sector never thermalises, the observed relic abundance can be obtained via the freeze-in mechanism~\cite{Hall:2009bx,Chu:2011be}.}

In this section we estimate the couplings necessary to obtain thermal equilibrium. For smaller couplings, it is no longer possible to reduce our model to three parameters by imposing the relic density constraint, so we will not consider such cases further. Fortunately, we will see that this requirement is not very constraining.

We impose the usual requirement for thermal equilibrium (see e.g.~\cite{Chu:2011be}), namely that the reaction rate is larger than the expansion rate of the Universe
\begin{equation}
 \langle \sigma v \rangle n_\text{eq} > H \; ,
\end{equation}
where as usual, $H = 1.66 \sqrt{g_\ast}\, T^2 / M_\text{P}$ and $n_\text{eq}$ is the equilibrium number density. In principle, the left-hand side can refer to both the production of SM particles from the dark sector or DM production from the visible sector (in thermal equilibrium both rates must be equal).

Using Maxwell-Boltzmann statistics for the equilibrium distributions, we obtain~\cite{Gondolo:1990dk,Chu:2011be}
\begin{equation}
 \langle \sigma v \rangle n_\text{eq} = \sum_f \langle \sigma v \rangle_{\bar{f}f \rightarrow \bar{\chi}\chi} \, n^{f}_\text{eq}
\end{equation}
with
\begin{equation}
  \langle \sigma v \rangle_{\bar{f}f \rightarrow \bar{\chi}\chi} \, n^{f}_\text{eq} = \frac{N_c}{8 \pi^2 \, m^2 \, K_2(m/T)} \int_{s_\text{min}}^{\infty} \mathrm{d}s \, \sigma_{\bar{f}f \to \bar{\chi}\chi} \, (s - 4 m^2) \, \sqrt{s} \, K_1(\sqrt{s}/T) \; ,
\end{equation}
where $K_i(s)$ are the modified Bessel functions of the second kind and
\begin{equation}
 \sigma_{\bar{f}f\rightarrow \bar{\chi}\chi} = \frac{g_f^2 \, g_\chi^2}{8 \pi} \frac{s}{(s - m_A^2)^2} \sqrt{\frac{s - 4 m_\chi^2}{s - 4 m_f^2}}
\end{equation}
is the annihilation cross section into two Dirac DM particles as a function of the centre-of-mass energy $\sqrt{s}$. Note that for definiteness we consider the production of DM particles from the visible sector, although the final expression is valid also for the inverse process.\footnote{Note in particular that the factor $N_c$ appears in the one case as part of the degrees of freedom that determine $n_\text{eq}$, while in the other case it appears in $\langle \sigma v \rangle$ as part of the available phase space.}

We find that the condition for thermal equilibrium is most easily satisfied for $T \sim \text{max}(m_f, m_\chi)$. The reason is that the cross section decreases proportional to~$1/s$ for large~$s$, so that at high temperatures $\langle \sigma v \rangle_{\bar{f}f \rightarrow \bar{\chi}\chi} \, n^{f}_\text{eq} = N_c \, g_f^2 \, g_\chi^2 \, T / (32 \pi^3) \propto T$, while $H \propto T^2$. For $T \ll m_\chi, m_f$, on the other hand, the reaction rate is exponentially suppressed. In practice, we check the condition for thermal equilibrium for a range of temperatures around the masses of the particles involved. Nevertheless, it is instructive to obtain a rough estimate by considering the case that \mbox{$T \sim m_\chi \gg m_f$}. In this approximation we find $\langle \sigma v \rangle_{\bar{f}f \rightarrow \bar{\chi}\chi} \, n^{f}_\text{eq} \approx 1.45 \, N_c \, g_f^2 \, g_\chi^2 \, m_{\chi} / (128 \pi^3)$ and hence the condition
\begin{equation}
 \sum_f \frac{1.45}{128 \pi^3} N_c \, g_f^2 \, g_\chi^2 > 1.66 \sqrt{g_\ast} \frac{m_\chi}{M_\text{P}} \; .
\end{equation}

If thermal equilibrium is achieved, we want the annihilation of DM into pairs of pseudoscalars to yield the observed relic abundance. As shown above, this gives the approximate relation $g_\chi^2 / m_\chi \approx 0.006 \: \text{GeV}^{-1}$. Substituting into the condition for thermal equilibrium and using $g_\ast \sim 80$ yields
\begin{equation}
\sum_f N_c \, g_f^2 \gtrsim 6 \cdot 10^{-13} \; .
\end{equation}
For annihilation of top quarks, we should evaluate both the reaction rate and the expansion rate at $T = m_t$. Assuming $m_\chi \ll m_t$, we then obtain 
\begin{equation}
3 \, g_t^2 \frac{m_\chi}{m_t} \gtrsim 6 \cdot 10^{-13} \; .
\end{equation}
For Yukawa-like couplings and $m_b \ll m_\chi \ll m_t$, the dominant contribution therefore comes from the top quark at $T \sim m_t$ giving $g_Y > 3 \cdot 10^{-7} v / \sqrt{m_t \, m_\chi}$. For universal couplings to all quarks, on the other hand, the contribution of the top quark is negligible, while all light quarks give an equal contribution at $T \sim m_\chi$, leading to the requirement $g_q \gtrsim 2 \cdot 10^{-7}$.

\begin{figure}[tb]
\centering
\includegraphics[width=0.48\textwidth, clip]{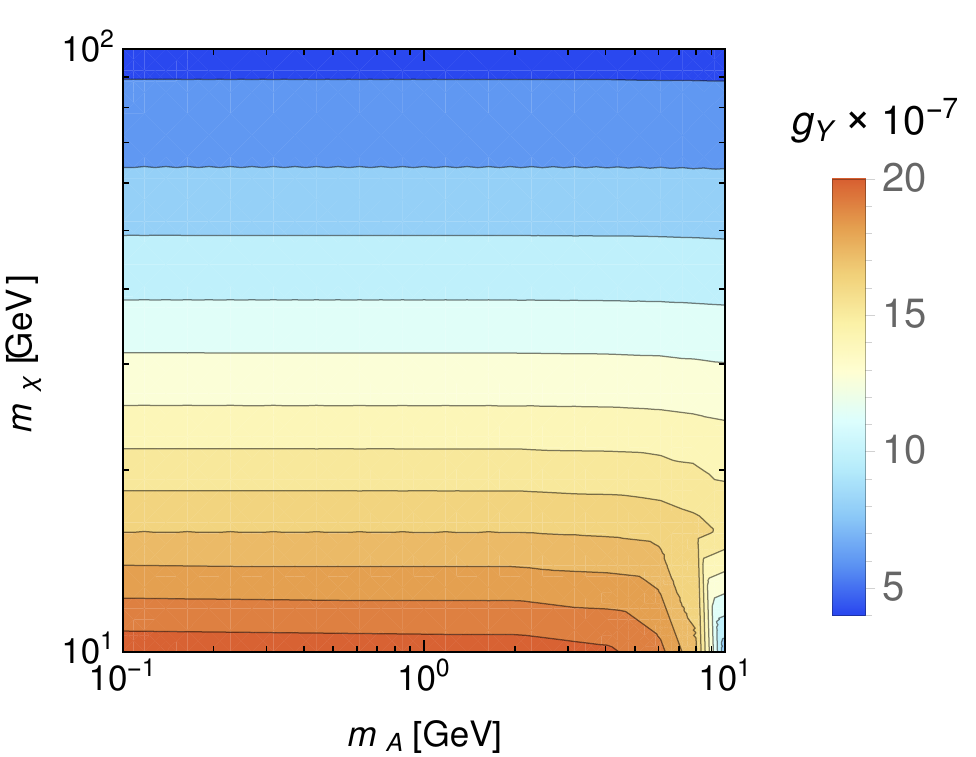}
\quad
\includegraphics[width=0.48\textwidth, clip]{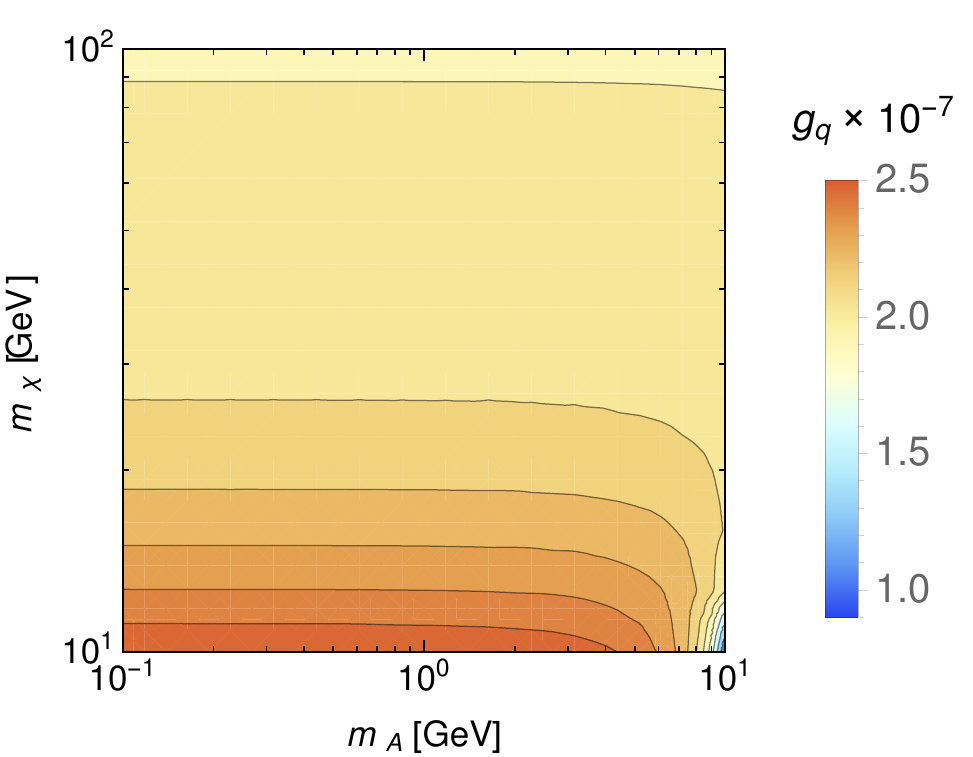}
\caption{Minimum value of $g_Y$ ($g_q$) required to achieve thermal equilibrium in the early Universe. The left (right) plot corresponds to the case of Yukawa-like couplings (universal quark couplings). }
\label{fig:TE}
\end{figure}

Fig.~\ref{fig:TE} shows the minimum coupling values required in order to obtain thermal equilibrium as a function of $m_A$ and $m_\chi$ (with $g_\chi$ fixed by the relic density requirement). As expected, for $m_A \ll m_\chi$ we find that this lower bound is independent of $m_A$. We also confirm the expectation that for universal quark couplings the bound is largely independent of~$m_\chi$ provided $m_b \ll m_\chi \ll m_t$, while for Yukawa-like couplings the bound is proportional to~$1/\sqrt{m_\chi}$.

\subsection{Big Bang Nucleosynthesis}

If DM annihilates predominantly into pseudoscalars,  these pseudoscalars must decay sufficiently quickly before Big Bang Nucleosynthesis (BBN) to avoid changes to the expansion rate and the entropy density during BBN or the destruction of certain elements. While the precise constraints are rather difficult to calculate (see e.g.~\cite{Tulin:2013teo}),  such constraints should not be important if the average lifetime of the pseudoscalar is less than 1 second. As discussed in Appendix~\ref{sec:lifetime} and illustrated in Fig.~\ref{fig:BRY}, however, the lifetime of the pseudoscalar can be large, in particular if tree-level decays are suppressed or forbidden. Consequently, BBN constraints are most severe for very light pseudoscalars and cease to be constraining for larger masses.

\begin{figure}[t]
\centering
\includegraphics[width=0.40\textwidth]{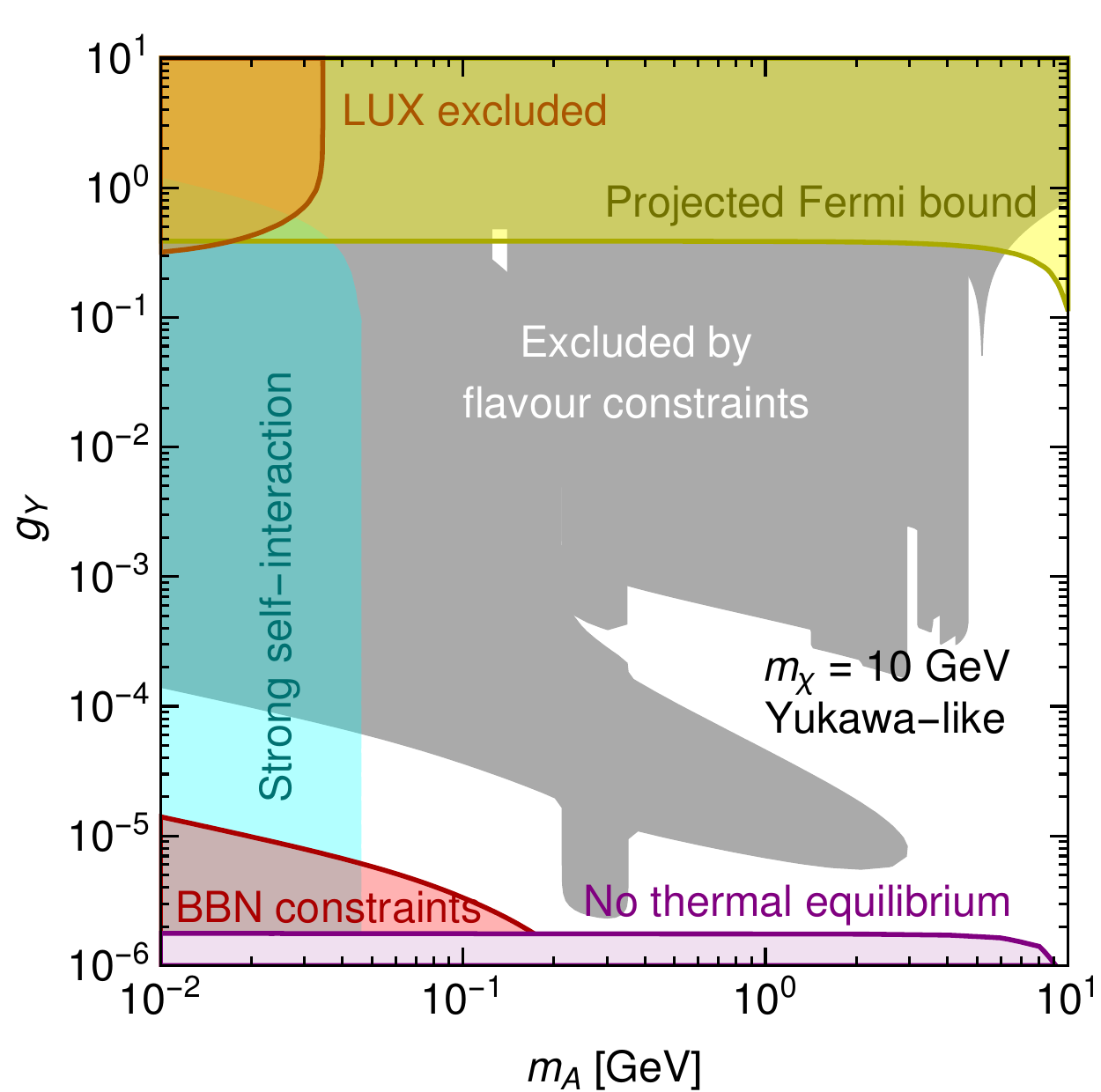}
\quad
\includegraphics[width=0.40\textwidth]{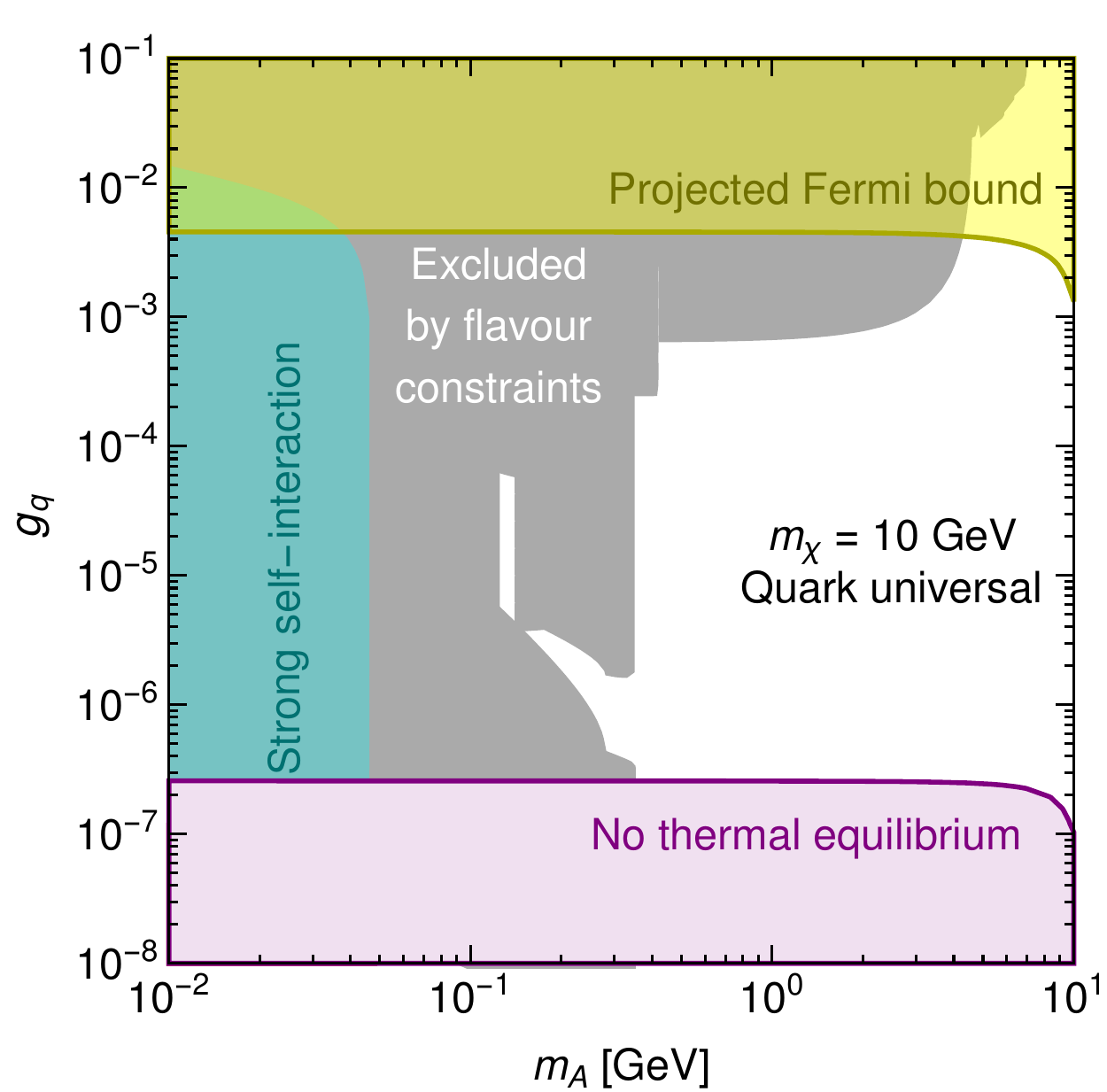}

\includegraphics[width=0.40\textwidth]{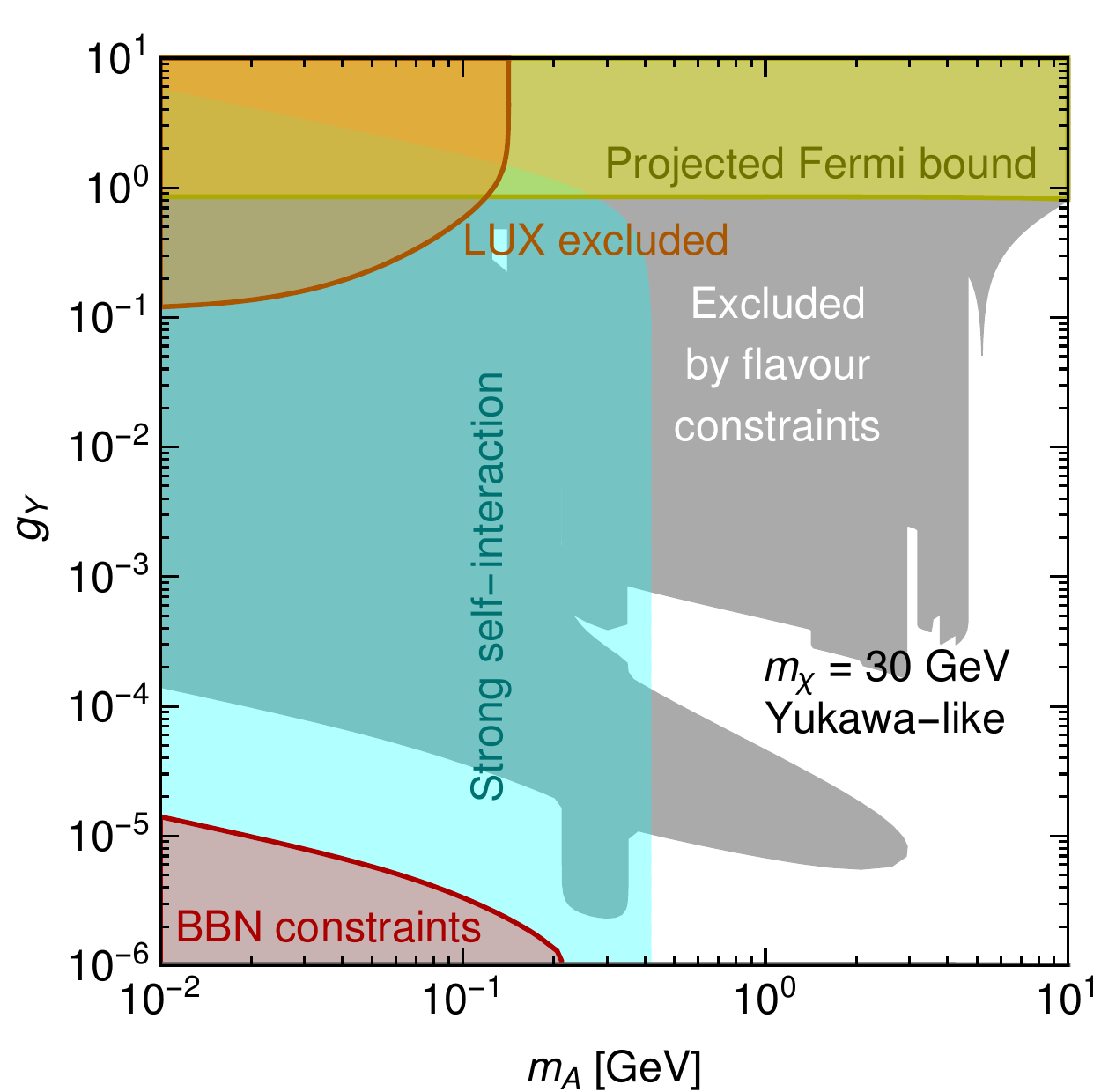}
\quad
\includegraphics[width=0.40\textwidth]{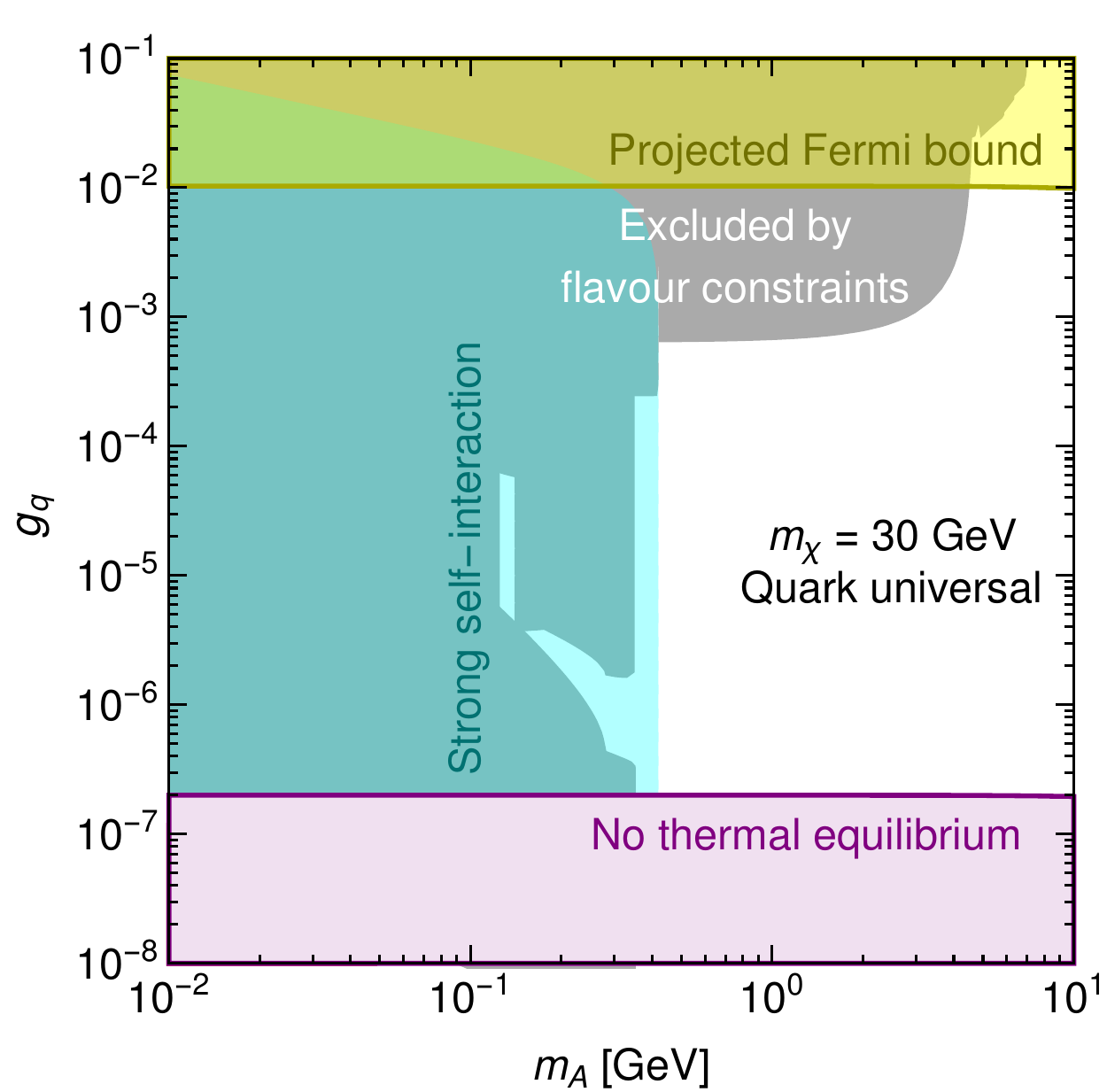}

\includegraphics[width=0.40\textwidth]{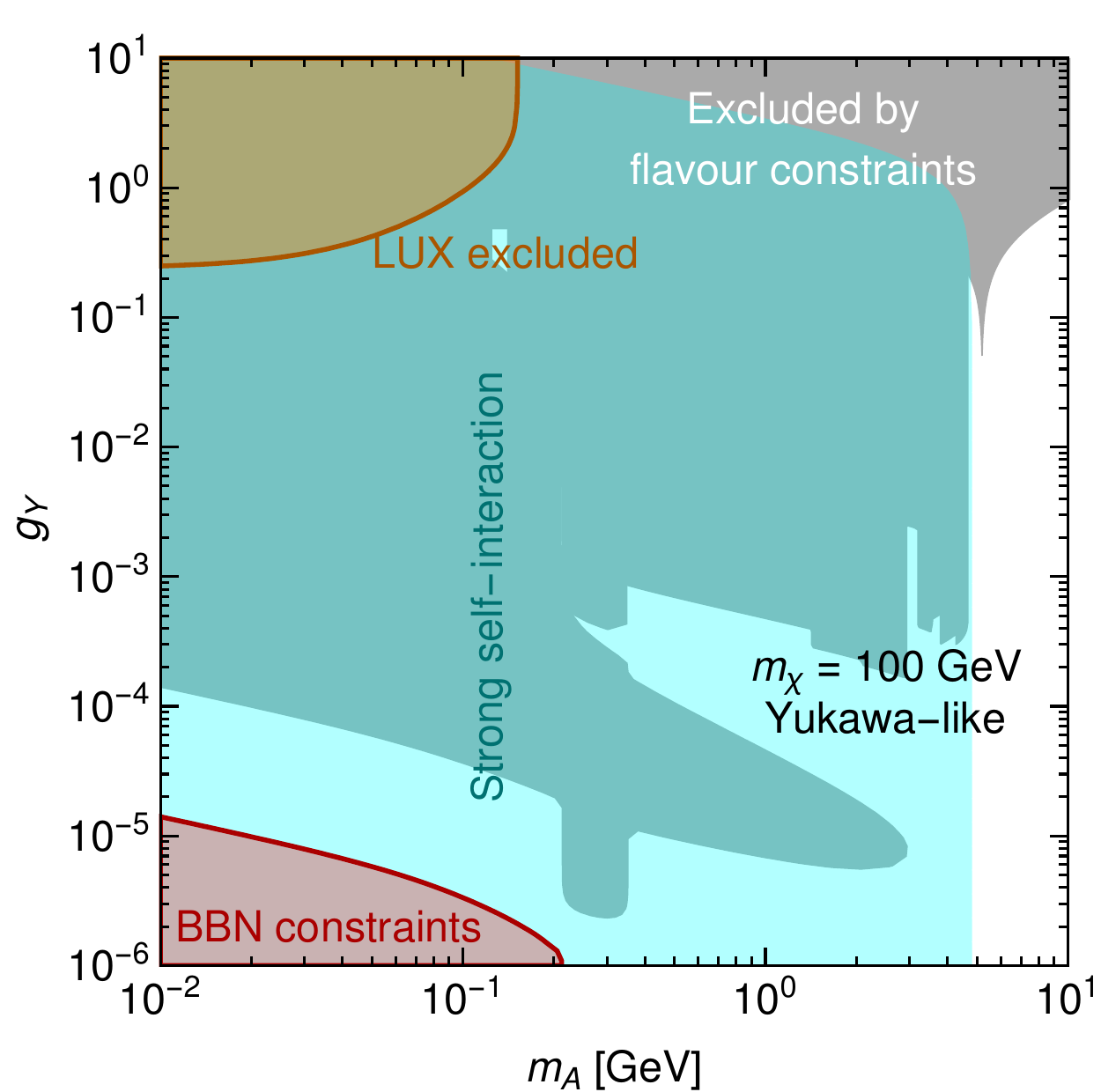}
\quad
\includegraphics[width=0.40\textwidth]{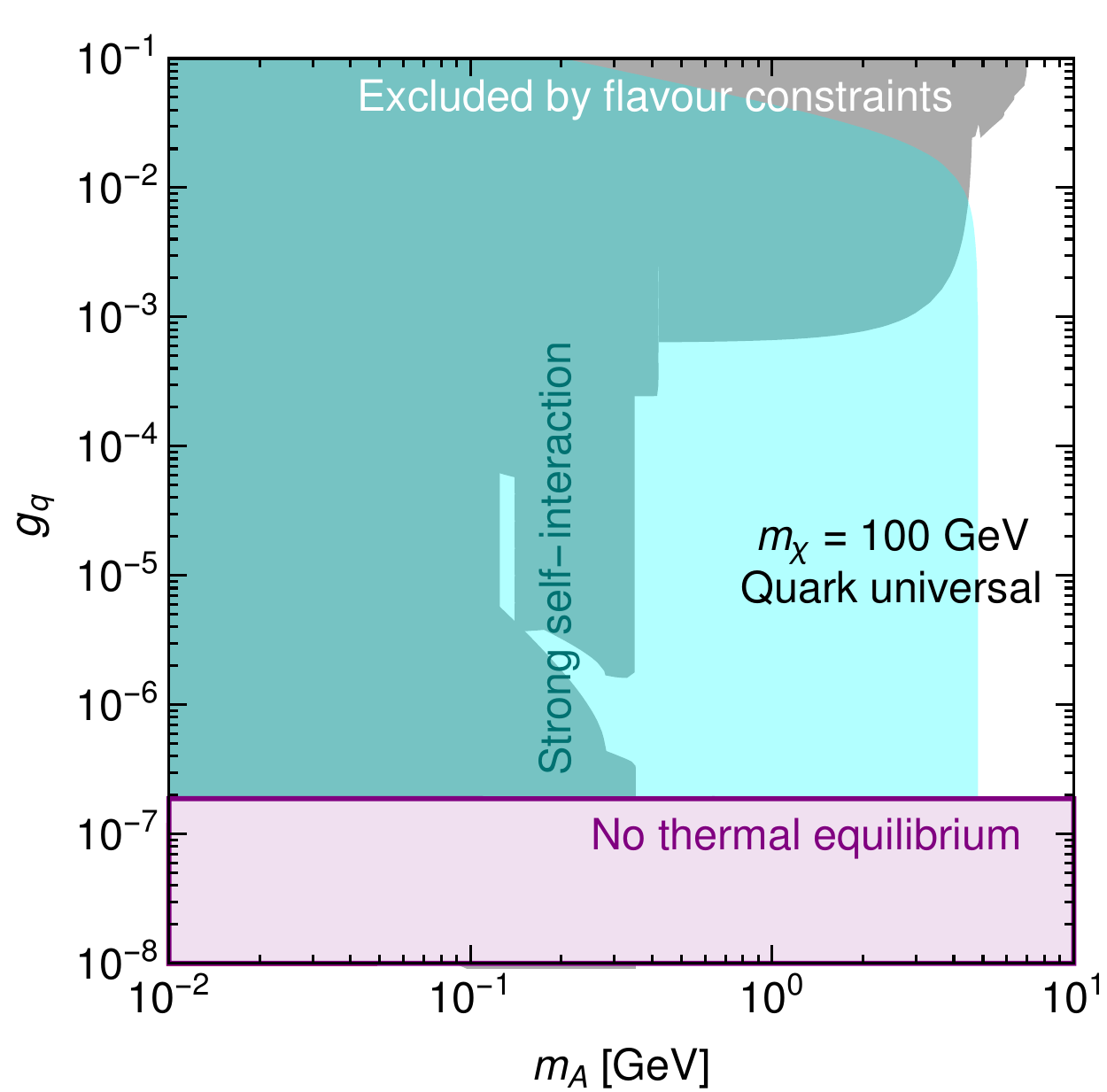}
\caption{ Bounds on $m_A$ and $g_f$ from flavour constraints, direct and indirect detection, and BBN for different values of $m_\chi$. Also shown are regions where the DM has strong self-interactions. Wherever thermal equilibrium can be achieved in the early Universe, $g_\chi$ has been fixed by the requirement to obtain the observed DM relic abundance. For the left (right) column, we have assumed Yukawa-like couplings to fermions (universal couplings to quarks). }
\label{fig:DMconstraints}
\end{figure}

We show the parameter region where BBN constraints are relevant in Fig.~\ref{fig:DMconstraints} as a function of $m_A$ and $g_f$ for different values of $m_\chi$, fixing $g_\chi$ such that the observed DM relic density is reproduced. We also show the parameter region where the dark sector does not reach thermal equilibrium with the visible sector. Moreover, Fig.~\ref{fig:DMconstraints} shows the parameter regions relevant for DM self-interactions, DM direct detection and DM indirect detection, which will be discussed next.

\subsection{Dark matter self-interactions}

As we have shown in Fig.~\ref{fig:relic} above, the relic density constraints require the coupling~$g_{\chi}$ between DM and the pseudoscalar to be rather large. For DM masses in the range $(10\text{--}100)\:\text{GeV}$, we typically find $\alpha_\chi \equiv g_\chi^2 / (4\pi) \sim \mathcal{O}(10^{-2})$. If the pseudoscalar is significantly lighter than the DM particle, DM scattering may therefore be in the strongly-coupled regime corresponding to $\alpha_\chi \, m_\chi / m_A > 1$, potentially leading to interesting effects and relevant constraints.

The quantity most relevant for DM self-interactions is the momentum transfer cross section\footnote{For a discussion of why it is important to weight the differential cross section with a factor $(1 - |\cos \theta|)$ rather than $(1 - \cos \theta)$ when discussing the scattering of identical particles, we refer to~\cite{Kahlhoefer:2013dca}.}
\begin{equation}
 \sigma_\mathrm{T} = \int \mathrm{d}\Omega (1 - |\cos \theta|) \frac{\mathrm{d}\sigma(v)}{\mathrm{d}\Omega} \; .
\end{equation}
In the weakly-coupled regime, we can calculate the momentum transfer cross section in the Born approximation. For particle-particle scattering, one finds approximately
\begin{equation}
 \sigma_T \simeq \left\{
     \begin{array}{cr}
       \frac{3\pi \, \alpha_\chi^2}{16 \, m_\chi^2} & \text{ for } m_A \ll m_\chi \, v\\
       \frac{19 \pi \alpha_\chi^2 \, m_\chi^2 \, v^4}{24 \, m_A^4} & \text{ for } m_A \gg m_\chi \, v\\
     \end{array}
   \right. \; ,
\end{equation}
while for particle-antiparticle scattering the low-velocity regime is dominated by $s$-channel pseudoscalar exchange, giving
\begin{equation}
 \sigma_T \simeq \left\{
     \begin{array}{cr}
       \frac{3\pi \, \alpha_\chi^2}{4 \, m_\chi^2} & \text{ for } m_A \ll m_\chi \, v\\
       \frac{4\pi \, \alpha_\chi^2 \, m_\chi^2}{(m_A^2 - 4 m_\chi^2)^2} & \text{ for } m_A \gg m_\chi \, v\\
     \end{array}
   \right. \; .
\end{equation}
In particular, the Born approximation predicts that for scattering via pseudoscalar exchange there is no enhancement of DM self-interactions for small velocities, which has been invoked to help reconcile observational constraints from galaxy clusters and elliptical galaxies with the self-interaction cross sections favoured by small-scale systems~\cite{Feng:2009mn,Buckley:2009in,Loeb:2010gj}. In addition, the absence of an enhancement at small velocities means that, as long as $\alpha_\chi \, m_\chi / m_A < 1$, the momentum transfer cross section is significantly too small to lead to any observable effects in astrophysical systems.

Nevertheless, a very different behaviour can be expected in the strongly-interacting regime~\cite{Buckley:2009in,Tulin:2013teo}. To find the momentum transfer cross section in this case, one needs to solve the Schroedinger equation for the potential created by pseudoscalar exchange. It is well known from pion-nucleon interactions that the relevant potential is the so-called one-pion-exchange potential~\cite{Ericson}:
\begin{equation}
\begin{split}
 V(\mathbf{r}) = & \frac{\alpha_\chi}{3} \left[\frac{e^{-m_A \, r}}{r}-\frac{4\pi}{m_A^2}\delta^3(\mathbf{r})\right] \mathbf{\sigma}_1 \cdot \mathbf{\sigma}_2 \\
& + \frac{\alpha_\chi}{3} \left(1 + \frac{3}{m_A \, r}+\frac{3}{m_A^2 r^2} \right) \frac{e^{-m_A \, r}}{r} S_{12}(\hat{\mathbf{r}}) \; ,
\end{split}
\label{eq:VPion}
\end{equation}
where
\begin{equation}
 S_{12}(\hat{\mathbf{r}}) = 3(\mathbf{\sigma}_1 \cdot \hat{\mathbf{r}})(\mathbf{\sigma}_2 \cdot \hat{\mathbf{r}})-\mathbf{\sigma}_1\cdot\mathbf{\sigma}_2 \; .
\end{equation}
This potential depends on the spin configuration of the two scattering DM particles through the spin variables $\sigma_{1,2}$. For a singlet configuration, the second line of Eq.~\eqref{eq:VPion} vanishes and we obtain an attractive Yukawa-like potential (together with a repulsive contact interaction). For a triplet configuration, on the other hand, the potential at small distances (i.e.\ $m_A \, r \ll 1$) is dominated by the tensor contribution in the second line, resulting in an attractive $1/r^3$ potential, which is singular at the origin.

The one-pion-exchange potential has been discussed in the context of DM both regarding a potential Sommerfeld enhancement of DM annihilation~\cite{Bedaque:2009ri} and an enhancement of DM self-interactions~\cite{Bellazzini:2013foa}, both references being careful to renormalise the divergences resulting from the singular potential. The conclusion is that resonances can significantly boost the interaction rates at low velocities, even though significant tuning may be required to obtain a sufficiently large momentum transfer cross section in small-velocity systems~\cite{Bellazzini:2013foa}.

Although DM self-interactions via pseudoscalar exchange are significantly more involved than for scalar exchange, the latter are strongly constrained when imposing relic density constraints and BBN constraints, because the predicted event rates in direct detection experiments are typically too large~\cite{Kaplinghat:2013yxa} (see~\cite{Kouvaris:2014uoa} for how to potentially evade these constraints). As we will see below, our scenario is much less constrained, since direct detection constraints are largely absent. Consequently, large DM self-interactions via pseudoscalar exchange are a viable and very interesting possibility. We leave a detailed study of the resulting effects to future work and simply show in Fig.~\ref{fig:DMconstraints} the parameter region corresponding to strong coupling as an indication of where DM self-interactions may become important.

We observe from Fig.~\ref{fig:DMconstraints} that for the case of Yukawa-like couplings and the values of $m_\chi$ that we consider, it is always possible to choose $g_Y$ such that strong DM self-interactions can be obtained while evading both flavour constraints and BBN constraints. For $m_A < 2 \, m_\mu$ the required value is typically $g_Y \sim 10^{-4}$. For $2 \, m_\mu < m_A < m_K - m_\pi$ the strong constraints from CHARM require $g_Y \sim 10^{-5}$, while for $m_A$ too heavy to be produced in kaon decays it is typically sufficient to have $g_Y \sim 10^{-3}$. For the case of quark universal couplings, on the other hand, we find that flavour constraints exclude the entire range of $g_q$ that lead to thermal equilibrium between the dark and the visible sector, unless $m_A > m_K - m_\pi$. Consequently, in order to obtain strong DM self-interactions, we must have $m_A \gtrsim 350\:\text{MeV}$, $m_\chi \gtrsim 30 \: \text{GeV}$ and $10^{-7} \lesssim g_q \lesssim 10^{-4}$.

\subsection{Direct detection}
\label{subsec:DD}

DM that scatters though the exchange of a pseudoscalar mediator leads to a spin-dependent interaction at direct detection experiments. As we show in Appendix~\ref{sec:direct}, the differential cross section to scatter off a nucleus is given by
\begin{equation}
\frac{\mathrm{d} \sigma}{\mathrm{d} E}=\frac{m_T}{32 \pi} \frac{1}{v^2} \frac{g_\chi^2}{(q^2 + m_A^2)^2} \frac{q^4}{m_N^2 \, m_{\chi}^2}\sum_{N,N'=p,n} g_N \, g_{N'} \, F_{\Sigma''}^{N,N'}\;,
\end{equation}
where $E$ is the nucleus recoil energy, $v$ is the DM speed, $m_N$ and $m_T$ are the mass of the nucleon and the mass of the target nucleus respectively, $F_{\Sigma''}^{N,N'}$ is the spin form-factor and the coefficients $g_N$ are functions of $g_f$. Crucially, the differential event rate also depends on the momentum transfer~$q$. This is in contrast to the cross section obtained from a scalar or vector mediator. Typically, the momentum transfer in DM-nucleus scattering is approximately $q \sim \mu \, v \sim 100\:\text{MeV}$, where $\mu$ is the reduced mass of the DM-nucleus system and $v \simeq 10^{-3} c$ is the typical speed of DM particles in the Galactic halo. For $m_A \gg q$, the differential cross section is proportional to $q^4 / (m_\chi^2 \, m_N^2)$,  leading to a suppression of differential event rates by up to twelve orders of magnitude. At face value, direct detection signatures with a pseudoscalar mediator are therefore unobservably small and as a result this interaction has frequently been neglected altogether.

For the typical parameters that we consider, however, the mass of the pseudoscalar can be comparable to~-- or smaller than~-- the momentum transfer. In this case, the $q$ dependence of the propagator must be correctly accounted for i.e.\ the factor $(m_A^2 + q^2)^{-2}$ cannot be neglected. It then becomes evident that for $m_A \lesssim q$ the momentum suppression in the numerator is cancelled by a small denominator and the differential event rate can become large enough to lead to observable signatures. 

To compare the current bounds from direct detection experiments with our other constraints, we consider the recent results from the LUX experiment~\cite{Akerib:2013tjd}.\footnote{The SIMPLE~\cite{Felizardo:2011uw}, PICASSO~\cite{Archambault:2012pm} and COUPP~\cite{Behnke:2012ys} experiments also give similar constraints.} Although there is no published limit for this case, it is straightforward to derive a bound at 90\% confidence level. We follow the analysis strategy described in~\cite{Buchmueller:2014yoa} based on the `pmax' statistic~\cite{Yellin:2002xd} and we have checked that the binned likelihood method from~\cite{Feldstein:2014ufa} gives very similar results. We take the xenon form factors $F_{\Sigma''}^{N,N'}$ from~\cite{Fitzpatrick:2012ix}, the Earth's velocity from~\cite{McCabe:2013kea,Lee:2013xxa}, assume the Standard Halo Model and use $v_0=230 \: \text{km}/\text{s}$, $v_{\rm{esc}}=550 \: \text{km}/\text{s}$ and $\rho=0.3\:\text{GeV/cm}^3$ for the astrophysical parameters~\cite{McCabe:2010zh}. 

The parameter regions excluded by LUX constraint are shown in Fig.~\ref{fig:DMconstraints}. As expected, we find that the LUX limit is more constraining for smaller values of $m_A$ because of the enhancement by a small denominator. For the case of Yukawa-like couplings (left panels), LUX excludes values of $g_Y\gtrsim\mathcal{O}(10^{-1})$. For the case of universal couplings to quarks (right panels), LUX provides no constraints i.e.~the LUX constraint on $g_q g_{\chi}$ is higher than the values required to obtain the observed DM abundance in all parameter regions. It is obvious from this figure that for all the scenarios we consider, LUX cannot probe the parameter region allowed by flavour constraints. In fact, the allowed parameter region is out of reach even for next-generation direct detection experiments.

\subsection{Indirect detection}

Indirect detection experiments are sensitive to the $s$-wave annihilation of DM into SM fermions $\bar{\chi}\chi\to \bar{f}f$. Our estimate of the sensitivity of near-future searches for gamma rays from DM annihilation in Dwarf Spheroidal Galaxies using Fermi-LAT data is shown in Fig.~\ref{fig:DMconstraints}. This estimate is based on the preliminary 95\% confidence level bound on $\langle \sigma v \rangle_{\bar{\chi} \chi \rightarrow  \bar{b} b}$ presented in~\cite{Fermi} and the additional assumption that a similar bound will apply for annihilation into light quarks, i.e.\ $\langle \sigma v \rangle_{\bar{\chi} \chi \rightarrow  \bar{q} q} \sim \langle \sigma v \rangle_{\bar{\chi} \chi \rightarrow  \bar{b} b}$. This assumption is in agreement with previous results from the Fermi-LAT collaboration~\cite{Ackermann:2013yva}. 

In Fig.~\ref{fig:DMconstraints} we have fixed $g_{\chi}$ by the requirement to obtain the observed DM relic abundance. It is therefore not a surprise that the Fermi-LAT bound constrains the parameter region where $s$-wave annihilation into SM fermions dominates (see~Fig.~\ref{fig:relic}), since the $p$-wave annihilation into two pseudoscalars is unobservably small. The annihilation cross section constrained by Fermi-LAT is larger than the thermal cross section (i.e.~$2.5 \cdot 10^{-26}\, \text{cm}^3/\text{s}$) for DM mass above approximately $100\:\text{GeV}$. This is why there is no Fermi-LAT constraint for the $m_{\chi}=100\:\text{GeV}$ cases in Fig.~\ref{fig:DMconstraints}.
We see clearly that the indirect detection constraints only become more constraining than the flavour constraints above the $B$ meson mass, when the flavour constraints lose sensitivity.

\section{Implications for dark matter signals}
\label{sec:detection}

In the previous section we have focussed on the general bounds that can be placed on thermal DM. However, there are two longstanding experimental signatures that are consistent with a DM origin. The first is the DAMA annual modulation signal~\cite{Bernabei:2010mq, Bernabei:2013xsa} and the second is the Galactic Centre excess of gamma rays observed by Fermi-LAT~\cite{Goodenough:2009gk,Hooper:2010mq,Hooper:2011ti,Abazajian:2012pn,Gordon:2013vta,Macias:2013vya,Daylan:2014rsa,Zhou:2014lva,Calore:2014xka,Calore:2014nla}. In both cases, pseudoscalar mediated interactions between the DM and SM fermions have been proposed to explain the putative signals. In this section we compare the flavour constraints discussed above to the respective preferred parameter regions.

\subsection{DAMA}

The DAMA modulation signal remains a longstanding puzzle. Under the standard assumptions of spin-independent contact interactions, many other direct detection experiments exclude the preferred DM scattering cross section by many orders of magnitude~\cite{Agnese:2013jaa,Akerib:2013tjd,Agnese:2014aze,Angloher:2014myn}.  It was pointed out in~\cite{Arina:2014yna} that pseudoscalar exchange may be one avenue for reconciling these direct detection experiments. For DM scattering via pseudoscalar exchange, both the shape of the recoil spectrum and the relevant nuclear matrix elements and form factors differ from the standard analysis of direct detection experiments under the assumption of spin-independent contact interactions. In particular, in the non-relativistic limit the pseudoscalar couples more strongly to the proton spin than to the neutron spin.\footnote{The precise value of the ratio of neutron coupling to proton coupling, $g_n / g_p$, depends sensitively on the assumed current quark masses and nuclear matrix elements. By varying these parameters within their $1\sigma$ range, we find for Yukawa-like couplings $-0.4 \lesssim g_n / g_p \lesssim 0$ and for universal couplings $-0.25 \lesssim g_n / g_p \lesssim 0.2$. In the following, we adopt the values from~\cite{Arina:2014yna} and take $g_p / g_n = -4.1$ for Yukawa-like couplings and $g_p / g_n = -16.4$ for universal couplings.} The resulting effects lead to a large enhancement of the signal expected in experiments such as DAMA, which have nuclei with  unpaired protons, compared to experiments such as LUX, which contain nuclei with  unpaired neutrons.

As emphasised in Sec.~\ref{subsec:DD}, in order to obtain observable scattering rates from pseudoscalar exchange in any direct detection experiment, the mediator mass must be comparable to the momentum transfer $q$. Once the mediator mass becomes smaller than the typical momentum exchange in DM scattering, the event rate becomes independent of $m_A$ and decreasing the mediator mass further does not lead to an additional enhancement.  In the case of DAMA, an observed energy of $E_\text{ee} = (2\text{--}4)\:\text{keV}$ corresponds to a momentum transfer for scattering off iodine and sodium of $q _{\rm{I}}= \sqrt{2 \, m_{\rm{I}} \, E_\text{ee} / Q_{\rm{I}}} = (70\text{--}100)\:\text{MeV}$ and $q _{\rm{Na}}= \sqrt{2 \, m_{\rm{Na}} \, E_\text{ee} / Q_{\rm{Na}}} = (17\text{--}24)\:\text{MeV}$ respectively, where we have assumed quenching factors of $Q_{\rm{I}}=0.09$ and $Q_{\rm{Na}}=0.3$ and ignored channeling~\cite{Bozorgnia:2010xy}. Consequently, the transition between contact interactions and long-range interactions for DM scattering off iodine occurs around $m_A \sim 100\:\text{MeV}$. This observation is in contrast with the treatment in~\cite{Arina:2014yna}, where contact interactions are assumed to remain valid down to mediator masses of around $30\:\text{MeV}$.

To analyse DAMA, we include the energy resolution from NaI and LIBRA weighted appropriately and perform a goodness-of-fit test to 12 equally spaced bins between $2$~keVee and $8$~keVee. We take the sodium and iodine form factors from~\cite{Fitzpatrick:2012ix} and use the astrophysical parameters mentioned in Sec.~\ref{subsec:DD}. We find that, in order to fit the annual modulation observed by DAMA, two mass values are preferred. For scattering dominantly off iodine, we require $m_\chi \approx 36\:\text{GeV}$, while for scattering dominantly off sodium, $m_\chi \approx 9\:\text{GeV}$. Although we focus our discussion on the case of scattering off iodine here for clarity the conclusions are the same for scattering off sodium. 
Furthermore, we focus on the case where the pseudoscalar couples either universally to all quarks or only to the third generation, since the suppression of LUX is strongest in this case~\cite{Arina:2014yna}.

For $m_\chi \approx 36\:\text{GeV}$ and universal quark couplings, we find that the pseudoscalar couplings must approximately satisfy the relation
\begin{equation}
{g_q \, g_\chi} \simeq
\begin{cases}
  0.06 \left(\frac{m_A}{100 \:\text{MeV}}\right)^2 & \text{ for } m_A \gg 100\:\text{MeV}\\
       0.05 & \text{ for } m_A \ll 100\:\text{MeV} \, .
\end{cases}
\end{equation}

Substituting these values into the expression for $\langle \sigma v\rangle_{\bar{\chi}\chi\rightarrow \bar{q}q}$ in Eq.~(\ref{eq:annqq}), we find that for any value of $m_A$, the couplings favoured by DAMA predict an annihilation cross section significantly larger than the thermal cross section, i.e.\ $\langle \sigma v \rangle_{\bar{\chi} \chi\rightarrow \bar{q} q} \gg 2.5 \cdot 10^{-26} \: \text{cm}^3/\text{s}$. In other words, to account for the DAMA modulation, the couplings must be so large that DM annihilates too efficiently in the early Universe and will be underproduced even when we neglect the direct annihilation into pseudoscalars. The presence of this additional annihilation channel will only make the tension worse. Assuming that the pseudoscalar couples only to the third generation of quarks (a scenario advocated in~\cite{Arina:2014yna}) also does also not help to reduce the tension.

We conclude that within our model the DAMA signal is incompatible with the assumption that DM is a thermal relic. Even if we were to invoke a different production mechanism for DM, the resulting annihilation cross section would still be so large that the favoured parameter region can be excluded by indirect detection experiments~\cite{Williams:2012pz}. Nevertheless, the indirect detection constraints can be avoided and the observed DM relic density can be obtained if the dark sector has an initial particle-antiparticle asymmetry similar to the baryon asymmetry of the visible sector. In the presence of such an asymmetry, all anti-particles annihilate away so that the final DM number density is entirely determined by the asymmetry~\cite{Nussinov:1985xr, Griest:1986yu, Barr:1990ca, Kaplan:1991ah, Kaplan:2009ag}.

\begin{figure}[tb]
\centering
\includegraphics[width=0.46\textwidth, clip]{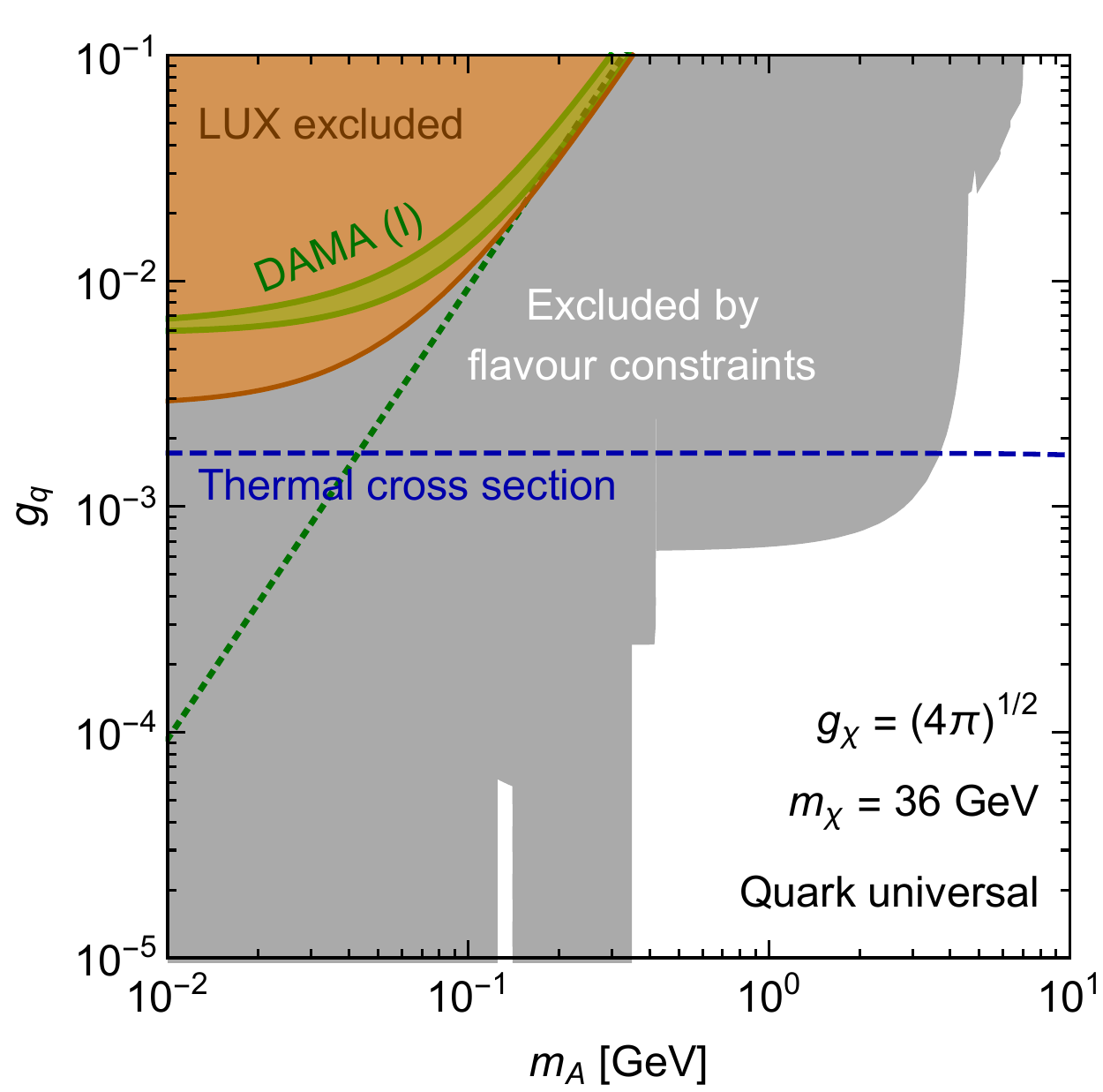}
\quad
\includegraphics[width=0.46\textwidth, clip]{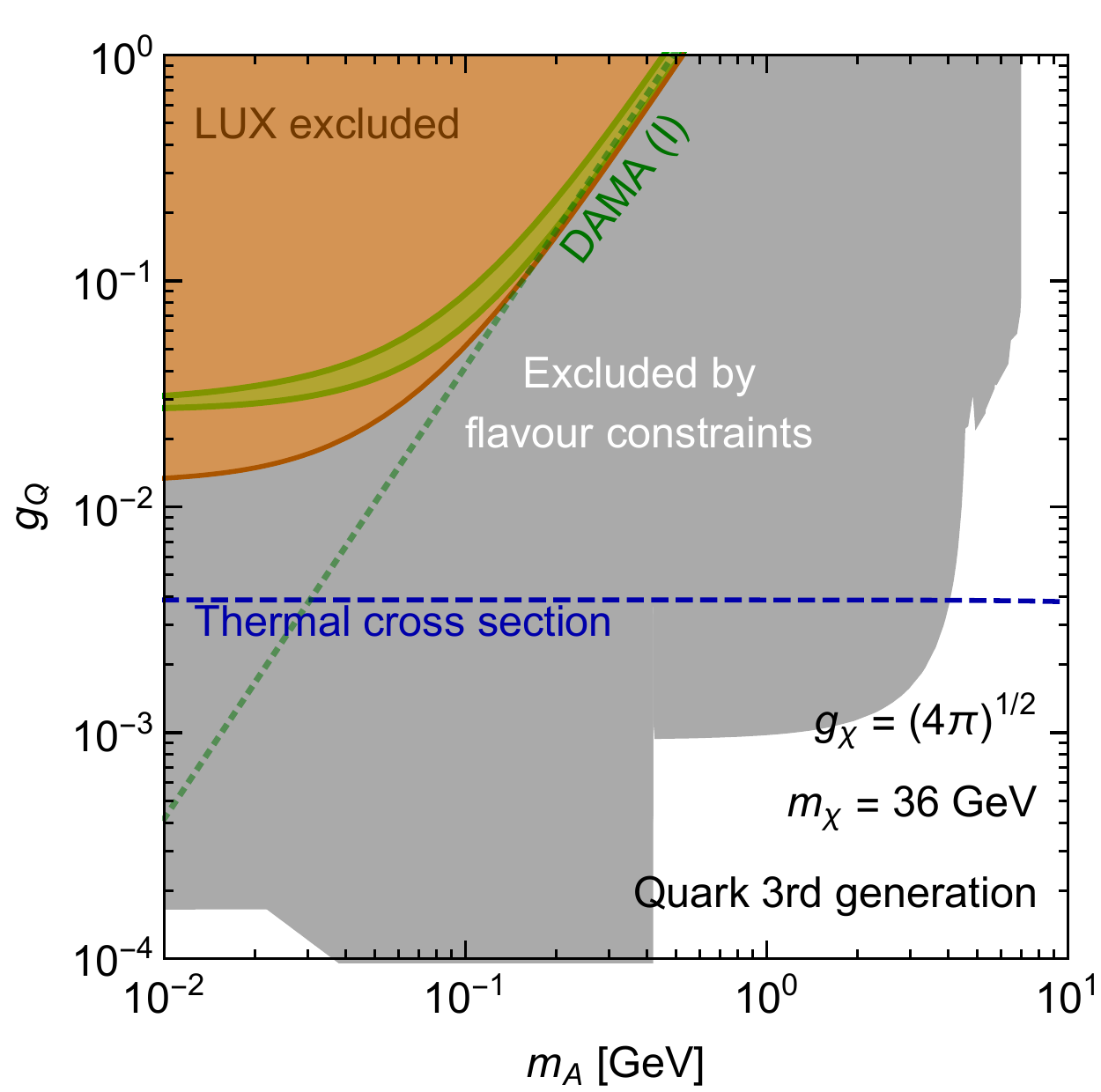}
\caption{The coupling $g_q$ favoured by DAMA as a function of $m_A$ for $m_\chi = 36\;\text{GeV}$ and $g_\chi = \sqrt{4\pi}$ together with the flavour constraints from Sec.~\ref{sec:results}. The flavour constraints exclude the DAMA region by several orders of magnitude. The (I) indicates that scattering in DAMA is off iodine. The blue dashed line indicates the value of $g_q$ that gives $\langle \sigma v \rangle_{\bar{\chi} \chi\rightarrow \bar{q} q} = 2.5 \cdot 10^{-26}\: \text{cm}^3/\text{s}$, the green dashed line represents the naive extrapolation of contact interactions down to small values of $m_A$. Additional constraints (e.g.\ from DM self-interactions) are not shown. Note the change of scale between the two panels.}
\label{fig:DAMA1}
\end{figure}

While the constraints on the DM annihilation cross section can thus be avoided, the flavour constraints presented in Sec.~\ref{sec:results} still apply. Of course, these constraints only probe the quark coupling $g_q$. Nevertheless, we cannot make $g_\chi$ arbitrarily large and therefore $g_q$ cannot be arbitrarily small if we want to keep the product $g_q \, g_\chi$ fixed to the value preferred by DAMA. In Fig.~\ref{fig:DAMA1} we consider the extreme case $g_\chi = \sqrt{4\pi}$ and show that even then, the values of $g_q$ required to account for the DAMA modulation are excluded by many orders of magnitude. Fig.~\ref{fig:DAMA1} also shows the naive extrapolation of contact interactions down to small mediator masses. In agreement with~\cite{Arina:2014yna} we find that this line crosses the relic density constraint around $m_A \sim (30\text{--}40)\:\text{MeV}$.

Let us briefly discuss whether other possible interactions can improve the agreement between DAMA and the flavour constraints. In this context it is important to note that the enhancement of DAMA compared to LUX is largely a result of how the pseudoscalar couples to SM quarks, i.e.\ it is mostly independent of the DM-pseudoscalar interactions. Therefore a possible modification of the dark sector is to consider a $CP$-violating coupling between DM and the pseudoscalar,  since the resulting coupling structure between DM and nuclei is the same as in the $CP$-preserving case~\cite{DelNobile:2013sia}. We therefore now introduce an additional $CP$-violating coupling between DM and the pseudoscalar:
\begin{equation}
\mathcal{L}_\text{DM} = i \, g_\chi^P \, A \, \bar{\chi} \gamma^5 \chi + g_\chi^S \, A \, \bar{\chi} \chi\; .
\end{equation}
Since the term proportional to $g_\chi^S$ violates $CP$, one would expect that $g_\chi^P \gg g_\chi^S$, so that this additional term does not significantly change the freeze-out of DM. In particular,~$g_\chi^S$ does not induce $s$-wave annihilation of DM into pseudoscalars. Moreover, the contribution of $g_\chi^S$ to the annihilation of DM into quarks is $p$-wave suppressed and hence completely negligible.

Nevertheless, even a small $CP$-violating coupling will significantly change the predictions for direct detection experiments~\cite{Dienes:2013xya}. The reason is that the additional term drastically reduces the momentum suppression. While we previously considered differential event rates proportional to $(g_\chi^P)^2 \, g_N^2 \, q^4 / (16 \, m_\chi^2 \, m_N^2)$, the new term will introduce an additional contribution proportional to $(g_\chi^S)^2 \, g_N^2 \, q^2 / (4 \, m_N^2)$ (see Appendix~\ref{sec:direct}). This new contribution will dominate, and therefore enhance the differential event rate, as soon as $g_\chi^S / g_\chi^P > q / (2 \, m_\chi) \sim 10^{-3}$. 
\begin{figure}[tb]
\centering
\includegraphics[width=0.46\textwidth, clip]{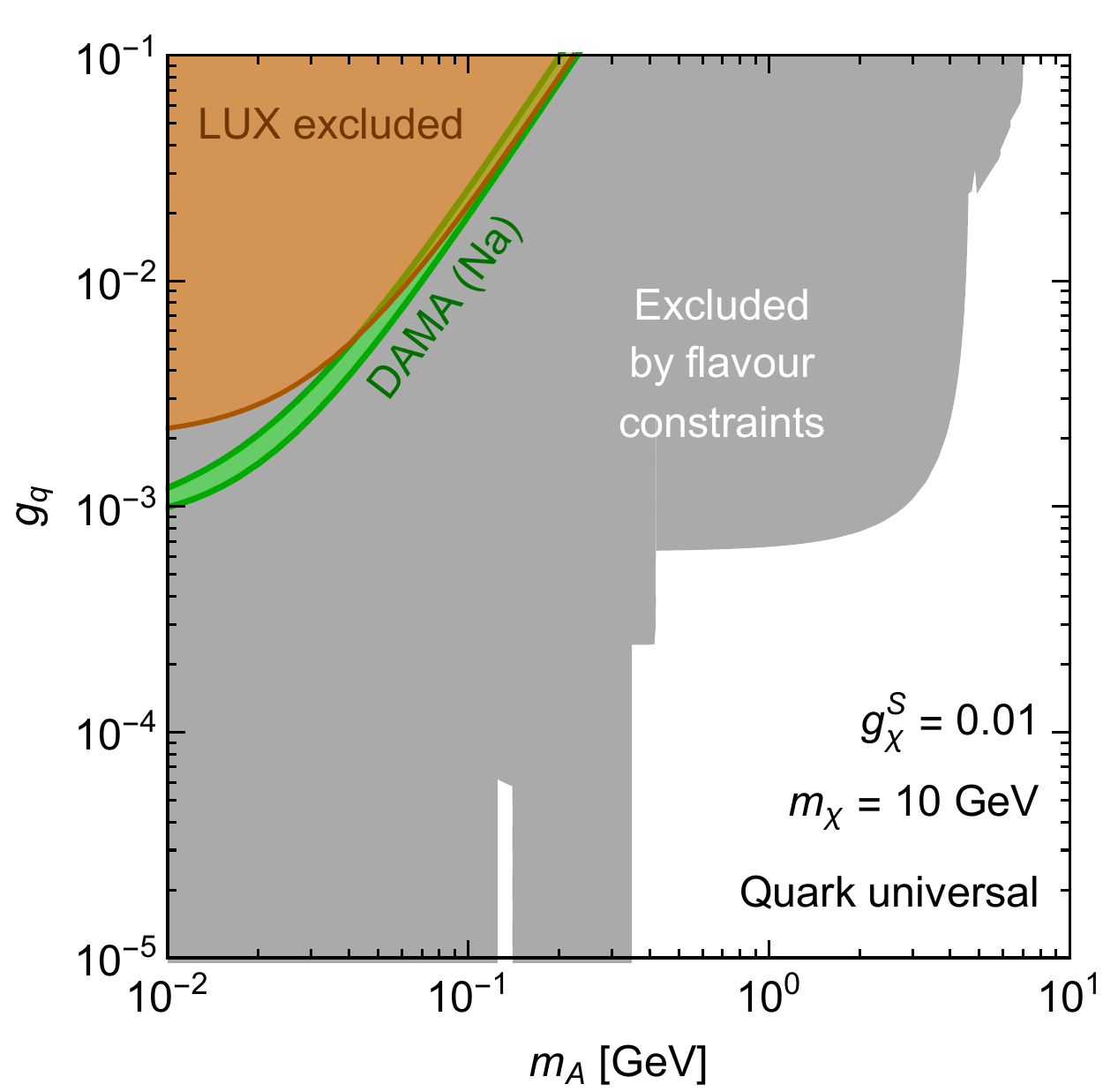}
\quad
\includegraphics[width=0.46\textwidth, clip]{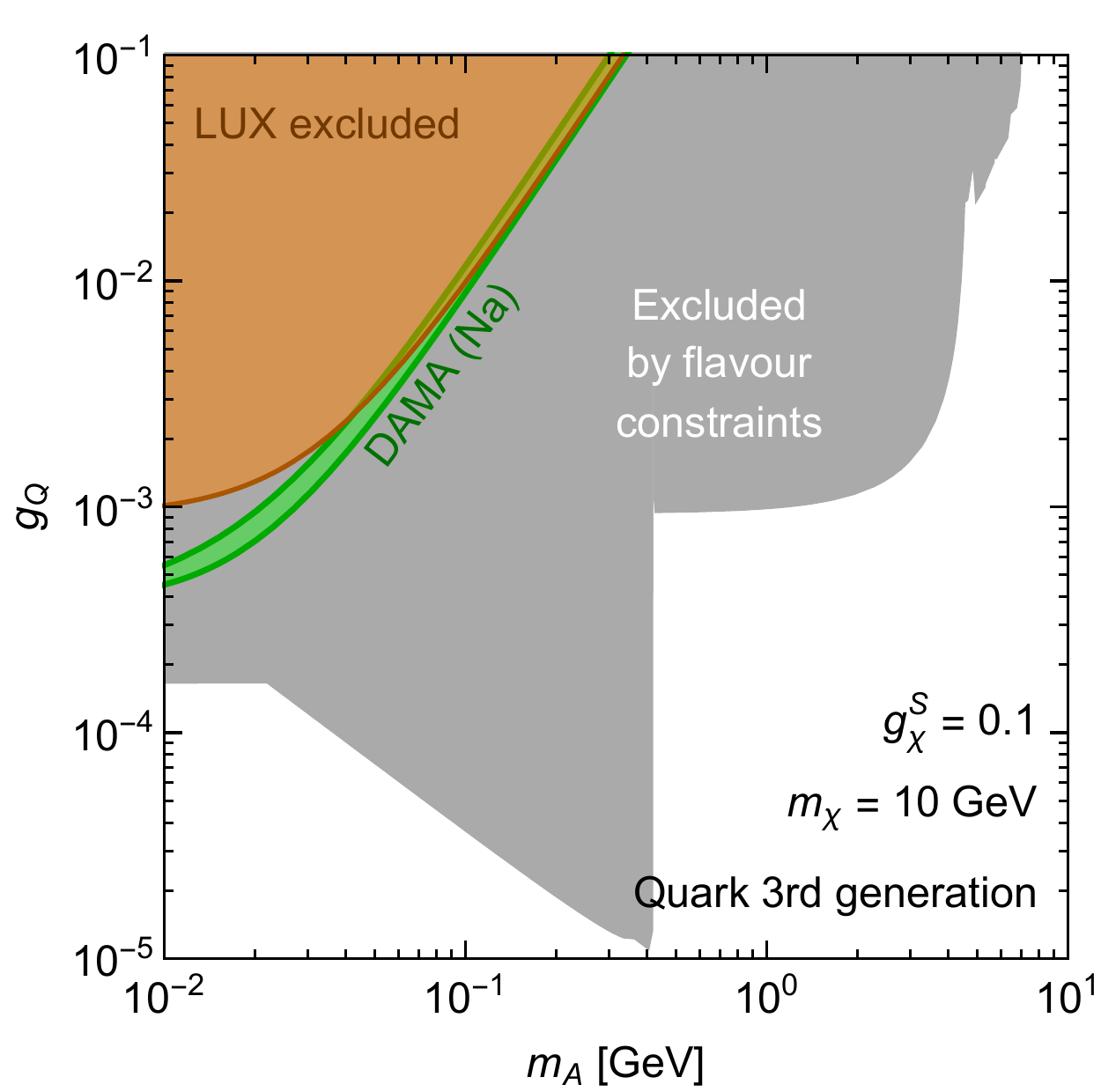}
\caption{Parameter region favoured by DAMA in the $g_q$-$m_A$ parameter plane in the presence of a $CP$-violating coupling $g_\chi^S$. The left-hand side assumes universal quark couplings and $g_\chi^S = 0.01$, the right-hand side assumes couplings only to the third generation and $g_\chi^S = 0.1$.  The (Na) indicates that scattering in DAMA is off sodium. The flavour constraints exclude the DAMA region by several orders of magnitude. }
\label{fig:DAMA2}
\end{figure}

The modified momentum dependence implies that experiments probing large momentum transfer receive a smaller enhancement than for the case $g_\chi^S = 0$. Indeed, for $m_A^2 \ll q^2$, the cross section is proportional to $1/q^2$, thus favouring experiments with low energy thresholds and light target materials. In the case of DAMA, we find that for small mediator masses, DM scattering on sodium always dominates over scattering on iodine, even for heavy DM. Consequently, we no longer obtain two separate best-fit DM masses corresponding to scattering on the two different isotopes, but rather there is now only the low-mass solution with $m_\chi \simeq 10\:\text{GeV}$.

In Fig.~\ref{fig:DAMA2} we show the DAMA best-fit region for this DM mass and two different scenarios. In the left panel, we fix $g_\chi^S = 0.01$, so that the $CP$-violating coupling plays no role in cosmology. Furthermore, we assume that the pseudoscalar has universal couplings to all quarks and that $CP$ is not violated in the visible sector. In the right panel of Fig.~\ref{fig:DAMA2}, we consider a more optimistic choice, namely $g_\chi^S = 0.1$. As long as $g_\chi^S \lesssim 0.3$, DM can still be a thermal relic, provided $g_\chi^P \sim g_\chi^S$, and there are no severe constraints from DM self-interactions. We make the interesting observation that for sufficiently low pseudoscalar masses, the enhancement for small momentum transfer is sufficient to reconcile DAMA and LUX. However, even if we make the additional optimistic assumption that the mediator couples only to the third generation of quarks, the parameter region favoured by DAMA remains solidly excluded by flavour constraints. Thus even a $CP$-violating coupling (together with additional optimistic assumptions) is insufficient to explain the DAMA modulation while at the same time evading flavour constraints.

\subsection{Fermi Galactic Centre excess}

\begin{figure}[tb]
\centering
\includegraphics[width=0.46\textwidth]{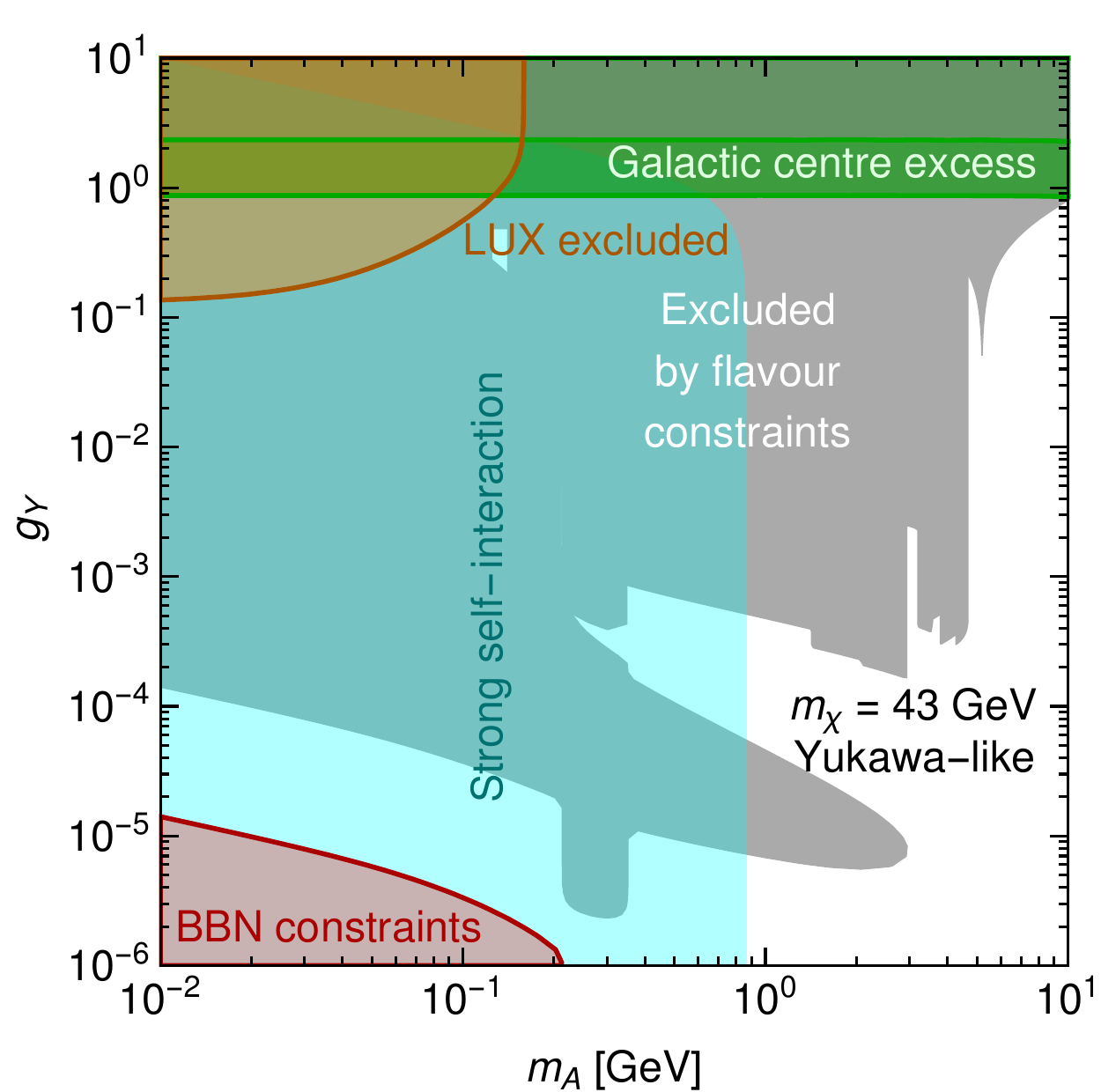}
\quad
\includegraphics[width=0.46\textwidth]{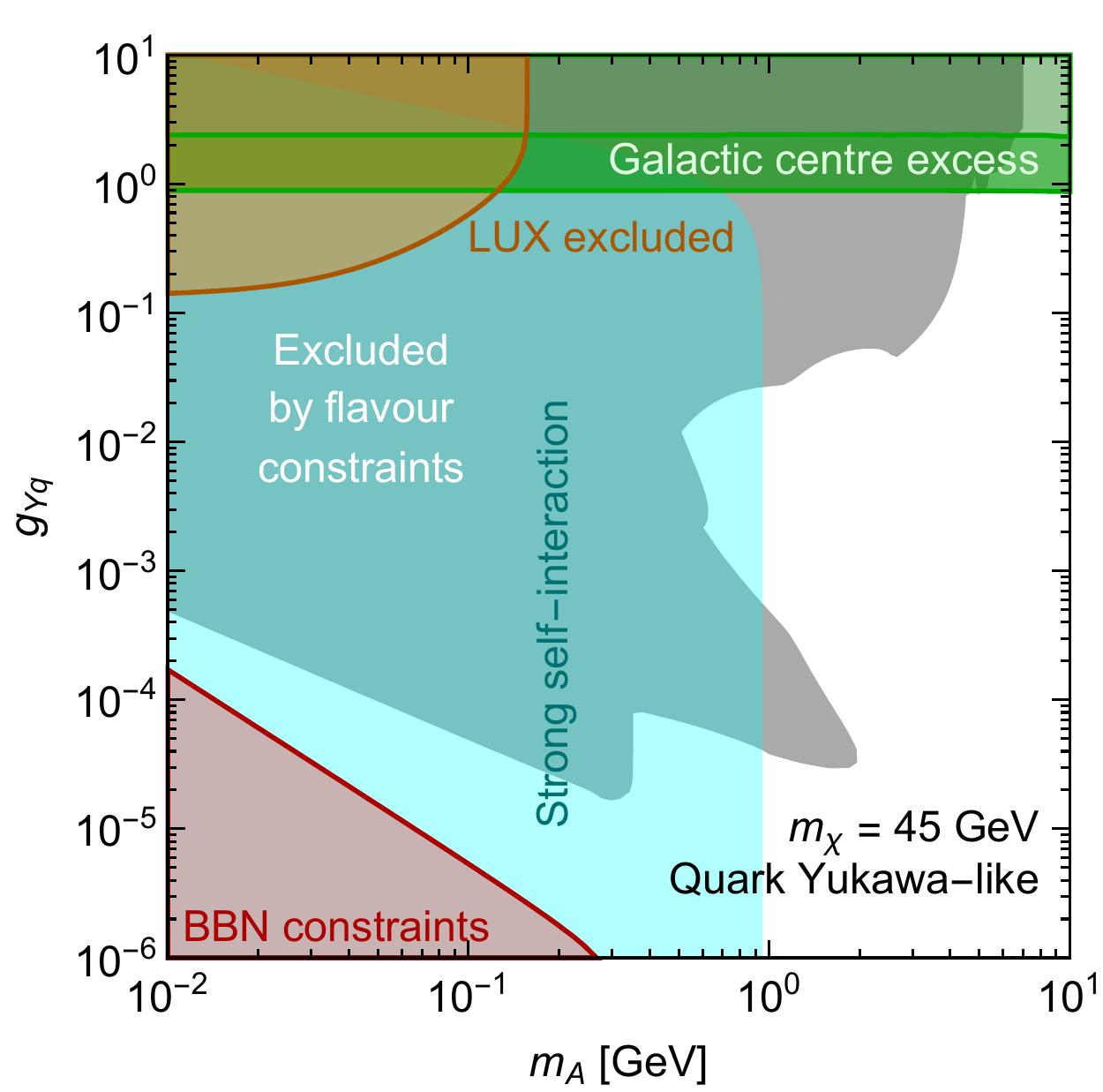}

\includegraphics[width=0.46\textwidth]{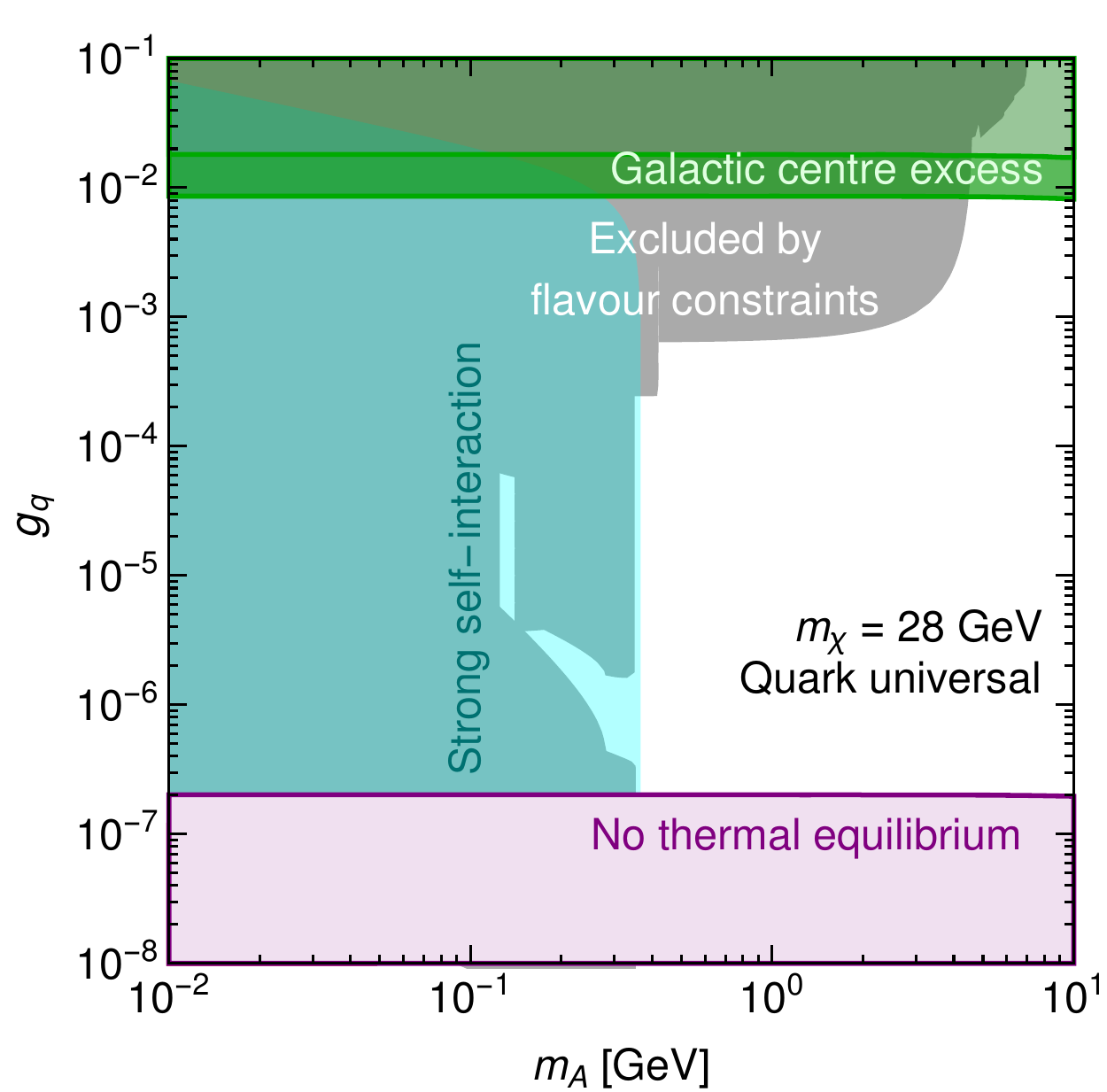}
\quad
\includegraphics[width=0.46\textwidth]{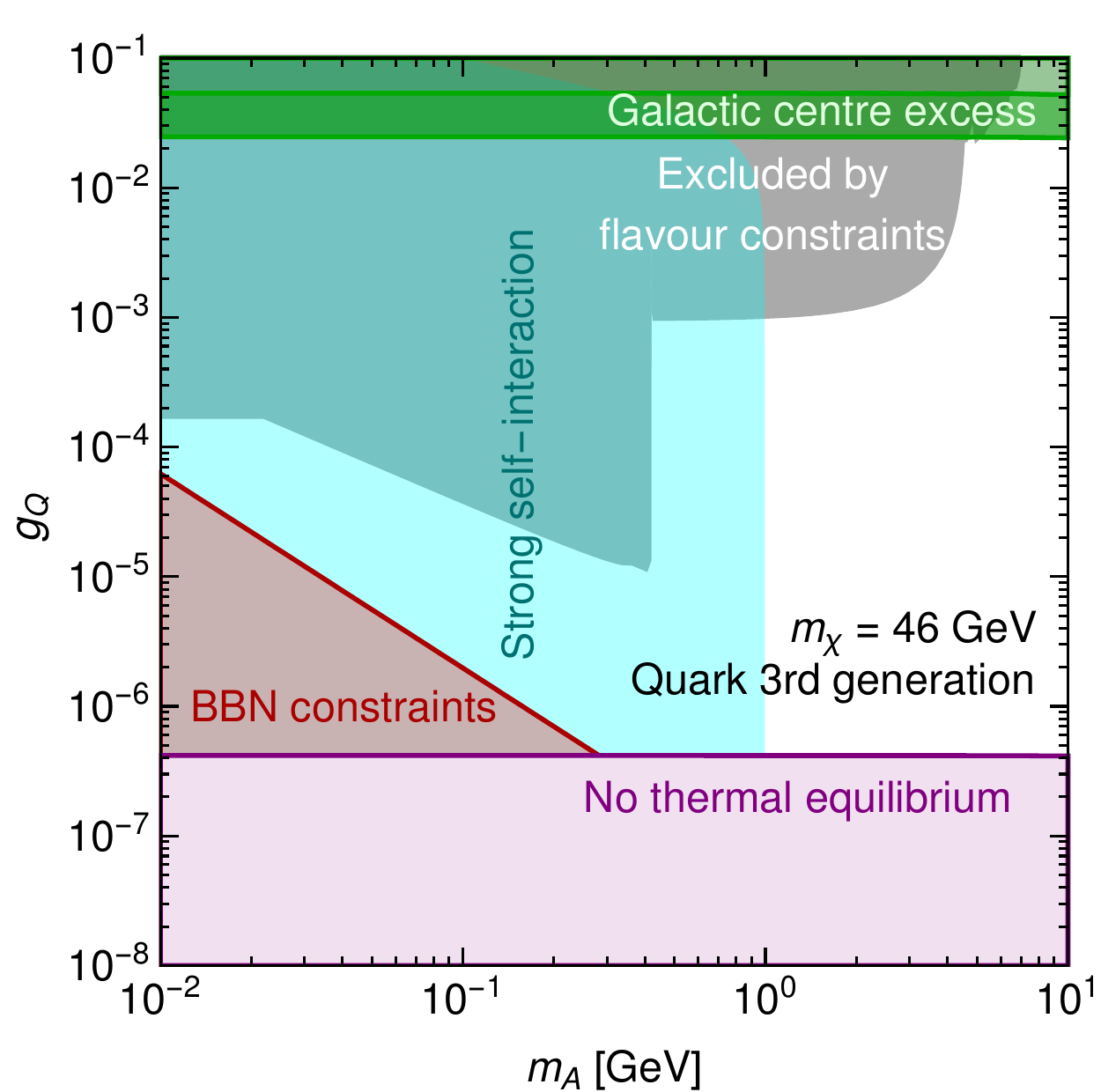}
\caption{The parameter region favoured by the Galactic Centre excess (green). As before, $g_\chi$ has been fixed by the requirement to obtain the correct DM relic abundance.  In all cases, an interpretation of the Galactic Centre excess with a pseudoscalar mediator requires $m_A\gtrsim5\;\text{GeV.}$  The lighter region corresponds to $\langle \sigma v \rangle_{\bar{\chi}\chi\rightarrow\bar{q}q} > 2 \cdot 10^{-26} \;\text{cm}^3/\text{s}$, the darker to $\langle \sigma v \rangle_{\bar{\chi}\chi\rightarrow\bar{q}q} > 10^{-26} \;\text{cm}^3/\text{s}$.}
\label{fig:GCE}
\end{figure}

Using data from Fermi-LAT, a number of independent groups have reported an excess in gamma rays at the Galactic Centre above the expected astrophysical emission~\cite{Goodenough:2009gk,Hooper:2010mq,Hooper:2011ti,Abazajian:2012pn,Gordon:2013vta,Macias:2013vya,Daylan:2014rsa,Zhou:2014lva,Calore:2014xka,Calore:2014nla}. The excess is largest for gamma-ray energies around a few GeV and it has many features, including the morphology and radial profile, expected from DM annihilating to SM fermions. Previous analyses have shown that the preferred annihilation cross section is around $10^{-26}\: \text{cm}^3/\text{s}$ for  annihilation into SM fermions. Uncertainties in the DM halo parameters translate to about a factor five uncertainty in the annihilation cross section~\cite{Calore:2014nla}, so we do not quote a more precise value.

A majority of the proposals to explain the Galactic Centre excess within a particle physics model have included DM annihilation into SM fermions mediated by a pseudoscalar mediator~\cite{Boehm:2014hva, Hektor:2014kga,Arina:2014yna,Alves:2014yha,Berlin:2014tja,Izaguirre:2014vva,Cerdeno:2014cda,Ipek:2014gua,Abdullah:2014lla,Martin:2014sxa,Berlin:2014pya,Han:2014nba,Cheung:2014lqa,Huang:2014cla,Ghorbani:2014qpa,Cahill-Rowley:2014ora,Guo:2014gra,Cao:2014efa,Freytsis:2014sua}. This annihilation process is $s$-wave and the annihilation cross section compatible with the excess are easily achievable. An advantage of a pseudoscalar mediator, which we have discussed above, is that scattering at direct detection searches is suppressed. It is therefore relatively straightforward to avoid constraints from these experiments. 

To determine the preferred DM mass for the different coupling structures, we generate the gamma rays from DM annihilation with Pythia 8.186~\cite{Sjostrand:2007gs} and fit to the spectral shape of the  Galactic Centre excess  data in~\cite{Calore:2014xka,Calore:2014nla}, which is sensitive to the mass, but not the normalisation of the signal. We find $m_\chi = 43.5^{+5.2}_{-4.5}\:\text{GeV}$, $m_\chi = 44.9^{+5.3}_{-4.6}\:\text{GeV}$, $m_\chi = 28.1^{+3.6}_{-2.8}\:\text{GeV}$ and $m_\chi = 45.9^{+5.4}_{-4.7}\:\text{GeV}$ for the case of Yukawa-like couplings, quark Yukawa-like couplings, universal quark couplings and universal couplings only to the quark third generation, respectively. 

\begin{emergency}{0.7em}
As already mentioned, owing to the large uncertainties in the DM halo we do not attempt to reproduce a precise value of the cross section. Instead, Fig.~\ref{fig:GCE} indicates the parameter regions that give $\langle \sigma v \rangle_{\bar{\chi}\chi\rightarrow\bar{q}q} > 10^{-26} \:\text{cm}^3/\text{s}$ (dark green) and \mbox{$\langle \sigma v \rangle_{\bar{\chi}\chi\rightarrow\bar{q}q} > 2 \cdot 10^{-26}\: \text{cm}^3/\text{s}$} (light green), both of which are consistent with the cross section required to fit the Galactic Centre excess. The upper value of the cross section is chosen because it is just excluded by the preliminary Fermi-LAT Dwarf Spheroidal limit presented in~\cite{Fermi}. In this figure we have again fixed $g_\chi$ by the requirement to obtain the observed relic abundance. Therefore the dark green region indicates parameter space that can fit the Galactic Centre excess, be consistent with the preliminary Fermi-LAT Dwarf Spheroidal limit and obtain the observed relic abundance. This demonstrates a second advantage of DM models with a pseudoscalar mediator. The $s$-wave annihilation of DM into SM fermions can be below the thermal value (i.e.~$2.5 \cdot 10^{-26}\; \text{cm}^3/\text{s}$) so that it is not in tension with the Fermi-LAT Dwarf limits, while the larger $p$-wave annihilation into pseudoscalars then ensures that the observed relic abundance is achieved.
\end{emergency}

Comparing the Galactic Centre excess region with the bounds from rare decays presented in Sec.~\ref{sec:results}, we see from Fig.~\ref{fig:GCE} that an interpretation of the Galactic Centre excess in terms of a pseudoscalar mediator requires $m_A \gtrsim 10\,\text{GeV}$. For mediator masses above this, we also see that it is impossible to obtain large enough DM self-interaction cross sections or large enough event rates in direct detection experiments to confirm these models with these search strategies.

\section{Conclusions}
\label{sec:discussion}

Pseudoscalar mediators coupling  the visible and dark sectors are interesting from both the model-building and phenomenological perspectives. In this article we have investigated constraints on light pseudoscalar mediators using results from flavour physics, which is less studied as a tool to constrain theories and properties of dark matter. In particular, we have calculated the branching ratios for various rare meson decays resulting from the loop-induced flavour-changing couplings of the pseudoscalar. In this context, we have also discussed general theoretical problems that can arise from assuming an arbitrary coupling structure not consistent with Minimal Flavour Violation.

We have then focussed on the various decay channels for the pseudoscalar  (see Fig.~\ref{fig:BRY}) and how the resulting experimental signatures can be constrained with existing searches (see Tab.~\ref{tab:PseudoSearches}). The limits we obtain on the couplings of the pseudoscalar to the Standard Model are summarised in Figs.~\ref{fig:yuk} and~\ref{fig:univqu} for a variety of pseudoscalar coupling structures: Yukawa-like (both including and excluding couplings to leptons), universal to quarks only and universal to third quark family only. The case of invisibly decaying pseudoscalars is also strongly constrained (see Fig.~\ref{fig:invisible}).

 Our results also suggest  new searches which could be performed to tighten the limits on the parameter space. In particular, a search for $B \rightarrow K \gamma\gamma$ would set constraints on models where the mediator couples only to quarks and has a mass below the three-pion threshold. Displaced vertex searches for $K_L \rightarrow \pi^0 \ell^+ \ell^-$ and $B\rightarrow K \ell^+ \ell^-$ would also help to improve limits for smaller mediator couplings which lead to longer decay lengths.

Cosmological and astrophysical measurements, on the other hand, allow us to set constraints on the direct couplings of such a pseudoscalar to dark matter and on the interactions between dark matter and Standard Model quarks mediated by it. In particular, it is generally possible to adjust the couplings such that dark matter freeze-out yields the observed dark matter relic density (see Fig.~\ref{fig:relic}). Doing so allows us to directly compare cosmological and astrophysical constraints with the bounds from rare decays and beam dump experiments and the results from traditional searches for dark matter in direct and indirect detection experiments. We present this comparison in Fig.~\ref{fig:DMconstraints} and find that the flavour constraints are substantially stronger. While we have focussed on Dirac dark matter, similar results also hold for the Majorana case.

From these results we may draw a number of interesting conclusions. Firstly, a light pseudoscalar mediator can induce strong dark matter self-interactions via non-perturbative effects. These effects will be largest for small mediator masses, which is exactly the parameter region most tightly constrained by rare decays. Consequently, large self-interactions will only be viable if the pseudoscalar couples very weakly to Standard Model particles, so that it does not seem possible to obtain both large self-interactions and at the same time a dark matter signal from direct or indirect detection experiments given current bounds (see Fig.~\ref{fig:DMconstraints}). Due to the complexity of the pseudoscalar potential, we have not attempted to fully characterise the regions in parameter space which may explain the various small-scale structure discrepancies but leave this problem to future work.

Secondly, pseudoscalar mediators have also been of interest since they naturally lead to suppressed event rates in direct detection experiments (`Coy Dark Matter'). While this suppression helps to evade the strong bounds from experiments such as LUX, observable event rates in direct detection experiments may still be obtained if the mediator is sufficiently light. In particular, it was recently suggested that a $\mathcal{O}(10^2)\:\text{MeV}$ pseudoscalar mediator could explain the DAMA modulation. Our results not only rule out such an interpretation, but indeed the possibility of any direct detection signal observed in the foreseeable future being due to pseudoscalar exchange (see Fig.~\ref{fig:DAMA1}). 

Finally, while the direct detection cross sections for pseudoscalar mediators are suppressed, the corresponding annihilation cross sections are not, making them of interest in explaining the Galactic Centre excess seen in the Fermi-LAT data. We find that in order for such an explanation to be consistent with flavour constraints, the mediator mass must be significantly greater than the mass of the $B$ mesons i.e. $m_A \gtrsim 10$~GeV (see Fig.~\ref{fig:GCE}). Improved measurements of $\mathrm{BR}(B_s \to \mu^+ \mu^-)$ provide an exciting opportunity for extending this bound to larger mediator masses.

While we have focussed in this article on pseudoscalar mediators, rare decays should also set strong constraints on light scalar mediators, which have also been studied in the context of self-interactions. Indeed, our work suggests that a detailed comparison of the two cases together with an analysis of the pseudoscalar potential will be very relevant in this context. A dedicated in-depth study of measurements from proton beam-dump experiments would also be highly interesting. Finally, it would certainly warrant further investigation how the particular coupling structures of the pseudoscalar considered in this paper may be obtained from a more complete high-energy theory and how our calculations would be modified in a given UV completion.

At first sight, flavour physics appears unrelated to the puzzle of dark matter. Whenever the mediator of the interactions between dark matter and Standard Model particles is light, however, constraints from rare decays can give very relevant and highly complementary information. In such cases our study  demonstrates the need to take these constraints into account when constructing phenomenological models in order to predict or explain signals in direct or indirect detection experiments. Clearly, input from flavour physics is a valuable and underappreciated tool in the challenge to identify the properties of dark matter.

\section*{Note added}

In the previous version of this study there was a sign mistake in the calculation of the flavour-changing amplitudes $b \rightarrow s \, A$ and $s \rightarrow d \, A$ in section~\ref{sec:penguin}, which has been corrected in the present version. As a result the vertex counter-term does not cancel the divergence of the one-loop diagram, so that the flavour-changing amplitudes $b \rightarrow s \, A$ and $s \rightarrow d \, A$ are in general divergent~\cite{Freytsis:2009ct}. We now find stronger constraints for the case of Yukawa-like couplings compared to the ones obtained previously. In particular, an interpretation of the Galactic Centre Excess in terms of a pseudoscalar mediator with Yukawa-like couplings is excluded for $m_A \lesssim 10\:\text{GeV}$, rather than for $m_A \lesssim 5\:\text{GeV}$. Apart from this correction, our conclusions remain unchanged.

\section*{Acknowledgements}

We thank John March-Russell, Andreas Ringwald and Martin W.~Winkler for helpful discussions and Ulrich Haisch for carefully reading the manuscript and providing a number of important comments. We thank Jesse Thaler for discovering a mistake in Eq.~(\ref{eq:oneloop}) and Brian Batell, Ulrich Haisch, Zoltan Ligeti and Maxim Pospelov for subsequent discussions. The work of FK and KSH is supported by the German Science Foundation (DFG) under the Collaborative Research Center (SFB) 676 Particles, Strings and the Early Universe. CM acknowledges support from the European Research Council through the ERC starting grant {\it WIMPs Kairos}, P.I.~G.~Bertone.

\appendix 

\section{Rare meson decays}
\label{sec:mesondecays}

In this appendix we provide the expressions for the partial decay widths of various rare meson decays in terms of the effective couplings $h^S_{qq'}$ and $h^P_{qq'}$ defined in Sec.~\ref{sec:penguin}.

\subsection*{Kaon decays}

Under the assumptions of CVC and PCAC, the decay width for $K^+ \rightarrow \pi^+ A$ is~\cite{Deshpande:2005mb}:
\begin{equation}
\Gamma(K^+ \rightarrow \pi^+ A) = \frac{1}{16 \pi \, m_{K^+}^3} \lambda^{1/2}(m_{K^+}^2, m_{\pi^+}^2, m_A^2) \left(\frac{m_{K^+}^2 - m_{\pi^+}^2}{m_s - m_d}\right)^2 |h^S_{ds}|^2 \; .
\label{eq:Kwidth}
\end{equation}
where the function $\lambda(a,b,c) =(a-b-c)^2-4\,b\,c$. For a more complete calculation, the expression above should be multiplied by a form factor $| f_0^{K^+}(m_A^2)|^2$ but this form factor is close to unity~\cite{Marciano:1996wy} so we neglect it. Note that the pseudoscalar coupling proportional to $h^P_{ds}$ does not contribute to this decay.

The decay width for $K_L \rightarrow \pi^0 A$ can be written as~\cite{Marciano:1996wy, Grossman:1997pa, He:2006uu}:\footnote{Note that~\cite{He:2006uu} states that $\mathcal{M}(K_L \rightarrow \pi^0 A) = -\text{Re}\left[\mathcal{M}(K^+ \rightarrow \pi^+ A)\right]$. However, given that the vertex $g^S_{ds} \, A \, \bar{d} s$ contributes a factor $i g^S_{ds}$ to the matrix element, this statement is consistent with the ones from~\cite{Marciano:1996wy,Grossman:1997pa}.}
\begin{equation}
\Gamma(K_L \rightarrow \pi^0 A) = \frac{1}{16 \pi \, m_{K_L}^3} \lambda^{1/2}(m_{K_L}^2, m_{\pi^0}^2, m_A^2) \left(\frac{m_{K_L}^2 - m_{\pi^0}^2}{m_s - m_d}\right)^2 \text{Im}(h^S_{ds})^2 \; .
\label{eq:KLwidth}
\end{equation}

\subsection*{$B$ meson decays}

There is a similar expression for the decay $B \rightarrow K \, A$~\cite{Bobeth:2001sq, Hiller:2004ii, Batell:2009jf}:
\begin{equation}
\Gamma(B \rightarrow K \, A) = \frac{1}{16 \pi \, m_{B}^3} \lambda^{1/2}(m_{B}^2, m_{K}^2, m_A^2) \left| f_0^{B^0}(m_A^2)\right|^2 \left(\frac{m_{B}^2 - m_{K}^2}{m_b - m_s}\right)^2 |h^S_{sb}|^2 \; .
\label{eq:Bwidth}
\end{equation}
This time, however, the form factor is approximately $f_0^{B^0} \sim 0.3$--$0.4$ and therefore cannot be neglected~\cite{Ali:1999mm}. Here we use the parameterisation from~\cite{Ball:2004ye}:
\begin{equation}
f_0^{B^0}(q^2) = \frac{0.33}{1-q^2/(38\:\text{GeV}^2)} \; ,
\end{equation}
which should be accurate within 10\%.

To avoid these theoretical uncertainties, it is possible to study the more inclusive decays $B \rightarrow X_s \, A$, where $X_s$ can be any strange meson. One then finds~\cite{Hiller:2004ii}
\begin{equation}
\Gamma(B \rightarrow X_s \, A) = \frac{1}{8 \pi} \frac{\left(m_{b}^2 - m_A^2\right)^2}{m_b^3} |h^S_{sb}|^2 \; .
\label{eq:BXwidth}
 \end{equation}

\subsection*{$B_s$ decays}

The decay $B_s \rightarrow A^* \rightarrow \mu^+ \mu^-$ will give relevant constraints for the case of non-vanishing coupling
to muons, e.g.\ for Yukawa-like couplings. In order to compare the contribution of the pseudoscalar mediator to the SM prediction, it is convenient to write the amplitude in the form
\begin{equation}
\mathcal{A} = \frac{-4 \, G_F}{\sqrt{2}} V_{tb} \, V^*_{ts} \, \frac{e^2}{16\pi^2} \, C^P \, m_b \, (\bar{s} \gamma^5 b) (\bar{\mu} \gamma^5 \mu) + \text{h.c.} \; ,
\end{equation}
where $C^P$ is given in terms of the effective coupling $h^P$ as
\begin{equation}
 C^P =  \frac{2 \pi \, g_\mu \, v^2 \, h^P_{sb}}{\alpha \, V_{tb} V^*_{ts} \, m_b \, (m_{B_s}^2-m_A^2+i \Gamma_A m_A)} \;.
\end{equation}
We then obtain the simple relation~\cite{Altmannshofer:2011gn}
\begin{equation}
\frac{\text{BR}(B_s \rightarrow \mu^+ \mu^-)_\text{NP}}{\text{BR}(B_s \rightarrow \mu^+ \mu^-)_\text{SM}} \simeq \frac{m_{B_s}^4}{4 \, m_\mu^2} \frac{|C^P|^2}{(C_{10}^\text{SM})^2} \; ,
\end{equation}
where $C_{10}^\text{SM} = -4.103$ and we have neglected terms due to the width difference in the $B_s$ system, which lead to a correction of about $10\%$~\cite{DeBruyn:2012wj}.

Substituting the expression for $C^P$ we therefore obtain
\begin{align}
\frac{\text{BR}(B_s \rightarrow \mu^+ \mu^-)_\text{NP}}{\text{BR}(B_s \rightarrow \mu^+ \mu^-)_\text{SM}} & \simeq  
\frac{\pi^2 \, g_\mu^2 \, v^4 \, m_{B_s}^4 \, |h^p_{sb}|^2}{\alpha^2 \,  m_\mu^2 \, m_b^2 \, |C_{10}^\text{SM}|^2 \, |V_{tb} V^*_{ts}|^2 \, \left((m_{B_s}^2-m_A^2)^2+ \Gamma_A^2 \, m_A^2\right)} \;.
\end{align}
For Yukawa-like couplings, $g_\mu$ should be replaced by $ g_Y\,\sqrt{2} \, m_\mu / v$. 

\subsection*{Radiative $\Upsilon$ decays}

\begin{figure}[!t]
\centering
\includegraphics[width=0.5\textwidth]{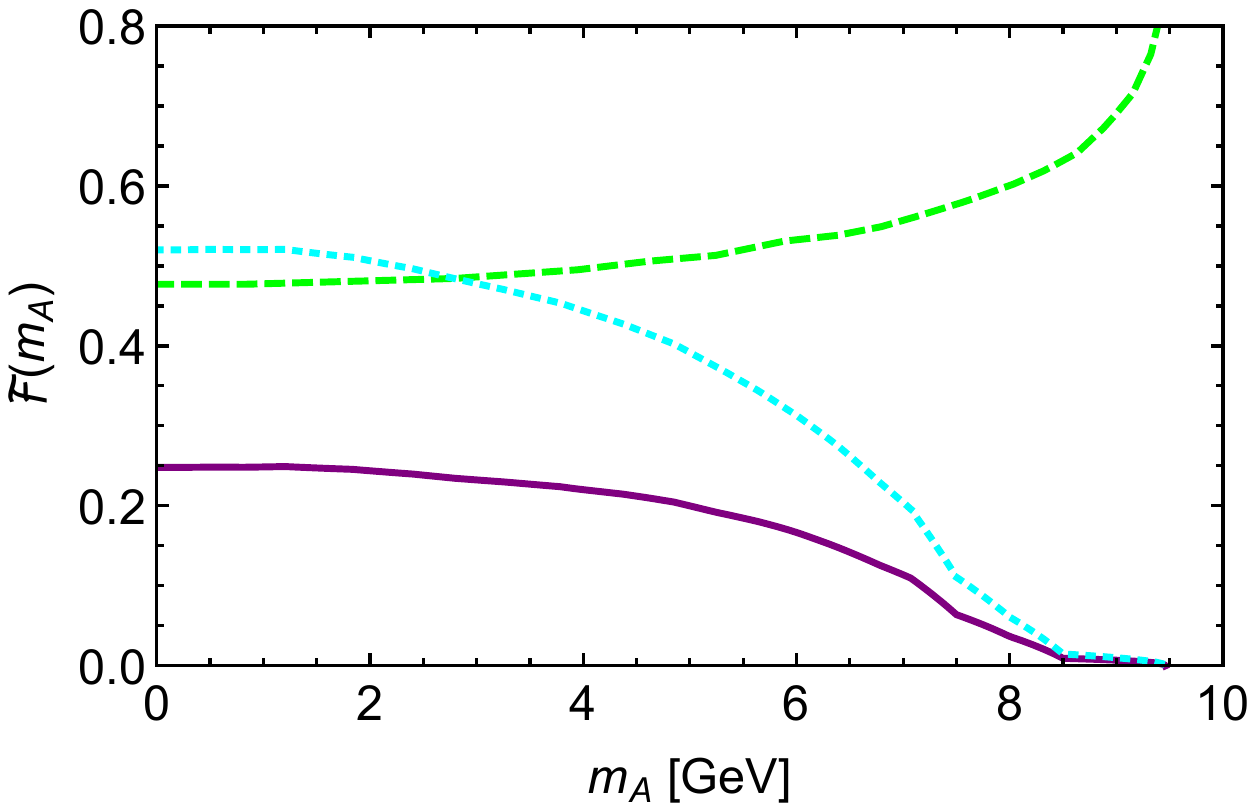}
\caption{The form factor $\mathcal{F}(m_A)$ for $\Upsilon$ decays (solid purple).  We assume that the QCD and relativistic effects factorise such that $\mathcal{F}(m_A)=\mathcal{F}_{\mathrm{QCD}} \cdot \mathcal{F}_{\mathrm{rel}}$.  Green dashed and cyan dotted show $\mathcal{F}_{\mathrm{QCD}} $ and $\mathcal{F}_{\mathrm{rel}}$ respectively.}
\label{fig:formfactor}	
\end{figure}

In contrast to the decays considered above the radiative decay $\Upsilon \rightarrow \gamma A$ does not probe the flavour-changing couplings $h^{S,P}_{q q'}$, but the tree-level coupling $g_b$, where $g_b =g_Y\, \sqrt{2} \, m_b  / v$ for Yukawa-like couplings and $g_b = g_q$ for universal quark couplings. The easiest way to remove theoretical uncertainties is by considering the ratio~\cite{Wilczek:1977zn}
\begin{equation}
\frac{\text{BR}(\Upsilon(nS) \rightarrow \gamma A)}{\text{BR}(\Upsilon(nS) \rightarrow \mu^+ \mu^-)} = \frac{g_f^2}{e^2} \left(1-\frac{m_A^2}{m_\Upsilon^2(nS)}\right) \mathcal{F}(m_A)\; ,
\end{equation}
where $\mathcal{F}(x)$ is a form factor which parameterises the effects of higher order QCD contributions~\cite{Nason:1986tr, Haber:1987ua} and relativistic corrections.\footnote{It was pointed out~\cite{Haber:1978jt, Ellis:1979jy} that bound state effects may also give an important contribution to $\mathcal{F}(x)$. Nevertheless, these effects are only important for $m_A \sim m_\Upsilon$, so we restrict ourselves to the other corrections.} We assume that the QCD effects and relativistic effects `factorise', i.e.\ that $\mathcal{F} = \mathcal{F}_{\mathrm{QCD}} \cdot \mathcal{F}_{\mathrm{rel}} $. 

We take the QCD corrections for a pseudoscalar from \cite{Nason:1986tr, McKeen:2008gd}, keeping in mind that these corrections are accurate only if $m_A$ is not too close to $m_\Upsilon$. Following~\cite{McKeen:2008gd} we write
\begin{equation}
\mathcal{F}_{\mathrm{QCD}}(x) = x \left(1 - \frac{\alpha_S \, C_F}{\pi} F_A(x)\right) \; ,
\end{equation}
where $C_F = (N^2 - 1)/(2N) = 4/3$, $\alpha_S(m_\Upsilon) \approx 0.184$~\cite{Brambilla:2007cz} and $F_A(x)$ can be extracted from Fig.~1 of~\cite{McKeen:2008gd} (see also~\cite{Nason:1986tr}). For $m_A \ll m_\Upsilon$ we find $\mathcal{F}_{\mathrm{QCD}}(x) \approx 0.5$, in agreement with~\cite{Domingo:2008rr}. 

For large pseudoscalar masses, on the other hand, the dominant effect comes from the relativistic corrections, which suppress the effective form factor significantly. We take these corrections from~\cite{Aznaurian:1986hi} for $m_A \lesssim 7\:\text{GeV}$, and from \cite{Faldt:1987zu} for $m_A \gtrsim 7\:\text{GeV}$. Note that these corrections were calculated for a scalar rather than a pseudoscalar, but lacking an explicit calculation for the latter we assume both cases to be similar. We show the individual contributions and the total form factor $\mathcal{F}(m_A)$ in Fig.~\ref{fig:formfactor}.

\section{Decays of the pseudoscalar mediator}
\label{sec:decays}

In this appendix we discuss the various decay channels of the pseudoscalar $A$ (see also~\cite{Hiller:2004ii, Andreas:2010ms}). As before, we use $g_f$ to denote the coupling of $A$ to SM fermions. For Yukawa-like couplings one has $g_f=g_Y\sqrt{2} m_f/v$, which depends on the fermion mass $m_f$.

\subsection*{Leptonic and Photonic Decays}

For $m_A \leq 3 \, m_\pi$, the pseudoscalar can only decay into pairs of leptons and pairs of photons. One finds~\cite{Djouadi:2005gj}
\begin{align}
\Gamma(A \rightarrow \ell^+ \ell^-) & = \frac{g_f^2}{8 \pi} \, m_A \, \sqrt{1 - \frac{1}{\tau_\ell}} \, ,\\
\Gamma(A \rightarrow \gamma \gamma) & = \frac{\alpha^2 \, m_A^3}{256 \pi^3}\left|\sum_f \frac{N_c \, Q_f^2 \, g_f}{m_f} F_A(\tau_f)\right|^2 \; ,
\end{align}
where $\tau_f = m_A^2 / (4 \, m_f^2)$ and
\begin{equation}
F_A(\tau)  = \frac{2}{\tau} \times \left\{ \begin{array}{lr} \text{arcsin}^2 \sqrt{\tau}
        \, , \ \ & \tau \leq 1 \\ -\frac{1}{4} \left[ \log \frac{1+\sqrt{1-\tau^{-1}}}{1-\sqrt{1-\tau^{-1}}} - i \pi\right]^2 \, , & \tau > 1 \end{array} \right. \; .
\end{equation}
For universal quark couplings and small pseudoscalar masses, light quarks give the dominant contribution to the decay of the pseudoscalar into photons. The partial decay width therefore depends sensitively on the assumed values of the quark masses in the loop (i.e.\ the values of $\tau$ used to evaluate $F_A$). Instead of using the current quark masses, we take \mbox{$\tau_u = \tau_d = m_A^2 / (4 \, m_\pi^2)$} and $\tau_s = m_A^2 / (4 \, m_K^2)$. We have checked that this choice approximately reproduces the results from chiral perturbation theory~\cite{Leutwyler:1989tn} for the scalar case.

\subsection*{Hadronic decays}

When $m_A > 2 \, m_\pi$, decays into hadrons become kinematically allowed. However, since the decay $A \rightarrow \pi\pi$ is forbidden by $CP$ symmetry, the pseudoscalar can only decay into three-body final states, such as $\pi\pi \pi$ for $m_A > 3 \, m_\pi$. This observation differs from the results of~\cite{Dermisek:2010mg}, which are based on NMHDECAY~\cite{Ellwanger:2004xm}. In these references, the pseudoscalar hadronic decay width is taken to be the same as the $A \to gg$ partial width even below $\Lambda_{\mathrm{QCD}}$, and in fact all the way down to $m_A=0\:\text{MeV}$.

Nevertheless, once $m_A > 3 \, m_\pi$, there will be hadronic decays. For the case of Yukawa-like couplings we use the perturbative spectator model to obtain an estimate of the partial decay widths. In contrast to the case of a scalar boson, the decay of a pseudoscalar proceeds via $s$-wave rather than $p$-wave. Consequently, we replace the factor $(1 - 4 m_\pi^2/m_A^2)^{3/2} / (8\pi)$ appearing for a SM-like Higgs boson by the phase space for isotropic three-body decays, $\rho(m_A, m_\pi, m_\pi, m_\pi)$, to obtain
\begin{equation}
 \Gamma(A \rightarrow u \bar{u}, d \bar{d} \rightarrow \pi\pi\pi) \sim 6 \, \frac{m_u^2+m_d^2}{v^2} g_Y^2 \, m_A \, \rho(m_A, m_\pi, m_\pi, m_\pi) \; .
\end{equation}
The parameters for the perturbative spectator model are inferred from matching to chiral perturbation theory at about $1\:\text{GeV}$. We follow~\cite{McKeen:2008gd} and use $m_d=m_u=50\:\text{MeV}$. An analogous expression is obtained for the decays $A \rightarrow K K \pi$ and $A \rightarrow D D \pi$, e.g.\
\begin{equation}
 \Gamma(A \rightarrow s \bar{s} \rightarrow K K \pi) \sim 6 \, \frac{m_s^2}{v^2} g_Y^2 \, m_A \, \rho(m_A, m_K, m_K, m_\pi) \; .
\end{equation}
where we take $m_s=450\:\text{MeV}$~\cite{McKeen:2008gd}.

Finally, there are also loop-induced decays into gluons, which again hadronise into (at least) three pions. To obtain an approximate expression for this decay, we consider $k$ heavy quarks, where we take $k = 4$ for $ 3 \, m_\pi < m_A < m_K$ and $k = 3$ for $m_A > m_K$. Taking furthermore into account that the pseudoscalar form factor asymptotically yields $F_A(0) = 2$, while the scalar form factor gives $F_S(0) = 4/3$ we obtain
\begin{equation}
 \Gamma(A \rightarrow g g \rightarrow \pi \pi \pi) \sim 2 \, k^2 \, g_Y^2 \left(\frac{\alpha_s}{2 \pi}\right)^2 \frac{m_A^3}{v^2} \, \rho(m_A, m_\pi, m_\pi, m_\pi) \; ,
\end{equation}
where we take $\alpha_s=0.47$~\cite{McKeen:2008gd}.

For the case of universal quark couplings, we do not attempt to derive corresponding expressions. Instead, we only assume that the partial width for decays into hadrons dominate over the one for decays into photons above the three-pion threshold. This assumption is fully sufficient, since our bounds are not sensitive to the total width of the pseudoscalar, as long as it is small compared to the mass so that the narrow width approximation is valid.

\section{Pseudoscalar total decay width and lifetime}
\label{sec:lifetime}

For sufficiently small values of $m_A$, more specifically $m_A < 2 \, m_\mu$ for Yukawa-like couplings and $m_A < 3 \, m_\pi$ for couplings only to quarks, the pseudoscalar will typically have a very small decay width, and consequently a very long lifetime. In addition, light pseudoscalars produced in meson decays will often be highly boosted so that their decay length can become comparable to the spatial dimensions of the detector. Such a long decay length can lead to two important effects: displaced vertices and escaping particles.

If a pseudoscalar is produced with momentum $\mathbf{p}_A$ in the laboratory frame, its decay length will be~\cite{Andreas:2010ms, Schmidt-Hoberg:2013hba}
\begin{equation}
l_d = \frac{|\mathbf{p}_A|}{m_A \, \Gamma_A} \; .
\end{equation}
Defining $l_\text{max}(\mathbf{p}_A)$ as the spatial extension of the detector from the interaction point in the direction of $\mathbf{p}_A$, the probability for such a particle to escape from the detector without decaying is then given by
\begin{equation}
P_\text{esc}(\mathbf{p}_A) =  \exp\left(-\frac{l_\text{max}(\mathbf{p}_A) \, m_A \, \Gamma_A}{|\mathbf{p}_A|}\right) \; .
\end{equation}
To determine the total probability for a pseudoscalar to escape, we need to multiply this expression with the predicted momentum distribution in the laboratory frame, $f(\mathbf{p}_A)$, and integrate over $\mathbf{p}_A$. 

If the mother particle (which can be $K$, $B$ or $\Upsilon$) is approximately at rest in the laboratory frame and for a two-body decay, the expression above can be significantly simplified, because $p'_A \equiv |\mathbf{p}'_A|$ is fixed. We then obtain
\begin{equation}
P_\text{esc} =  \exp \left(-\frac{l_\text{max} \, m_A \, \Gamma_A}{p'_A} \right) \; .
\label{eq:escaperest}
\end{equation}
In many of the experiments that we consider, however, the mother particles will travel in a collimated beam with high energy. In this case, Eq.~(\ref{eq:escaperest}) will remain valid in the transverse direction, with $l_\text{max}$ replaced by the transverse dimension $l_{T,\text{max}}$. In the direction of the beam there is an additional factor $1/\gamma_M$ in the exponent reflecting the boost of the mother particle. For typical forward detectors, such as NA48/2, this additional factor is precisely cancelled by the larger spatial extend of the detector in the direction of the beam. Consequently, we can use the same expression also for decays along the direction of the beam, taking the transverse dimensions of the detector $l_{T,\text{max}}$ instead of $l_\text{max}$ everywhere. Nevertheless, the approximation in Eq.~(\ref{eq:escaperest}) is not sufficient for high-energy colliders like LHCb, where we have to include the factor $1/\gamma_M$ explicitly. We will determine an appropriate value for $\gamma_M$ below.

If the pseudoscalar produced in a meson decay escapes from the detector without decaying, one would typically only observe a single decay product with large unbalanced momentum. These kinds of decays are strongly constrained, giving us a unique opportunity to test pseudoscalars with very small couplings. 
Another possibility is that the pseudoscalar decays inside the detector, but at a visible distance from the interaction point. While such displaced vertices are in principle a very promising way to search for weakly-coupled pseudoscalars, most existing searches will discard these events as background, since the event reconstruction would yield a vertex with too low quality. 

If $\mathbf{p}_A$ is the momentum of the pseudoscalar in the laboratory frame and $l_\text{min}(\mathbf{p}_A)$ is the vertex resolution in the direction of $\mathbf{p}_A$, the probability for the decay of $A$ to yield a sufficiently high quality vertex is approximately
\begin{equation}
P_\text{prompt}(\mathbf{p}_A) =  1 - \exp\left(-\frac{l_\text{min}(\mathbf{p}_A) \, m_A \, \Gamma_A}{|\mathbf{p}_A|}\right) \; .
\end{equation}
As before, we will evaluate this expression in the rest frame of the mother particle. Again, it is a reasonable approximation for forward detectors to use the transverse vertex resolution $l_{T,\text{min}}$ instead of $l_\text{min}(\mathbf{p}_A)$.\footnote{For example, NA48/2 has a transverse vertex resolution of a few millimetres, while the longitudinal vertex resolution is only around $50\:\text{cm}$, corresponding to the larger boost factor in that direction.} We then find
\begin{equation}
P_\text{prompt} =  1 - \exp\left(-\frac{l_{T,\text{min}} \, m_A \, \Gamma_A}{p'_A} \cdot \frac{1}{\gamma_M} \right) \; .
\end{equation}

To check our approximation, we have compared our results for $P_\text{prompt}$ with the more detailed study of displaced vertices in~\cite{Schmidt-Hoberg:2013hba}. We find that for BELLE the predictions agree well without the need to introduce an additional factor $\gamma_M > 1$. In the case of LHCb the $B$ mesons are much more boosted, so our approximation is no longer valid. Nevertheless, we can recover the values of $P_\text{prompt}$ shown in Fig.~4 of~\cite{Schmidt-Hoberg:2013hba} by setting $\gamma_M\sim20$, which corresponds roughly to the average boost of the $B$ mesons.

Although it is possible that the lifetime of the pseudoscalar is such that the fraction of decays from displaced vertices is large, there will always be either a sizeable fraction of prompt decays or a sizeable fraction of escaping particles. Obviously, the two requirements are highly complementary in the sense that it is impossible to reduce the fraction of escaping particles (e.g.\ by increasing the coupling strength) without at the same time increasing the fraction of prompt decays. By combining searches for both prompt decays and missing momentum, we are able to probe many orders of magnitude in coupling strength.

\section{Direct detection from pseudoscalar exchange}
\label{sec:direct}

In this appendix we derive the scattering cross section at direct detection experiments for a pseudoscalar $A$ that interacts with the DM and SM particles through the interaction terms in Eqs.~\eqref{eq:ADM} and~\eqref{eq:AgSM}.
Let us for the moment make the usual assumption that the pseudoscalar is heavy compared to the typical momentum transfer $q$ in DM direct detection experiments, i.e.~$m_A^2 \gg q^2$, where the momentum transfer is related to the target nucleus mass $m_T$ and the recoil energy $E$ by $q=\sqrt{2 m_T E}\sim100$~MeV. In this case, the pseudoscalar can be integrated out and the effective DM-quark interaction is
\begin{equation}
\mathcal{L}_q = c_q\, \bar{\chi}i\gamma^5 \chi \, \bar{q} i\gamma^5 q\;,
\end{equation}
where $c_q=g_\chi \, g_q / m_A^2$. Here we make no assumption on the coupling structure of $g_q$. 

Following the notation from~\cite{Fitzpatrick:2012ix}, the DM-nucleon interaction is given by
\begin{equation}
\mathcal{L}_N=c_N \, \bar{\chi}i\gamma^5 \chi \, \bar{N} i\gamma^5 N\equiv c_N \, \mathcal{O}_4^N\;,
\label{eq:LNDD}
\end{equation}
where $N=\{p,n\}$ is the nucleon field with mass $m_N$ and
\begin{equation}
c_N=\sum_{q=u,d,s}\frac{m_N}{m_q}\left[c_q-\sum_{q'=u,...,t}c_{q'}\frac{\overline{m}}{m_{q'}} \right] \Delta_{q}^{(N)}\;.
\end{equation}
Here $\overline{m}=\left[\sum_{q=u,d,s}m_q^{-1} \right]^{-1}$ and $\Delta_{q}^{(N)}$ give the quark spin content of a nucleon: we use the values $\Delta_{u}^{(p)}=\Delta_{d}^{(n)}=0.84$, $\Delta_{d}^{(p)}=\Delta_{u}^{(n)}=-0.44$, $\Delta_{s}^{(p)}=\Delta_{s}^{(n)}=-0.03$~\cite{Cheng:2012qr}.

Taking the non-relativistic limit of Eq.~\eqref{eq:LNDD} yields
\begin{equation}
\mathcal{L}_N=4 \, c_N (\vec{s}_{\chi}\cdot\vec{q})\; (\vec{s}_{N}\cdot\vec{q})\equiv 4 \, c_N \, \mathcal{O}_6^{\rm{NR}}\;.
\end{equation}
The spin-averaged matrix element can therefore be written as
\begin{equation}
\overline{|\mathcal{M}|^2_T}=\frac{m_T^2}{m_N^2}\sum_{N,N'=p,n}16 \, c_N \, c_{N'} \, F_{66}^{N N'}(q^2)\;,
\end{equation}
where
\begin{equation}
F_{66}^{N N'}(q^2)=C(j_{\chi})\frac{q^4}{16}F_{\Sigma''}^{N,N'}\;
\end{equation}
and $C(j_{\chi}=1/2)=1$. The from factors $F_{\Sigma''}^{N,N'}$ for various target nuclei are tabulated in the appendices of~\cite{Fitzpatrick:2012ix}. Bringing these results together, we obtain our final result for the scattering cross section:
\begin{equation}
\frac{\mathrm{d} \sigma}{\mathrm{d} E}=\frac{m_T}{32 \pi}\frac{1}{v^2} \frac{q^4}{m_N^2 \, m_{\chi}^2}\sum_{N,N'=p,n} c_N \, c_{N'} \, F_{\Sigma''}^{N,N'}\;,
\end{equation}
which agrees with the result in~\cite{Arina:2014yna}.

For the parameters we consider, the mass of the pseudoscalar can be comparable to (or even smaller than) the typical momentum transfer. In this case, the overall pre-factor $m_A^{-4}$ (absorbed into the coefficients $c_N$) must be replaced by $(m_A^2 + q^2)^{-2}$. Defining
\begin{equation}
g_N=\sum_{q=u,d,s}\frac{m_N}{m_q}\left[g_q-\sum_{q'=u,...,t}g_{q'}\frac{\overline{m}}{m_{q'}} \right] \Delta_{q}^{(N)}
\label{eq:app:gn}
\end{equation}
in analogy to the definition of $c_N$ above, we can then write
\begin{equation}
\frac{\mathrm{d} \sigma}{\mathrm{d} E}=\frac{m_T}{32 \pi} \frac{1}{v^2} \frac{g_\chi^2}{(q^2 + m_A^2)^2} \frac{q^4}{m_N^2 \, m_{\chi}^2}\sum_{N,N'=p,n} g_N g_{N'}F_{\Sigma''}^{N,N'}\;,
\end{equation}
which is identical to the expression above for $m_A^2 \gg q^2$, but becomes independent of both $q^2$ and $m_A^2$ for $m_A^2 \ll q^2$. 

\subsection*{$CP$ violating interaction}

We also briefly  discuss the case where the DM and pseudoscalar are coupled with a $CP$-violating coupling. Integrating out the mediator leads to the effective DM-quark interaction
\begin{equation}
\mathcal{L}_{\cancel{CP}} = d_q\, \bar{\chi} \chi \, \bar{q} i\gamma^5 q\;,
\end{equation}
where $d_q=g^S_\chi \, g_q / m_A^2$. Performing the same procedure as in the case above, we obtain the final result
\begin{equation}
\frac{\mathrm{d} \sigma_{\cancel{CP}}}{\mathrm{d} E}=\frac{m_T}{8 \pi} \frac{1}{v^2} \frac{(g^S_\chi)^2}{(q^2 + m_A^2)^2} \frac{q^2}{m_N^2}\sum_{N,N'=p,n} g_N g_{N'}F_{\Sigma''}^{N,N'}\;,
\end{equation}
where the effective nucleon coupling $g_N$ is again given by Eq.~\eqref{eq:app:gn}.


\begin{thebibliography}{100}

\bibitem{Agnese:2013jaa}
{\bf SuperCDMS Collaboration}, R.~Agnese et~al.,
  \href{http://dx.doi.org/10.1103/PhysRevLett.112.041302}{{\it {Search for
  Low-Mass Weakly Interacting Massive Particles Using Voltage-Assisted
  Calorimetric Ionization Detection in the SuperCDMS Experiment}}, } {\em
  Phys.Rev.Lett.} {\bf 112} (2014) 041302,
  [\href{http://arxiv.org/abs/1309.3259}{{\tt arXiv:1309.3259}}].

\bibitem{Akerib:2013tjd}
{\bf LUX Collaboration}, D.~Akerib et~al.,
  \href{http://dx.doi.org/10.1103/PhysRevLett.112.091303}{{\it {First results
  from the LUX dark matter experiment at the Sanford Underground Research
  Facility}}, } {\em Phys.Rev.Lett.} {\bf 112} (2014) 091303,
  [\href{http://arxiv.org/abs/1310.8214}{{\tt arXiv:1310.8214}}].

\bibitem{Agnese:2014aze}
{\bf SuperCDMS Collaboration}, R.~Agnese et~al.,
  \href{http://dx.doi.org/10.1103/PhysRevLett.112.241302}{{\it {Search for
  Low-Mass WIMPs with SuperCDMS}}, } {\em Phys.Rev.Lett.} {\bf 112} (2014)
  241302, [\href{http://arxiv.org/abs/1402.7137}{{\tt arXiv:1402.7137}}].

\bibitem{Angloher:2014myn}
{\bf CRESST-II Collaboration}, G.~Angloher et~al.,
  \href{http://dx.doi.org/10.1140/epjc/s10052-014-3184-9}{{\it {Results on low
  mass WIMPs using an upgraded CRESST-II detector}}, } {\em Eur.Phys.J} {\bf
  C74} (2014) 3184, [\href{http://arxiv.org/abs/1407.3146}{{\tt
  arXiv:1407.3146}}].

\bibitem{Aad:2014iia}
{\bf ATLAS Collaboration}, G.~Aad et~al.,
  \href{http://dx.doi.org/10.1103/PhysRevLett.112.201802}{{\it {Search for
  Invisible Decays of a Higgs Boson Produced in Association with a Z Boson in
  ATLAS}}, } {\em Phys.Rev.Lett.} {\bf 112} (2014) 201802,
  [\href{http://arxiv.org/abs/1402.3244}{{\tt arXiv:1402.3244}}].

\bibitem{Chatrchyan:2014tja}
{\bf CMS Collaboration}, S.~Chatrchyan et~al.,
  \href{http://dx.doi.org/10.1140/epjc/s10052-014-2980-6}{{\it {Search for
  invisible decays of Higgs bosons in the vector boson fusion and associated ZH
  production modes}}, } {\em Eur.Phys.J.} {\bf C74} (2014) 2980,
  [\href{http://arxiv.org/abs/1404.1344}{{\tt arXiv:1404.1344}}].

\bibitem{Gunion:1989we}
J.~F. Gunion, H.~E. Haber, G.~L. Kane, and S.~Dawson, {\it {The Higgs Hunter's
  Guide}},  {\em Front.Phys.} {\bf 80} (2000) 1--448.

\bibitem{Martin:1997ns}
S.~P. Martin, \href{http://dx.doi.org/10.1142/9789814307505_0001}{{\it {A
  Supersymmetry primer}}, } {\em Adv.Ser.Direct.High Energy Phys.} {\bf 21}
  (2010) 1--153, [\href{http://arxiv.org/abs/hep-ph/9709356}{{\tt
  hep-ph/9709356}}].

\bibitem{Freytsis:2010ne}
M.~Freytsis and Z.~Ligeti,
  \href{http://dx.doi.org/10.1103/PhysRevD.83.115009}{{\it {On dark matter
  models with uniquely spin-dependent detection possibilities}}, } {\em
  Phys.Rev.} {\bf D83} (2011) 115009,
  [\href{http://arxiv.org/abs/1012.5317}{{\tt arXiv:1012.5317}}].

\bibitem{Dienes:2013xya}
K.~R. Dienes, J.~Kumar, B.~Thomas, and D.~Yaylali,
  \href{http://dx.doi.org/10.1103/PhysRevD.90.015012}{{\it {Overcoming Velocity
  Suppression in Dark-Matter Direct-Detection Experiments}}, } {\em Phys.Rev.}
  {\bf D90} (2014) 015012, [\href{http://arxiv.org/abs/1312.7772}{{\tt
  arXiv:1312.7772}}].

\bibitem{Boehm:2014hva}
C.~Boehm, M.~J. Dolan, C.~McCabe, M.~Spannowsky, and C.~J. Wallace,
  \href{http://dx.doi.org/10.1088/1475-7516/2014/05/009}{{\it {Extended
  gamma-ray emission from Coy Dark Matter}}, } {\em JCAP} {\bf 1405} (2014)
  009, [\href{http://arxiv.org/abs/1401.6458}{{\tt arXiv:1401.6458}}].

\bibitem{Hektor:2014kga}
A.~Hektor and L.~Marzola,
  \href{http://dx.doi.org/10.1103/PhysRevD.90.053007}{{\it {Coy Dark Matter and
  the anomalous magnetic moment}}, } {\em Phys.Rev.} {\bf D90} (2014) 053007,
  [\href{http://arxiv.org/abs/1403.3401}{{\tt arXiv:1403.3401}}].

\bibitem{Arina:2014yna}
C.~Arina, E.~Del~Nobile, and P.~Panci,
  \href{http://dx.doi.org/10.1103/PhysRevLett.114.011301}{{\it {Dark Matter
  with Pseudoscalar-Mediated Interactions Explains the DAMA Signal and the
  Galactic Center Excess}}, } {\em Phys.Rev.Lett.} {\bf 114} (2015) 011301,
  [\href{http://arxiv.org/abs/1406.5542}{{\tt arXiv:1406.5542}}].

\bibitem{Goodenough:2009gk}
L.~Goodenough and D.~Hooper, {\it {Possible Evidence For Dark Matter
  Annihilation In The Inner Milky Way From The Fermi Gamma Ray Space
  Telescope}},  \href{http://arxiv.org/abs/0910.2998}{{\tt arXiv:0910.2998}}.

\bibitem{Hooper:2010mq}
D.~Hooper and L.~Goodenough,
  \href{http://dx.doi.org/10.1016/j.physletb.2011.02.029}{{\it {Dark Matter
  Annihilation in The Galactic Center As Seen by the Fermi Gamma Ray Space
  Telescope}}, } {\em Phys.Lett.} {\bf B697} (2011) 412--428,
  [\href{http://arxiv.org/abs/1010.2752}{{\tt arXiv:1010.2752}}].

\bibitem{Hooper:2011ti}
D.~Hooper and T.~Linden,
  \href{http://dx.doi.org/10.1103/PhysRevD.84.123005}{{\it {On The Origin Of
  The Gamma Rays From The Galactic Center}}, } {\em Phys.Rev.} {\bf D84} (2011)
  123005, [\href{http://arxiv.org/abs/1110.0006}{{\tt arXiv:1110.0006}}].

\bibitem{Abazajian:2012pn}
K.~N. Abazajian and M.~Kaplinghat,
  \href{http://dx.doi.org/10.1103/PhysRevD.86.083511}{{\it {Detection of a
  Gamma-Ray Source in the Galactic Center Consistent with Extended Emission
  from Dark Matter Annihilation and Concentrated Astrophysical Emission}}, }
  {\em Phys.Rev.} {\bf D86} (2012) 083511,
  [\href{http://arxiv.org/abs/1207.6047}{{\tt arXiv:1207.6047}}].

\bibitem{Gordon:2013vta}
C.~Gordon and O.~Macias, \href{http://dx.doi.org/10.1103/PhysRevD.88.083521,
  10.1103/PhysRevD.89.049901}{{\it {Dark Matter and Pulsar Model Constraints
  from Galactic Center Fermi-LAT Gamma Ray Observations}}, } {\em Phys.Rev.}
  {\bf D88} (2013), no.~4 083521, [\href{http://arxiv.org/abs/1306.5725}{{\tt
  arXiv:1306.5725}}].

\bibitem{Macias:2013vya}
O.~Macias and C.~Gordon,
  \href{http://dx.doi.org/10.1103/PhysRevD.89.063515}{{\it {The Contribution of
  Cosmic Rays Interacting With Molecular Clouds to the Galactic Center
  Gamma-Ray Excess}}, } {\em Phys.Rev.} {\bf D89} (2014) 063515,
  [\href{http://arxiv.org/abs/1312.6671}{{\tt arXiv:1312.6671}}].

\bibitem{Daylan:2014rsa}
T.~Daylan, D.~P. Finkbeiner, D.~Hooper, T.~Linden, S.~K.~N. Portillo, et~al.,
  {\it {The Characterization of the Gamma-Ray Signal from the Central Milky
  Way: A Compelling Case for Annihilating Dark Matter}},
  \href{http://arxiv.org/abs/1402.6703}{{\tt arXiv:1402.6703}}.

\bibitem{Zhou:2014lva}
B.~Zhou, Y.-F. Liang, X.~Huang, X.~Li, Y.-Z. Fan, et~al., {\it {GeV excess in
  the Milky Way: Depending on Diffuse Galactic gamma ray Emission template?}},
  \href{http://arxiv.org/abs/1406.6948}{{\tt arXiv:1406.6948}}.

\bibitem{Calore:2014xka}
F.~Calore, I.~Cholis, and C.~Weniger, {\it {Background model systematics for
  the Fermi GeV excess}},  \href{http://arxiv.org/abs/1409.0042}{{\tt
  arXiv:1409.0042}}.

\bibitem{Agrawal:2014oha}
P.~Agrawal, B.~Batell, P.~J. Fox, and R.~Harnik, {\it {WIMPs at the Galactic
  Center}},  \href{http://arxiv.org/abs/1411.2592}{{\tt arXiv:1411.2592}}.

\bibitem{Calore:2014nla}
F.~Calore, I.~Cholis, C.~McCabe, and C.~Weniger, {\it {A Tale of Tails: Dark
  Matter Interpretations of the Fermi GeV Excess in Light of Background Model
  Systematics}},  \href{http://arxiv.org/abs/1411.4647}{{\tt arXiv:1411.4647}}.

\bibitem{Alves:2014yha}
A.~Alves, S.~Profumo, F.~S. Queiroz, and W.~Shepherd,
  \href{http://dx.doi.org/10.1103/PhysRevD.90.115003}{{\it {The Effective
  Hooperon}}, } {\em Phys.Rev.} {\bf D90} (2014) 115003,
  [\href{http://arxiv.org/abs/1403.5027}{{\tt arXiv:1403.5027}}].

\bibitem{Berlin:2014tja}
A.~Berlin, D.~Hooper, and S.~D. McDermott,
  \href{http://dx.doi.org/10.1103/PhysRevD.89.115022}{{\it {Simplified Dark
  Matter Models for the Galactic Center Gamma-Ray Excess}}, } {\em Phys.Rev.}
  {\bf D89} (2014) 115022, [\href{http://arxiv.org/abs/1404.0022}{{\tt
  arXiv:1404.0022}}].

\bibitem{Izaguirre:2014vva}
E.~Izaguirre, G.~Krnjaic, and B.~Shuve,
  \href{http://dx.doi.org/10.1103/PhysRevD.90.055002}{{\it {The Galactic Center
  Excess from the Bottom Up}}, } {\em Phys.Rev.} {\bf D90} (2014) 055002,
  [\href{http://arxiv.org/abs/1404.2018}{{\tt arXiv:1404.2018}}].

\bibitem{Cerdeno:2014cda}
D.~Cerdeno, M.~Peiro, and S.~Robles,
  \href{http://dx.doi.org/10.1088/1475-7516/2014/08/005}{{\it {Low-mass
  right-handed sneutrino dark matter: SuperCDMS and LUX constraints and the
  Galactic Centre gamma-ray excess}}, } {\em JCAP} {\bf 1408} (2014) 005,
  [\href{http://arxiv.org/abs/1404.2572}{{\tt arXiv:1404.2572}}].

\bibitem{Ipek:2014gua}
S.~Ipek, D.~McKeen, and A.~E. Nelson,
  \href{http://dx.doi.org/10.1103/PhysRevD.90.055021}{{\it {A Renormalizable
  Model for the Galactic Center Gamma Ray Excess from Dark Matter
  Annihilation}}, } {\em Phys.Rev.} {\bf D90} (2014) 055021,
  [\href{http://arxiv.org/abs/1404.3716}{{\tt arXiv:1404.3716}}].

\bibitem{Abdullah:2014lla}
M.~Abdullah, A.~DiFranzo, A.~Rajaraman, T.~M. Tait, P.~Tanedo, et~al.,
  \href{http://dx.doi.org/10.1103/PhysRevD.90.035004}{{\it {Hidden On-Shell
  Mediators for the Galactic Center Gamma-Ray Excess}}, } {\em Phys.Rev.} {\bf
  D90} (2014) 035004, [\href{http://arxiv.org/abs/1404.6528}{{\tt
  arXiv:1404.6528}}].

\bibitem{Martin:2014sxa}
A.~Martin, J.~Shelton, and J.~Unwin,
  \href{http://dx.doi.org/10.1103/PhysRevD.90.103513}{{\it {Fitting the
  Galactic Center Gamma-Ray Excess with Cascade Annihilations}}, } {\em
  Phys.Rev.} {\bf D90} (2014), no.~10 103513,
  [\href{http://arxiv.org/abs/1405.0272}{{\tt arXiv:1405.0272}}].

\bibitem{Berlin:2014pya}
A.~Berlin, P.~Gratia, D.~Hooper, and S.~D. McDermott,
  \href{http://dx.doi.org/10.1103/PhysRevD.90.015032}{{\it {Hidden Sector Dark
  Matter Models for the Galactic Center Gamma-Ray Excess}}, } {\em Phys.Rev.}
  {\bf D90} (2014) 015032, [\href{http://arxiv.org/abs/1405.5204}{{\tt
  arXiv:1405.5204}}].

\bibitem{Han:2014nba}
T.~Han, Z.~Liu, and S.~Su,
  \href{http://dx.doi.org/10.1007/JHEP08(2014)093}{{\it {Light Neutralino Dark
  Matter: Direct/Indirect Detection and Collider Searches}}, } {\em JHEP} {\bf
  1408} (2014) 093, [\href{http://arxiv.org/abs/1406.1181}{{\tt
  arXiv:1406.1181}}].

\bibitem{Cheung:2014lqa}
C.~Cheung, M.~Papucci, D.~Sanford, N.~R. Shah, and K.~M. Zurek,
  \href{http://dx.doi.org/10.1103/PhysRevD.90.075011}{{\it {NMSSM
  Interpretation of the Galactic Center Excess}}, } {\em Phys.Rev.} {\bf D90}
  (2014) 075011, [\href{http://arxiv.org/abs/1406.6372}{{\tt
  arXiv:1406.6372}}].

\bibitem{Huang:2014cla}
J.~Huang, T.~Liu, L.-T. Wang, and F.~Yu,
  \href{http://dx.doi.org/10.1103/PhysRevD.90.115006}{{\it {Supersymmetric
  Sub-Electroweak Scale Dark Matter, the Galactic Center Gamma-ray Excess, and
  Exotic Decays of the 125 GeV Higgs Boson}}, } {\em Phys.Rev.} {\bf D90}
  (2014) 115006, [\href{http://arxiv.org/abs/1407.0038}{{\tt
  arXiv:1407.0038}}].

\bibitem{Ghorbani:2014qpa}
K.~Ghorbani, \href{http://dx.doi.org/10.1088/1475-7516/2015/01/015}{{\it
  {Fermionic dark matter with pseudo-scalar Yukawa interaction}}, } {\em JCAP}
  {\bf 1501} (2015) 015, [\href{http://arxiv.org/abs/1408.4929}{{\tt
  arXiv:1408.4929}}].

\bibitem{Cahill-Rowley:2014ora}
M.~Cahill-Rowley, J.~Gainer, J.~Hewett, and T.~Rizzo, {\it {Towards a
  Supersymmetric Description of the Fermi Galactic Center Excess}},
  \href{http://arxiv.org/abs/1409.1573}{{\tt arXiv:1409.1573}}.

\bibitem{Guo:2014gra}
J.~Guo, J.~Li, T.~Li, and A.~G. Williams, {\it {NMSSM Explanations of the
  Galactic Gamma Ray Excess and Promising LHC Searches}},
  \href{http://arxiv.org/abs/1409.7864}{{\tt arXiv:1409.7864}}.

\bibitem{Cao:2014efa}
J.~Cao, L.~Shang, P.~Wu, J.~M. Yang, and Y.~Zhang, {\it {SUSY explanation of
  the Fermi Galactic Center Excess and its test at LHC Run-II}},
  \href{http://arxiv.org/abs/1410.3239}{{\tt arXiv:1410.3239}}.

\bibitem{Freytsis:2014sua}
M.~Freytsis, D.~J. Robinson, and Y.~Tsai, {\it {Galactic Center Gamma-Ray
  Excess through a Dark Shower}},  \href{http://arxiv.org/abs/1410.3818}{{\tt
  arXiv:1410.3818}}.

\bibitem{Buckley:2014fba}
M.~R. Buckley, D.~Feld, and D.~Goncalves, {\it {Scalar Simplified Models for
  Dark Matter}},  \href{http://arxiv.org/abs/1410.6497}{{\tt arXiv:1410.6497}}.

\bibitem{MarchRussell:2012hi}
J.~March-Russell, J.~Unwin, and S.~M. West,
  \href{http://dx.doi.org/10.1007/JHEP08(2012)029}{{\it {Closing in on
  Asymmetric Dark Matter I: Model independent limits for interactions with
  quarks}}, } {\em JHEP} {\bf 1208} (2012) 029,
  [\href{http://arxiv.org/abs/1203.4854}{{\tt arXiv:1203.4854}}].

\bibitem{Spergel:1999mh}
D.~N. Spergel and P.~J. Steinhardt,
  \href{http://dx.doi.org/10.1103/PhysRevLett.84.3760}{{\it {Observational
  evidence for selfinteracting cold dark matter}}, } {\em Phys.Rev.Lett.} {\bf
  84} (2000) 3760--3763, [\href{http://arxiv.org/abs/astro-ph/9909386}{{\tt
  astro-ph/9909386}}].

\bibitem{Markevitch:2003at}
M.~Markevitch, A.~Gonzalez, D.~Clowe, A.~Vikhlinin, L.~David, et~al.,
  \href{http://dx.doi.org/10.1086/383178}{{\it {Direct constraints on the dark
  matter self-interaction cross-section from the merging galaxy cluster
  1E0657-56}}, } {\em Astrophys.J.} {\bf 606} (2004) 819--824,
  [\href{http://arxiv.org/abs/astro-ph/0309303}{{\tt astro-ph/0309303}}].

\bibitem{Kahlhoefer:2013dca}
F.~Kahlhoefer, K.~Schmidt-Hoberg, M.~T. Frandsen, and S.~Sarkar,
  \href{http://dx.doi.org/10.1093/mnras/stt2097}{{\it {Colliding clusters and
  dark matter self-interactions}}, } {\em Mon.Not.Roy.Astron.Soc.} {\bf 437}
  (2014) 2865--2881, [\href{http://arxiv.org/abs/1308.3419}{{\tt
  arXiv:1308.3419}}].

\bibitem{Gnedin:2000ea}
O.~Y. Gnedin and J.~P. Ostriker, \href{http://dx.doi.org/10.1086/323211}{{\it
  {Limits on collisional dark matter from elliptical galaxies in clusters}}, }
  {\em Astrophys.J.} {\bf 561} (2001) 61,
  [\href{http://arxiv.org/abs/astro-ph/0010436}{{\tt astro-ph/0010436}}].

\bibitem{Rocha:2012jg}
M.~Rocha, A.~H. Peter, J.~S. Bullock, M.~Kaplinghat, S.~Garrison-Kimmel,
  et~al., \href{http://dx.doi.org/10.1093/mnras/sts514}{{\it {Cosmological
  Simulations with Self-Interacting Dark Matter I: Constant Density Cores and
  Substructure}}, } {\em Mon.Not.Roy.Astron.Soc.} {\bf 430} (2013) 81--104,
  [\href{http://arxiv.org/abs/1208.3025}{{\tt arXiv:1208.3025}}].

\bibitem{Feng:2009mn}
J.~L. Feng, M.~Kaplinghat, H.~Tu, and H.-B. Yu,
  \href{http://dx.doi.org/10.1088/1475-7516/2009/07/004}{{\it {Hidden Charged
  Dark Matter}}, } {\em JCAP} {\bf 0907} (2009) 004,
  [\href{http://arxiv.org/abs/0905.3039}{{\tt arXiv:0905.3039}}].

\bibitem{Buckley:2009in}
M.~R. Buckley and P.~J. Fox,
  \href{http://dx.doi.org/10.1103/PhysRevD.81.083522}{{\it {Dark Matter
  Self-Interactions and Light Force Carriers}}, } {\em Phys.Rev.} {\bf D81}
  (2010) 083522, [\href{http://arxiv.org/abs/0911.3898}{{\tt
  arXiv:0911.3898}}].

\bibitem{Peter:2012jh}
A.~H. Peter, M.~Rocha, J.~S. Bullock, and M.~Kaplinghat,
  \href{http://dx.doi.org/10.1093/mnras/sts535}{{\it {Cosmological Simulations
  with Self-Interacting Dark Matter II: Halo Shapes vs. Observations}}, } {\em
  Mon.Not.Roy.Astron.Soc.} {\bf 430} (2012) 105--120,
  [\href{http://arxiv.org/abs/1208.3026}{{\tt arXiv:1208.3026}}].

\bibitem{Vogelsberger:2012ku}
M.~Vogelsberger, J.~Zavala, and A.~Loeb,
  \href{http://dx.doi.org/10.1111/j.1365-2966.2012.21182.x}{{\it {Subhaloes in
  Self-Interacting Galactic Dark Matter Haloes}}, } {\em
  Mon.Not.Roy.Astron.Soc.} {\bf 423} (2012) 3740,
  [\href{http://arxiv.org/abs/1201.5892}{{\tt arXiv:1201.5892}}].

\bibitem{Zavala:2012us}
J.~Zavala, M.~Vogelsberger, and M.~G. Walker,
  \href{http://dx.doi.org/10.1093/mnrasl/sls053}{{\it {Constraining
  Self-Interacting Dark Matter with the Milky Way's dwarf spheroidals}}, } {\em
  Mon.Not.Roy.Astron.Soc.: Letters} {\bf 431} (2013) L20--L24,
  [\href{http://arxiv.org/abs/1211.6426}{{\tt arXiv:1211.6426}}].

\bibitem{Loeb:2010gj}
A.~Loeb and N.~Weiner,
  \href{http://dx.doi.org/10.1103/PhysRevLett.106.171302}{{\it {Cores in Dwarf
  Galaxies from Dark Matter with a Yukawa Potential}}, } {\em Phys.Rev.Lett.}
  {\bf 106} (2011) 171302, [\href{http://arxiv.org/abs/1011.6374}{{\tt
  arXiv:1011.6374}}].

\bibitem{Aarssen:2012fx}
L.~G. van~den Aarssen, T.~Bringmann, and C.~Pfrommer,
  \href{http://dx.doi.org/10.1103/PhysRevLett.109.231301}{{\it {Is dark matter
  with long-range interactions a solution to all small-scale problems of
  $\Lambda$ CDM cosmology?}}, } {\em Phys.Rev.Lett.} {\bf 109} (2012) 231301,
  [\href{http://arxiv.org/abs/1205.5809}{{\tt arXiv:1205.5809}}].

\bibitem{Tulin:2013teo}
S.~Tulin, H.-B. Yu, and K.~M. Zurek,
  \href{http://dx.doi.org/10.1103/PhysRevD.87.115007}{{\it {Beyond
  Collisionless Dark Matter: Particle Physics Dynamics for Dark Matter Halo
  Structure}}, } {\em Phys.Rev.} {\bf D87} (2013) 115007,
  [\href{http://arxiv.org/abs/1302.3898}{{\tt arXiv:1302.3898}}].

\bibitem{Hiller:2004ii}
G.~Hiller, \href{http://dx.doi.org/10.1103/PhysRevD.70.034018}{{\it {B physics
  signals of the lightest CP odd Higgs in the NMSSM at large tan beta}}, } {\em
  Phys.Rev.} {\bf D70} (2004) 034018,
  [\href{http://arxiv.org/abs/hep-ph/0404220}{{\tt hep-ph/0404220}}].

\bibitem{Andreas:2010ms}
S.~Andreas, O.~Lebedev, S.~Ramos-Sanchez, and A.~Ringwald,
  \href{http://dx.doi.org/10.1007/JHEP08(2010)003}{{\it {Constraints on a very
  light CP-odd Higgs of the NMSSM and other axion-like particles}}, } {\em
  JHEP} {\bf 1008} (2010) 003, [\href{http://arxiv.org/abs/1005.3978}{{\tt
  arXiv:1005.3978}}].

\bibitem{Fayet:2006sp}
P.~Fayet, \href{http://dx.doi.org/10.1103/PhysRevD.74.054034}{{\it {Constraints
  on Light Dark Matter and U bosons, from $\psi$, $\Upsilon$, $K^+$, $\pi^0$,
  $\eta$ and $\eta'$ decays}}, } {\em Phys.Rev.} {\bf D74} (2006) 054034,
  [\href{http://arxiv.org/abs/hep-ph/0607318}{{\tt hep-ph/0607318}}].

\bibitem{Fayet:2007ua}
P.~Fayet, \href{http://dx.doi.org/10.1103/PhysRevD.75.115017}{{\it {U-boson
  production in $e^+ e^-$ annihilations, $\psi$ and $\Upsilon$ decays, and
  Light Dark Matter}}, } {\em Phys.Rev.} {\bf D75} (2007) 115017,
  [\href{http://arxiv.org/abs/hep-ph/0702176}{{\tt hep-ph/0702176}}].

\bibitem{Batell:2011tc}
B.~Batell, J.~Pradler, and M.~Spannowsky,
  \href{http://dx.doi.org/10.1007/JHEP08(2011)038}{{\it {Dark Matter from
  Minimal Flavor Violation}}, } {\em JHEP} {\bf 1108} (2011) 038,
  [\href{http://arxiv.org/abs/1105.1781}{{\tt arXiv:1105.1781}}].

\bibitem{Agrawal:2014una}
P.~Agrawal, B.~Batell, D.~Hooper, and T.~Lin,
  \href{http://dx.doi.org/10.1103/PhysRevD.90.063512}{{\it {Flavored Dark
  Matter and the Galactic Center Gamma-Ray Excess}}, } {\em Phys.Rev.} {\bf
  D90} (2014) 063512, [\href{http://arxiv.org/abs/1404.1373}{{\tt
  arXiv:1404.1373}}].

\bibitem{D'Ambrosio:2002ex}
G.~D'Ambrosio, G.~Giudice, G.~Isidori, and A.~Strumia,
  \href{http://dx.doi.org/10.1016/S0550-3213(02)00836-2}{{\it {Minimal flavor
  violation: An Effective field theory approach}}, } {\em Nucl.Phys.} {\bf
  B645} (2002) 155--187, [\href{http://arxiv.org/abs/hep-ph/0207036}{{\tt
  hep-ph/0207036}}].

\bibitem{Deshpande:2005mb}
N.~Deshpande, G.~Eilam, and J.~Jiang,
  \href{http://dx.doi.org/10.1016/j.physletb.2005.10.050}{{\it {On the
  possibility of a new boson $X_0$ (214 MeV) in $\Sigma^+ \rightarrow p \mu^+
  \mu^-$}}, } {\em Phys.Lett.} {\bf B632} (2006) 212--214,
  [\href{http://arxiv.org/abs/hep-ph/0509081}{{\tt hep-ph/0509081}}].

\bibitem{Hall:1981bc}
L.~J. Hall and M.~B. Wise,
  \href{http://dx.doi.org/10.1016/0550-3213(81)90469-7}{{\it {Flavor changing
  Higgs boson couplings}}, } {\em Nucl.Phys.} {\bf B187} (1981) 397.

\bibitem{Bobeth:2001sq}
C.~Bobeth, T.~Ewerth, F.~Kruger, and J.~Urban,
  \href{http://dx.doi.org/10.1103/PhysRevD.64.074014}{{\it {Analysis of neutral
  Higgs boson contributions to the decays $\bar{B}_{s} \to \ell^{+} \ell^{-}$
  and $\bar{B} \to K \ell^{+} \ell^{-}$}}, } {\em Phys.Rev.} {\bf D64} (2001)
  074014, [\href{http://arxiv.org/abs/hep-ph/0104284}{{\tt hep-ph/0104284}}].

  
  
  \bibitem{Logan:2000iv}
H.~E.~Logan and U.~Nierste, \href{http://dx.doi.org/10.1016/S0550-3213(00)00417-X}{{\it
  {$B(s, d) \to \ell^+ \ell^-$ in a two Higgs doublet model}}, } {\em
  Nucl.\ Phys.\ B} {\bf 586} (2000) 39,
  [\href{http://arxiv.org/abs/hep-ph/0004139}{{\tt hep-ph/0004139}}].

  
  
  
\bibitem{Hahn:1998yk}
T.~Hahn and M.~Perez-Victoria,
  \href{http://dx.doi.org/10.1016/S0010-4655(98)00173-8}{{\it {Automatized one
  loop calculations in four-dimensions and D-dimensions}}, } {\em
  Comput.Phys.Commun.} {\bf 118} (1999) 153--165,
  [\href{http://arxiv.org/abs/hep-ph/9807565}{{\tt hep-ph/9807565}}].

\bibitem{Hahn:2000kx}
T.~Hahn, \href{http://dx.doi.org/10.1016/S0010-4655(01)00290-9}{{\it
  {Generating Feynman diagrams and amplitudes with FeynArts 3}}, } {\em
  Comput.Phys.Commun.} {\bf 140} (2001) 418--431,
  [\href{http://arxiv.org/abs/hep-ph/0012260}{{\tt hep-ph/0012260}}].

  
 
  \bibitem{Freytsis:2009ct}
M.~Freytsis, Z.~Ligeti and J.~Thaler, \href{http://dx.doi.org/10.1103/PhysRevD.81.034001}{{\it
  {Constraining the Axion Portal with $B \to K \ell^+ \ell^-$}}, } {\em
  Phys.\ Rev.\ D} {\bf 81} (2010) 034001,
  [\href{http://arxiv.org/abs/0911.5355}{{\tt 0911.5355}}].

  \bibitem{Batell:2009jf}
B.~Batell, M.~Pospelov, and A.~Ritz,
  \href{http://dx.doi.org/10.1103/PhysRevD.83.054005}{{\it {Multi-lepton
  Signatures of a Hidden Sector in Rare B Decays}}, } {\em Phys.Rev.} {\bf D83}
  (2011) 054005, [\href{http://arxiv.org/abs/0911.4938}{{\tt
  arXiv:0911.4938}}].


  
\bibitem{McKeen:2008gd}
D.~McKeen, \href{http://dx.doi.org/10.1103/PhysRevD.79.015007}{{\it
  {Constraining Light Bosons with Radiative $\Upsilon(1S)$ Decays}}, } {\em
  Phys.Rev.} {\bf D79} (2009) 015007,
  [\href{http://arxiv.org/abs/0809.4787}{{\tt arXiv:0809.4787}}].

\bibitem{Domingo:2008rr}
F.~Domingo, U.~Ellwanger, E.~Fullana, C.~Hugonie, and M.-A. Sanchis-Lozano,
  \href{http://dx.doi.org/10.1088/1126-6708/2009/01/061}{{\it {Radiative
  $\Upsilon$ decays and a light pseudoscalar Higgs in the NMSSM}}, } {\em JHEP}
  {\bf 0901} (2009) 061, [\href{http://arxiv.org/abs/0810.4736}{{\tt
  arXiv:0810.4736}}].

\bibitem{Dermisek:2010mg}
R.~Dermisek and J.~F. Gunion,
  \href{http://dx.doi.org/10.1103/PhysRevD.81.075003}{{\it {New constraints on
  a light CP-odd Higgs boson and related NMSSM Ideal Higgs Scenarios}}, } {\em
  Phys.Rev.} {\bf D81} (2010) 075003,
  [\href{http://arxiv.org/abs/1002.1971}{{\tt arXiv:1002.1971}}].

\bibitem{Donoghue:1990xh}
J.~F. Donoghue, J.~Gasser, and H.~Leutwyler,
  \href{http://dx.doi.org/10.1016/0550-3213(90)90474-R}{{\it {The Decay of a
  Light Higgs Boson}}, } {\em Nucl.Phys.} {\bf B343} (1990) 341--368.

\bibitem{Clarke:2013aya}
J.~D. Clarke, R.~Foot, and R.~R. Volkas,
  \href{http://dx.doi.org/10.1007/JHEP02(2014)123}{{\it {Phenomenology of a
  very light scalar (100 MeV $ < m_h < $ 10 GeV) mixing with the SM Higgs}}, }
  {\em JHEP} {\bf 1402} (2014) 123, [\href{http://arxiv.org/abs/1310.8042}{{\tt
  arXiv:1310.8042}}].

\bibitem{Anisimovsky:2004hr}
{\bf E949 Collaboration}, V.~Anisimovsky et~al.,
  \href{http://dx.doi.org/10.1103/PhysRevLett.93.031801}{{\it {Improved
  measurement of the $K^+ \rightarrow \pi^+ \nu \bar{\nu}$ branching ratio}}, }
  {\em Phys.Rev.Lett.} {\bf 93} (2004) 031801,
  [\href{http://arxiv.org/abs/hep-ex/0403036}{{\tt hep-ex/0403036}}].

\bibitem{Artamonov:2009sz}
{\bf BNL-E949 Collaboration}, A.~Artamonov et~al.,
  \href{http://dx.doi.org/10.1103/PhysRevD.79.092004}{{\it {Study of the decay
  $K^+ \rightarrow \pi^+ \nu \bar{\nu}$ in the momentum region $140 < P(\pi) <
  199$ MeV/c}}, } {\em Phys.Rev.} {\bf D79} (2009) 092004,
  [\href{http://arxiv.org/abs/0903.0030}{{\tt arXiv:0903.0030}}].

\bibitem{Adler:2004hp}
{\bf E787 Collaboration}, S.~Adler et~al.,
  \href{http://dx.doi.org/10.1103/PhysRevD.70.037102}{{\it {Further search for
  the decay $K^+ \rightarrow \pi^+ \nu \bar{\nu}$ in the momentum region $P <
  195$ MeV/c}}, } {\em Phys.Rev.} {\bf D70} (2004) 037102,
  [\href{http://arxiv.org/abs/hep-ex/0403034}{{\tt hep-ex/0403034}}].

\bibitem{Artamonov:2005cu}
{\bf E949 Collaboration}, A.~Artamonov et~al.,
  \href{http://dx.doi.org/10.1103/PhysRevD.72.091102}{{\it {Upper limit on the
  branching ratio for the decay $\pi^0 \rightarrow \nu \bar{\nu}$}}, } {\em
  Phys.Rev.} {\bf D72} (2005) 091102,
  [\href{http://arxiv.org/abs/hep-ex/0506028}{{\tt hep-ex/0506028}}].

\bibitem{Batley:2009aa}
{\bf NA48/2 Collaboration}, J.~Batley et~al.,
  \href{http://dx.doi.org/10.1016/j.physletb.2009.05.040}{{\it {Precise
  measurement of the $K^{+-} \rightarrow \pi^{+-}e^+e^-$ decay}}, } {\em
  Phys.Lett.} {\bf B677} (2009) 246--254,
  [\href{http://arxiv.org/abs/0903.3130}{{\tt arXiv:0903.3130}}].

\bibitem{AlaviHarati:2003mr}
{\bf KTeV Collaboration}, A.~Alavi-Harati et~al.,
  \href{http://dx.doi.org/10.1103/PhysRevLett.93.021805}{{\it {Search for the
  rare decay $K_L \rightarrow \pi^0 e^+ e^-$}}, } {\em Phys.Rev.Lett.} {\bf 93}
  (2004) 021805, [\href{http://arxiv.org/abs/hep-ex/0309072}{{\tt
  hep-ex/0309072}}].

\bibitem{Batley:2011zz}
{\bf NA48/2 Collaboration}, J.~Batley et~al.,
  \href{http://dx.doi.org/10.1016/j.physletb.2011.01.042}{{\it {New measurement
  of the $K^{+-} \rightarrow \pi^{+-}\mu^+\mu^-$ decay}}, } {\em Phys.Lett.}
  {\bf B697} (2011) 107--115, [\href{http://arxiv.org/abs/1011.4817}{{\tt
  arXiv:1011.4817}}].

\bibitem{AlaviHarati:2000hs}
{\bf KTEV Collaboration}, A.~Alavi-Harati et~al.,
  \href{http://dx.doi.org/10.1103/PhysRevLett.84.5279}{{\it {Search for the
  Decay $K_L \to \pi^0 \mu^+ \mu^-$}}, } {\em Phys.Rev.Lett.} {\bf 84} (2000)
  5279--5282, [\href{http://arxiv.org/abs/hep-ex/0001006}{{\tt
  hep-ex/0001006}}].

\bibitem{Abouzaid:2008xm}
{\bf KTeV Collaboration}, E.~Abouzaid et~al.,
  \href{http://dx.doi.org/10.1103/PhysRevD.77.112004}{{\it {Final Results from
  the KTeV Experiment on the Decay $K_{L} \to \pi^0 \gamma \gamma$}}, } {\em
  Phys.Rev.} {\bf D77} (2008) 112004,
  [\href{http://arxiv.org/abs/0805.0031}{{\tt arXiv:0805.0031}}].

\bibitem{Alexopoulos:2004sx}
{\bf KTeV Collaboration}, T.~Alexopoulos et~al.,
  \href{http://dx.doi.org/10.1103/PhysRevD.70.092006}{{\it {Measurements of
  $K_L$ branching fractions and the CP violation parameter $|\eta_{+-}|$}}, }
  {\em Phys.Rev.} {\bf D70} (2004) 092006,
  [\href{http://arxiv.org/abs/hep-ex/0406002}{{\tt hep-ex/0406002}}].

\bibitem{Yamazaki:1984vg}
T.~Yamazaki, T.~Ishikawa, T.~Taniguchi, T.~Yamanaka, T.~Tanimori, et~al.,
  \href{http://dx.doi.org/10.1103/PhysRevLett.52.1089}{{\it {Search for a
  neutral boson in a two-body decay of $K^+ \rightarrow \pi^+ X_0$}}, } {\em
  Phys.Rev.Lett.} {\bf 52} (1984) 1089--1091.

\bibitem{Ammar:2001gi}
{\bf CLEO Collaboration}, R.~Ammar et~al.,
  \href{http://dx.doi.org/10.1103/PhysRevLett.87.271801}{{\it {Search for the
  familon via $B^{+-} \rightarrow \pi^{+-} X_0$, $B^{+-} \rightarrow K^{+-}
  X_0$, and $B_0 \rightarrow K^0_{S} X_0$ decays}}, } {\em Phys.Rev.Lett.} {\bf
  87} (2001) 271801, [\href{http://arxiv.org/abs/hep-ex/0106038}{{\tt
  hep-ex/0106038}}].

\bibitem{Aubert:2003cm}
{\bf BaBar Collaboration}, B.~Aubert et~al.,
  \href{http://dx.doi.org/10.1103/PhysRevLett.91.221802}{{\it {Evidence for the
  rare decay $B \to K^* \ell^+ \ell^-$ and measurement of the $B \to K \ell^+
  \ell^-$ branching fraction}}, } {\em Phys.Rev.Lett.} {\bf 91} (2003) 221802,
  [\href{http://arxiv.org/abs/hep-ex/0308042}{{\tt hep-ex/0308042}}].

\bibitem{Wei:2009zv}
{\bf BELLE Collaboration}, J.-T. Wei et~al.,
  \href{http://dx.doi.org/10.1103/PhysRevLett.103.171801}{{\it {Measurement of
  the Differential Branching Fraction and Forward-Backword Asymmetry for $B
  \rightarrow K^{(*)} \ell^+ \ell^-$}}, } {\em Phys.Rev.Lett.} {\bf 103} (2009)
  171801, [\href{http://arxiv.org/abs/0904.0770}{{\tt arXiv:0904.0770}}].

\bibitem{Aaij:2012vr}
{\bf LHCb Collaboration}, R.~Aaij et~al.,
  \href{http://dx.doi.org/10.1007/JHEP02(2013)105}{{\it {Differential branching
  fraction and angular analysis of the $B^{+} \rightarrow K^{+}\mu^{+}\mu^{-}$
  decay}}, } {\em JHEP} {\bf 1302} (2013) 105,
  [\href{http://arxiv.org/abs/1209.4284}{{\tt arXiv:1209.4284}}].

\bibitem{Brodzicka:2012jm}
{\bf Belle Collaboration}, J.~Brodzicka et~al.,
  \href{http://dx.doi.org/10.1093/ptep/pts072}{{\it {Physics Achievements from
  the Belle Experiment}}, } {\em PTEP} {\bf 2012} (2012) 04D001,
  [\href{http://arxiv.org/abs/1212.5342}{{\tt arXiv:1212.5342}}].

\bibitem{Coan:1997ye}
{\bf CLEO Collaboration}, T.~Coan et~al.,
  \href{http://dx.doi.org/10.1103/PhysRevLett.80.1150}{{\it {Flavor - specific
  inclusive B decays to charm}}, } {\em Phys.Rev.Lett.} {\bf 80} (1998)
  1150--1155, [\href{http://arxiv.org/abs/hep-ex/9710028}{{\tt
  hep-ex/9710028}}].

\bibitem{Aaij:2013aka}
{\bf LHCb Collaboration}, R.~Aaij et~al.,
  \href{http://dx.doi.org/10.1103/PhysRevLett.111.101805}{{\it {Measurement of
  the $B^0_s \to \mu^+ \mu^-$ branching fraction and search for $B^0 \to \mu^+
  \mu^-$ decays at the LHCb experiment}}, } {\em Phys.Rev.Lett.} {\bf 111}
  (2013) 101805, [\href{http://arxiv.org/abs/1307.5024}{{\tt
  arXiv:1307.5024}}].

\bibitem{Chatrchyan:2013bka}
{\bf CMS Collaboration}, S.~Chatrchyan et~al.,
  \href{http://dx.doi.org/10.1103/PhysRevLett.111.101804}{{\it {Measurement of
  the B(s) to mu+ mu- branching fraction and search for B0 to mu+ mu- with the
  CMS Experiment}}, } {\em Phys.Rev.Lett.} {\bf 111} (2013) 101804,
  [\href{http://arxiv.org/abs/1307.5025}{{\tt arXiv:1307.5025}}].

\bibitem{Lees:2012te}
{\bf BaBar Collaboration}, J.~Lees et~al.,
  \href{http://dx.doi.org/10.1103/PhysRevD.88.071102}{{\it {Search for a
  low-mass scalar Higgs boson decaying to a tau pair in single-photon decays of
  $\Upsilon(1S)$}}, } {\em Phys.Rev.} {\bf D88} (2013) 071102,
  [\href{http://arxiv.org/abs/1210.5669}{{\tt arXiv:1210.5669}}].

\bibitem{Lees:2012iw}
{\bf BaBar Collaboration}, J.~Lees et~al.,
  \href{http://dx.doi.org/10.1103/PhysRevD.87.031102}{{\it {Search for di-muon
  decays of a low-mass Higgs boson in radiative decays of the $\Upsilon(1S)$}},
  } {\em Phys.Rev.} {\bf D87} (2013) 031102,
  [\href{http://arxiv.org/abs/1210.0287}{{\tt arXiv:1210.0287}}].

\bibitem{Lees:2011wb}
{\bf BaBar Collaboration}, J.~Lees et~al.,
  \href{http://dx.doi.org/10.1103/PhysRevLett.107.221803}{{\it {Search for
  hadronic decays of a light Higgs boson in the radiative decay $\Upsilon \to
  \gamma A^0$}}, } {\em Phys.Rev.Lett.} {\bf 107} (2011) 221803,
  [\href{http://arxiv.org/abs/1108.3549}{{\tt arXiv:1108.3549}}].

\bibitem{Bergsma:1985qz}
{\bf CHARM Collaboration}, F.~Bergsma et~al.,
  \href{http://dx.doi.org/10.1016/0370-2693(85)90400-9}{{\it {Search for Axion
  Like Particle Production in 400 GeV Proton - Copper Interactions}}, } {\em
  Phys.Lett.} {\bf B157} (1985) 458.

\bibitem{Agashe:2014kda}
{\bf Particle Data Group}, K.~Olive et~al.,
  \href{http://dx.doi.org/10.1088/1674-1137/38/9/090001}{{\it {Review of
  Particle Physics}}, } {\em Chin.Phys.} {\bf C38} (2014) 090001.

\bibitem{D'Ambrosio:1996sw}
G.~D'Ambrosio and J.~Portoles,
  \href{http://dx.doi.org/10.1016/S0550-3213(97)00116-8}{{\it {Vector meson
  exchange contributions to $K \rightarrow \pi \gamma \gamma$ and $K_L
  \rightarrow \gamma \ell^+ \ell^-$}}, } {\em Nucl.Phys.} {\bf B492} (1997)
  417--454, [\href{http://arxiv.org/abs/hep-ph/9610244}{{\tt hep-ph/9610244}}].

\bibitem{Feldman:1997qc}
G.~J. Feldman and R.~D. Cousins,
  \href{http://dx.doi.org/10.1103/PhysRevD.57.3873}{{\it {A Unified approach to
  the classical statistical analysis of small signals}}, } {\em Phys.Rev.} {\bf
  D57} (1998) 3873--3889, [\href{http://arxiv.org/abs/physics/9711021}{{\tt
  physics/9711021}}].

\bibitem{Schmidt-Hoberg:2013hba}
K.~Schmidt-Hoberg, F.~Staub, and M.~W. Winkler,
  \href{http://dx.doi.org/10.1016/j.physletb.2013.11.015}{{\it {Constraints on
  light mediators: confronting dark matter searches with B physics}}, } {\em
  Phys.Lett.} {\bf B727} (2013) 506--510,
  [\href{http://arxiv.org/abs/1310.6752}{{\tt arXiv:1310.6752}}].

\bibitem{CMS:2014xfa}
{\bf CMS Collaboration, LHCb Collaboration}, V.~Khachatryan et~al., {\it
  {Observation of the rare $B^0_s\to\mu^+\mu^-$ decay from the combined
  analysis of CMS and LHCb data}},  \href{http://arxiv.org/abs/1411.4413}{{\tt
  arXiv:1411.4413}}.

\bibitem{Buras:2013uqa}
A.~J. Buras, R.~Fleischer, J.~Girrbach, and R.~Knegjens,
  \href{http://dx.doi.org/10.1007/JHEP07(2013)077}{{\it {Probing New Physics
  with the $B_s \rightarrow {\mu}^+ {\mu}^-$ Time-Dependent Rate}}, } {\em
  JHEP} {\bf 1307} (2013) 77, [\href{http://arxiv.org/abs/1303.3820}{{\tt
  arXiv:1303.3820}}].

\bibitem{Blumlein:1990ay}
J.~Blumlein, J.~Brunner, H.~Grabosch, P.~Lanius, S.~Nowak, et~al.,
  \href{http://dx.doi.org/10.1007/BF01548556}{{\it {Limits on neutral light
  scalar and pseudoscalar particles in a proton beam dump experiment}}, } {\em
  Z.Phys.} {\bf C51} (1991) 341--350.

\bibitem{Duffy:1988rw}
M.~Duffy, G.~Fanourakis, R.~Loveless, D.~Reeder, E.~Smith, et~al.,
  \href{http://dx.doi.org/10.1103/PhysRevD.38.2032}{{\it {Neutrino Production
  by 400-{GeV}/c Protons in a Beam-dump Experiment}}, } {\em Phys.Rev.} {\bf
  D38} (1988) 2032.

\bibitem{Kahn:2014sra}
Y.~Kahn, G.~Krnjaic, J.~Thaler, and M.~Toups, {\it {DAEdALUS and Dark Matter}},
   \href{http://arxiv.org/abs/1411.1055}{{\tt arXiv:1411.1055}}.

\bibitem{Bezrukov:2009yw}
F.~Bezrukov and D.~Gorbunov,
  \href{http://dx.doi.org/10.1007/JHEP05(2010)010}{{\it {Light inflaton
  Hunter's Guide}}, } {\em JHEP} {\bf 1005} (2010) 010,
  [\href{http://arxiv.org/abs/0912.0390}{{\tt arXiv:0912.0390}}].

\bibitem{Blumlein:2011mv}
J.~Blumlein and J.~Brunner,
  \href{http://dx.doi.org/10.1016/j.physletb.2011.05.046}{{\it {New Exclusion
  Limits for Dark Gauge Forces from Beam-Dump Data}}, } {\em Phys.Lett.} {\bf
  B701} (2011) 155--159, [\href{http://arxiv.org/abs/1104.2747}{{\tt
  arXiv:1104.2747}}].

\bibitem{Blumlein:2013cua}
J.~Bl�mlein and J.~Brunner,
  \href{http://dx.doi.org/10.1016/j.physletb.2014.02.029}{{\it {New Exclusion
  Limits on Dark Gauge Forces from Proton Bremsstrahlung in Beam-Dump Data}}, }
  {\em Phys.Lett.} {\bf B731} (2014) 320--326,
  [\href{http://arxiv.org/abs/1311.3870}{{\tt arXiv:1311.3870}}].

\bibitem{Soper:2014ska}
D.~E. Soper, M.~Spannowsky, C.~J. Wallace, and T.~M.~P. Tait,
  \href{http://dx.doi.org/10.1103/PhysRevD.90.115005}{{\it {Scattering of Dark
  Particles with Light Mediators}}, } {\em Phys.Rev.} {\bf D90} (2014) 115005,
  [\href{http://arxiv.org/abs/1407.2623}{{\tt arXiv:1407.2623}}].

\bibitem{Essig:2010gu} 
  R.~Essig, R.~Harnik, J.~Kaplan and N.~Toro,
\href{ http://dx.doi.org/10.1103/PhysRevD.82.113008} { {\it  {Discovering New Light States at Neutrino Experiments}}, }
  {\em Phys.Rev.} {\bf D82}, 113008 (2010),
  [\href{http://arxiv.org/abs/1008.0636}{{\tt arXiv:1008.0636}}].
  
\bibitem{Belanger:2013oya}
G.~Belanger, F.~Boudjema, A.~Pukhov, and A.~Semenov,
  \href{http://dx.doi.org/10.1016/j.cpc.2013.10.016}{{\it {micrOMEGAs\_3: A
  program for calculating dark matter observables}}, } {\em
  Comput.Phys.Commun.} {\bf 185} (2014) 960--985,
  [\href{http://arxiv.org/abs/1305.0237}{{\tt arXiv:1305.0237}}].

\bibitem{Gondolo:1990dk}
P.~Gondolo and G.~Gelmini,
  \href{http://dx.doi.org/10.1016/0550-3213(91)90438-4}{{\it {Cosmic abundances
  of stable particles: Improved analysis}}, } {\em Nucl.Phys.} {\bf B360}
  (1991) 145--179.

\bibitem{Chu:2011be}
X.~Chu, T.~Hambye, and M.~H. Tytgat,
  \href{http://dx.doi.org/10.1088/1475-7516/2012/05/034}{{\it {The Four Basic
  Ways of Creating Dark Matter Through a Portal}}, } {\em JCAP} {\bf 1205}
  (2012) 034, [\href{http://arxiv.org/abs/1112.0493}{{\tt arXiv:1112.0493}}].

\bibitem{Hall:2009bx}
L.~J. Hall, K.~Jedamzik, J.~March-Russell, and S.~M. West,
  \href{http://dx.doi.org/10.1007/JHEP03(2010)080}{{\it {Freeze-In Production
  of FIMP Dark Matter}}, } {\em JHEP} {\bf 1003} (2010) 080,
  [\href{http://arxiv.org/abs/0911.1120}{{\tt arXiv:0911.1120}}].

\bibitem{Ericson}
T.~Ericson and W.~Weise, {\em {Pions and Nuclei}}.
\newblock Clarendon Press, 1988.

\bibitem{Bedaque:2009ri}
P.~F. Bedaque, M.~I. Buchoff, and R.~K. Mishra,
  \href{http://dx.doi.org/10.1088/1126-6708/2009/11/046}{{\it {Sommerfeld
  enhancement from Goldstone pseudo-scalar exchange}}, } {\em JHEP} {\bf 0911}
  (2009) 046, [\href{http://arxiv.org/abs/0907.0235}{{\tt arXiv:0907.0235}}].

\bibitem{Bellazzini:2013foa}
B.~Bellazzini, M.~Cliche, and P.~Tanedo,
  \href{http://dx.doi.org/10.1103/PhysRevD.88.083506}{{\it {Effective theory of
  self-interacting dark matter}}, } {\em Phys.Rev.} {\bf D88} (2013) 083506,
  [\href{http://arxiv.org/abs/1307.1129}{{\tt arXiv:1307.1129}}].

\bibitem{Kaplinghat:2013yxa}
M.~Kaplinghat, S.~Tulin, and H.-B. Yu,
  \href{http://dx.doi.org/10.1103/PhysRevD.89.035009}{{\it {Direct Detection
  Portals for Self-interacting Dark Matter}}, } {\em Phys.Rev.} {\bf D89}
  (2014) 035009, [\href{http://arxiv.org/abs/1310.7945}{{\tt
  arXiv:1310.7945}}].

\bibitem{Kouvaris:2014uoa}
C.~Kouvaris, I.~M. Shoemaker, and K.~Tuominen, {\it {Self-Interacting Dark
  Matter through the Higgs Portal}},
  \href{http://arxiv.org/abs/1411.3730}{{\tt arXiv:1411.3730}}.

\bibitem{Felizardo:2011uw}
M.~Felizardo, T.~Girard, T.~Morlat, A.~Fernandes, A.~Ramos, et~al.,
  \href{http://dx.doi.org/10.1103/PhysRevLett.108.201302}{{\it {Final Analysis
  and Results of the Phase II SIMPLE Dark Matter Search}}, } {\em
  Phys.Rev.Lett.} {\bf 108} (2012) 201302,
  [\href{http://arxiv.org/abs/1106.3014}{{\tt arXiv:1106.3014}}].

\bibitem{Archambault:2012pm}
{\bf PICASSO Collaboration}, S.~Archambault et~al.,
  \href{http://dx.doi.org/10.1016/j.physletb.2012.03.078}{{\it {Constraints on
  Low-Mass WIMP Interactions on $^{19}F$ from PICASSO}}, } {\em Phys.Lett.}
  {\bf B711} (2012) 153--161, [\href{http://arxiv.org/abs/1202.1240}{{\tt
  arXiv:1202.1240}}].

\bibitem{Behnke:2012ys}
{\bf COUPP Collaboration}, E.~Behnke et~al.,
  \href{http://dx.doi.org/10.1103/PhysRevD.86.052001,
  10.1103/PhysRevD.90.079902}{{\it {First Dark Matter Search Results from a
  4-kg CF$_3$I Bubble Chamber Operated in a Deep Underground Site}}, } {\em
  Phys.Rev.} {\bf D86} (2012) 052001,
  [\href{http://arxiv.org/abs/1204.3094}{{\tt arXiv:1204.3094}}].

\bibitem{Buchmueller:2014yoa}
O.~Buchmueller, M.~J. Dolan, S.~A. Malik, and C.~McCabe,
  \href{http://dx.doi.org/10.1007/JHEP01(2015)037}{{\it {Characterising dark
  matter searches at colliders and direct detection experiments: Vector
  mediators}}, } {\em JHEP} {\bf 1501} (2015) 037,
  [\href{http://arxiv.org/abs/1407.8257}{{\tt arXiv:1407.8257}}].

\bibitem{Yellin:2002xd}
S.~Yellin, \href{http://dx.doi.org/10.1103/PhysRevD.66.032005}{{\it {Finding an
  upper limit in the presence of unknown background}}, } {\em Phys.Rev.} {\bf
  D66} (2002) 032005, [\href{http://arxiv.org/abs/physics/0203002}{{\tt
  physics/0203002}}].

\bibitem{Feldstein:2014ufa}
B.~Feldstein and F.~Kahlhoefer, {\it {Quantifying (dis)agreement between direct
  detection experiments in a halo-independent way}},
  \href{http://arxiv.org/abs/1409.5446}{{\tt arXiv:1409.5446}}.

\bibitem{Fitzpatrick:2012ix}
A.~L. Fitzpatrick, W.~Haxton, E.~Katz, N.~Lubbers, and Y.~Xu,
  \href{http://dx.doi.org/10.1088/1475-7516/2013/02/004}{{\it {The Effective
  Field Theory of Dark Matter Direct Detection}}, } {\em JCAP} {\bf 1302}
  (2013) 004, [\href{http://arxiv.org/abs/1203.3542}{{\tt arXiv:1203.3542}}].

\bibitem{McCabe:2013kea}
C.~McCabe, \href{http://dx.doi.org/10.1088/1475-7516/2014/02/027}{{\it {The
  Earth's velocity for direct detection experiments}}, } {\em JCAP} {\bf 1402}
  (2014) 027, [\href{http://arxiv.org/abs/1312.1355}{{\tt arXiv:1312.1355}}].

\bibitem{Lee:2013xxa}
S.~K. Lee, M.~Lisanti, and B.~R. Safdi,
  \href{http://dx.doi.org/10.1088/1475-7516/2013/11/033}{{\it {Dark-Matter
  Harmonics Beyond Annual Modulation}}, } {\em JCAP} {\bf 1311} (2013) 033,
  [\href{http://arxiv.org/abs/1307.5323}{{\tt arXiv:1307.5323}}].

\bibitem{McCabe:2010zh}
C.~McCabe, \href{http://dx.doi.org/10.1103/PhysRevD.82.023530}{{\it {The
  Astrophysical Uncertainties Of Dark Matter Direct Detection Experiments}}, }
  {\em Phys.Rev.} {\bf D82} (2010) 023530,
  [\href{http://arxiv.org/abs/1005.0579}{{\tt arXiv:1005.0579}}].

\bibitem{Fermi}
B.~Anderson, ``{A Search for Dark Matter Annihilation in Dwarf Spheroidal
  Galaxies with Pass 8 Data}.'' {Presented at the 5th Fermi Symposium, Nagoya,
  2014}.

\bibitem{Ackermann:2013yva}
{\bf Fermi-LAT Collaboration}, M.~Ackermann et~al.,
  \href{http://dx.doi.org/10.1103/PhysRevD.89.042001}{{\it {Dark matter
  constraints from observations of 25 Milky Way satellite galaxies with the
  Fermi Large Area Telescope}}, } {\em Phys.Rev.} {\bf D89} (2014) 042001,
  [\href{http://arxiv.org/abs/1310.0828}{{\tt arXiv:1310.0828}}].

\bibitem{Bernabei:2010mq}
{\bf DAMA Collaboration, LIBRA Collaboration}, R.~Bernabei et~al.,
  \href{http://dx.doi.org/10.1140/epjc/s10052-010-1303-9}{{\it {New results
  from DAMA/LIBRA}}, } {\em Eur.Phys.J.} {\bf C67} (2010) 39--49,
  [\href{http://arxiv.org/abs/1002.1028}{{\tt arXiv:1002.1028}}].

\bibitem{Bernabei:2013xsa}
R.~Bernabei, P.~Belli, F.~Cappella, V.~Caracciolo, S.~Castellano, et~al.,
  \href{http://dx.doi.org/10.1140/epjc/s10052-013-2648-7}{{\it {Final model
  independent result of DAMA/LIBRA-phase1}}, } {\em Eur.Phys.J.} {\bf C73}
  (2013) 2648, [\href{http://arxiv.org/abs/1308.5109}{{\tt arXiv:1308.5109}}].

\bibitem{Bozorgnia:2010xy}
N.~Bozorgnia, G.~B. Gelmini, and P.~Gondolo,
  \href{http://dx.doi.org/10.1088/1475-7516/2010/11/019}{{\it {Channeling in
  direct dark matter detection I: channeling fraction in NaI (Tl) crystals}}, }
  {\em JCAP} {\bf 1011} (2010) 019, [\href{http://arxiv.org/abs/1006.3110}{{\tt
  arXiv:1006.3110}}].

\bibitem{Williams:2012pz}
A.~J. Williams, C.~Boehm, S.~M. West, and D.~A. Vasquez,
  \href{http://dx.doi.org/10.1103/PhysRevD.86.055018}{{\it {Regenerating WIMPs
  in the Light of Direct and Indirect Detection}}, } {\em Phys.Rev.} {\bf D86}
  (2012) 055018, [\href{http://arxiv.org/abs/1204.3727}{{\tt
  arXiv:1204.3727}}].

\bibitem{Nussinov:1985xr}
S.~Nussinov, \href{http://dx.doi.org/10.1016/0370-2693(85)90689-6}{{\it
  {Technocosmology: Could a technibaryon excess provide a 'natural' missing
  mass candidate?}}, } {\em Phys.Lett.} {\bf B165} (1985) 55.

\bibitem{Griest:1986yu}
K.~Griest and D.~Seckel,
  \href{http://dx.doi.org/10.1016/0550-3213(87)90293-8}{{\it {Cosmic Asymmetry,
  Neutrinos and the Sun}}, } {\em Nucl.Phys.} {\bf B283} (1987) 681.

\bibitem{Barr:1990ca}
S.~M. Barr, R.~S. Chivukula, and E.~Farhi,
  \href{http://dx.doi.org/10.1016/0370-2693(90)91661-T}{{\it {Electroweak
  Fermion Number Violation and the Production of Stable Particles in the Early
  Universe}}, } {\em Phys.Lett.} {\bf B241} (1990) 387--391.

\bibitem{Kaplan:1991ah}
D.~B. Kaplan, \href{http://dx.doi.org/10.1103/PhysRevLett.68.741}{{\it {A
  Single explanation for both the baryon and dark matter densities}}, } {\em
  Phys.Rev.Lett.} {\bf 68} (1992) 741--743.

\bibitem{Kaplan:2009ag}
D.~E. Kaplan, M.~A. Luty, and K.~M. Zurek,
  \href{http://dx.doi.org/10.1103/PhysRevD.79.115016}{{\it {Asymmetric Dark
  Matter}}, } {\em Phys.Rev.} {\bf D79} (2009) 115016,
  [\href{http://arxiv.org/abs/0901.4117}{{\tt arXiv:0901.4117}}].

\bibitem{DelNobile:2013sia}
M.~Cirelli, E.~Del~Nobile, and P.~Panci,
  \href{http://dx.doi.org/10.1088/1475-7516/2013/10/019}{{\it {Tools for
  model-independent bounds in direct dark matter searches}}, } {\em JCAP} {\bf
  1310} (2013) 019, [\href{http://arxiv.org/abs/1307.5955}{{\tt
  arXiv:1307.5955}}].

\bibitem{Sjostrand:2007gs}
T.~Sjostrand, S.~Mrenna, and P.~Z. Skands,
  \href{http://dx.doi.org/10.1016/j.cpc.2008.01.036}{{\it {A Brief Introduction
  to PYTHIA 8.1}}, } {\em Comput.Phys.Commun.} {\bf 178} (2008) 852--867,
  [\href{http://arxiv.org/abs/0710.3820}{{\tt arXiv:0710.3820}}].

\bibitem{Marciano:1996wy}
W.~Marciano and Z.~Parsa, \href{http://dx.doi.org/10.1103/PhysRevD.53.R1}{{\it
  {Rare kaon decays with 'missing energy'}}, } {\em Phys.Rev.} {\bf D53} (1996)
  1--5.

\bibitem{Grossman:1997pa}
Y.~Grossman, Y.~Nir, and R.~Rattazzi,
  \href{http://dx.doi.org/10.1142/9789812812667_0011}{{\it {CP violation beyond
  the standard model}}, } {\em Adv.Ser.Direct.High Energy Phys.} {\bf 15}
  (1998) 755--794, [\href{http://arxiv.org/abs/hep-ph/9701231}{{\tt
  hep-ph/9701231}}].

\bibitem{He:2006uu}
X.-G. He, J.~Tandean, and G.~Valencia,
  \href{http://dx.doi.org/10.1103/PhysRevD.74.115015}{{\it {Light Higgs
  production in hyperon decay}}, } {\em Phys.Rev.} {\bf D74} (2006) 115015,
  [\href{http://arxiv.org/abs/hep-ph/0610274}{{\tt hep-ph/0610274}}].


\bibitem{Ali:1999mm}
A.~Ali, P.~Ball, L.~Handoko, and G.~Hiller,
  \href{http://dx.doi.org/10.1103/PhysRevD.61.074024}{{\it {A Comparative study
  of the decays $B \to (K, \, K^{*}) \ell^+ \ell^-$ in standard model and
  supersymmetric theories}}, } {\em Phys.Rev.} {\bf D61} (2000) 074024,
  [\href{http://arxiv.org/abs/hep-ph/9910221}{{\tt hep-ph/9910221}}].

\bibitem{Ball:2004ye}
P.~Ball and R.~Zwicky, \href{http://dx.doi.org/10.1103/PhysRevD.71.014015}{{\it
  {New results on $B \rightarrow \pi, \, K, \, \eta$ decay formfactors from
  light-cone sum rules}}, } {\em Phys.Rev.} {\bf D71} (2005) 014015,
  [\href{http://arxiv.org/abs/hep-ph/0406232}{{\tt hep-ph/0406232}}].

\bibitem{Altmannshofer:2011gn}
W.~Altmannshofer, P.~Paradisi, and D.~M. Straub,
  \href{http://dx.doi.org/10.1007/JHEP04(2012)008}{{\it {Model-Independent
  Constraints on New Physics in $b \rightarrow s$ Transitions}}, } {\em JHEP}
  {\bf 1204} (2012) 008, [\href{http://arxiv.org/abs/1111.1257}{{\tt
  arXiv:1111.1257}}].

\bibitem{DeBruyn:2012wj}
K.~De~Bruyn, R.~Fleischer, R.~Knegjens, P.~Koppenburg, M.~Merk, et~al.,
  \href{http://dx.doi.org/10.1103/PhysRevD.86.014027}{{\it {Branching Ratio
  Measurements of $B_s$ Decays}}, } {\em Phys.Rev.} {\bf D86} (2012) 014027,
  [\href{http://arxiv.org/abs/1204.1735}{{\tt arXiv:1204.1735}}].

\bibitem{Wilczek:1977zn}
F.~Wilczek, \href{http://dx.doi.org/10.1103/PhysRevLett.39.1304}{{\it {Decays
  of Heavy Vector Mesons Into Higgs Particles}}, } {\em Phys.Rev.Lett.} {\bf
  39} (1977) 1304.

\bibitem{Nason:1986tr}
P.~Nason, \href{http://dx.doi.org/10.1016/0370-2693(86)90721-5}{{\it {{QCD}
  Radiative Corrections to $\Upsilon$ Decay Into Scalar$+\gamma$ and
  Pseudoscalar$+\gamma$}}, } {\em Phys.Lett.} {\bf B175} (1986) 223.

\bibitem{Haber:1987ua}
H.~E. Haber, A.~S. Schwarz, and A.~E. Snyder,
  \href{http://dx.doi.org/10.1016/0550-3213(87)90584-0}{{\it {Hunting the Higgs
  in $B$ Decays}}, } {\em Nucl.Phys.} {\bf B294} (1987) 301.

\bibitem{Haber:1978jt}
H.~Haber, G.~L. Kane, and T.~Sterling,
  \href{http://dx.doi.org/10.1016/0550-3213(79)90225-6}{{\it {The Fermion Mass
  Scale and Possible Effects of Higgs Bosons on Experimental Observables}}, }
  {\em Nucl.Phys.} {\bf B161} (1979) 493.

\bibitem{Ellis:1979jy}
J.~R. Ellis, M.~Gaillard, D.~V. Nanopoulos, and C.~T. Sachrajda,
  \href{http://dx.doi.org/10.1016/0370-2693(79)91122-5}{{\it {Is the Mass of
  the Higgs Boson About 10 GeV?}}, } {\em Phys.Lett.} {\bf B83} (1979) 339.

\bibitem{Brambilla:2007cz}
N.~Brambilla, X.~Garcia~i Tormo, J.~Soto, and A.~Vairo,
  \href{http://dx.doi.org/10.1103/PhysRevD.75.074014}{{\it {Extraction of
  $\alpha_s$ from radiative $\Upsilon(1S)$ decays}}, } {\em Phys.Rev.} {\bf
  D75} (2007) 074014, [\href{http://arxiv.org/abs/hep-ph/0702079}{{\tt
  hep-ph/0702079}}].

\bibitem{Aznaurian:1986hi}
I.~Aznaurian, S.~Grigorian, and S.~G. Matinyan, {\it {Relativistic Effects in
  $V \rightarrow H_0 \gamma$ Decay}},  {\em JETP Lett.} {\bf 43} (1986) 646.

\bibitem{Faldt:1987zu}
G.~Faldt, P.~Osland, and T.~T. Wu,
  \href{http://dx.doi.org/10.1103/PhysRevD.38.164}{{\it {Relativistic Theory of
  the Decay of $\Upsilon$ Into Higgs + Photon}}, } {\em Phys.Rev.} {\bf D38}
  (1988) 164.

\bibitem{Djouadi:2005gj}
A.~Djouadi, \href{http://dx.doi.org/10.1016/j.physrep.2007.10.005}{{\it {The
  Anatomy of electro-weak symmetry breaking. II. The Higgs bosons in the
  minimal supersymmetric model}}, } {\em Phys.Rept.} {\bf 459} (2008) 1--241,
  [\href{http://arxiv.org/abs/hep-ph/0503173}{{\tt hep-ph/0503173}}].

\bibitem{Leutwyler:1989tn}
H.~Leutwyler and M.~A. Shifman,
  \href{http://dx.doi.org/10.1016/0370-2693(89)91730-9}{{\it {Goldstone bosons
  generate peculiar conformal anomalies}}, } {\em Phys.Lett.} {\bf B221} (1989)
  384.

\bibitem{Ellwanger:2004xm}
U.~Ellwanger, J.~F. Gunion, and C.~Hugonie,
  \href{http://dx.doi.org/10.1088/1126-6708/2005/02/066}{{\it {NMHDECAY: A
  Fortran code for the Higgs masses, couplings and decay widths in the NMSSM}},
  } {\em JHEP} {\bf 0502} (2005) 066,
  [\href{http://arxiv.org/abs/hep-ph/0406215}{{\tt hep-ph/0406215}}].

\bibitem{Cheng:2012qr}
H.-Y. Cheng and C.-W. Chiang,
  \href{http://dx.doi.org/10.1007/JHEP07(2012)009}{{\it {Revisiting Scalar and
  Pseudoscalar Couplings with Nucleons}}, } {\em JHEP} {\bf 1207} (2012) 009,
  [\href{http://arxiv.org/abs/1202.1292}{{\tt arXiv:1202.1292}}].

\end{thebibliography}

\providecommand{\href}[2]{#2}\begingroup\raggedright\endgroup

\end{document}